\documentclass[final]{jfp-epi}
\usepackage{tikz-cd}
\pgfrealjobname{main}
\usepackage{quoting}
\usepackage{hyperref}
\usepackage[utf8]{inputenc}
\usepackage{cleveref}

\makeatletter
\@ifundefined{lhs2tex.lhs2tex.sty.read}%
  {\@namedef{lhs2tex.lhs2tex.sty.read}{}%
   \newcommand\SkipToFmtEnd{}%
   \newcommand\EndFmtInput{}%
   \long\def\SkipToFmtEnd#1\EndFmtInput{}%
  }\SkipToFmtEnd

\newcommand\ReadOnlyOnce[1]{\@ifundefined{#1}{\@namedef{#1}{}}\SkipToFmtEnd}
\usepackage{amstext}
\usepackage{stmaryrd}
\DeclareFontFamily{OT1}{cmtex}{}
\DeclareFontShape{OT1}{cmtex}{m}{n}
  {<5><6><7><8>cmtex8
   <9>cmtex9
   <10><10.95><12><14.4><17.28><20.74><24.88>cmtex10}{}
\DeclareFontShape{OT1}{cmtex}{m}{it}
  {<-> ssub * cmtt/m/it}{}

\DeclareFontShape{OT1}{cmtt}{bx}{n}
  {<5><6><7><8>cmtt8
   <9>cmbtt9
   <10><10.95><12><14.4><17.28><20.74><24.88>cmbtt10}{}
\DeclareFontShape{OT1}{cmtex}{bx}{n}
  {<-> ssub * cmtt/bx/n}{}

\newcommand{\Conid}[1]{\mathit{#1}}
\newcommand{\Varid}[1]{\mathit{#1}}
\newcommand{\anonymous}{\kern0.06em \vbox{\hrule\@width.5em}}

\renewcommand{\leq}{\leqslant}

\usepackage{polytable}

\@ifundefined{mathindent}%
  {\newdimen\mathindent\mathindent\leftmargini}%
  {}%

\def\resethooks{%
  \global\let\SaveRestoreHook\empty
  \global\let\ColumnHook\empty}
\newcommand*{\savecolumns}[1][default]%
  {\g@addto@macro\SaveRestoreHook{\savecolumns[#1]}}
\newcommand*{\restorecolumns}[1][default]%
  {\g@addto@macro\SaveRestoreHook{\restorecolumns[#1]}}
\newcommand*{\aligncolumn}[2]%
  {\g@addto@macro\ColumnHook{\column{#1}{#2}}}

\resethooks

\newcommand{\onelinecommentchars}{\quad-{}- }
\newcommand{\commentbeginchars}{\enskip\{-}
\newcommand{\commentendchars}{-\}\enskip}

\newcommand{\visiblecomments}{%
  \let\onelinecomment=\onelinecommentchars
  \let\commentbegin=\commentbeginchars
  \let\commentend=\commentendchars}

\newcommand{\invisiblecomments}{%
  \let\onelinecomment=\empty
  \let\commentbegin=\empty
  \let\commentend=\empty}

\visiblecomments

\newlength{\blanklineskip}
\newcommand{\hsindent}[1]{\quad}
\let\hspre\empty
\let\hspost\empty

\EndFmtInput
\makeatother
\ReadOnlyOnce{polycode.fmt}%
\makeatletter

\newcommand{\hsnewpar}[1]%
  {{\parskip=0pt\parindent=0pt\par\vskip #1\noindent}}

\newcommand{\hscodestyle}{}

\newcommand{\sethscode}[1]%
  {\expandafter\let\expandafter\hscode\csname #1\endcsname
   \expandafter\let\expandafter\endhscode\csname end#1\endcsname}

  {\par\noindent
   \advance\leftskip\mathindent
   \hscodestyle
   \let\\=\@normalcr
   \let\hspre\(\let\hspost\)%
   \pboxed}%
  {\endpboxed\)%
   \par\noindent
   \ignorespacesafterend}

  {\hsnewpar\abovedisplayskip
   \advance\leftskip\mathindent
   \hscodestyle
   \let\hspre\(\let\hspost\)%
   \pboxed}%
  {\endpboxed%
   \hsnewpar\belowdisplayskip
   \ignorespacesafterend}

  {\hsnewpar\abovedisplayskip
   \advance\leftskip\mathindent
   \hscodestyle
   \let\\=\@normalcr
   \(\pboxed}%
  {\endpboxed\)%
   \hsnewpar\belowdisplayskip
   \ignorespacesafterend}

\newcommand{\plainhs}{\sethscode{plainhscode}}

\plainhs

  {\hsnewpar\abovedisplayskip
   \advance\leftskip\mathindent
   \hscodestyle
   \let\\=\@normalcr
   \(\parray}%
  {\endparray\)%
   \hsnewpar\belowdisplayskip
   \ignorespacesafterend}

  {\parray}{\endparray}

  {\(\parray}{\endparray\)}

\def\codeframewidth{\arrayrulewidth}
\RequirePackage{calc}

  {\parskip=\abovedisplayskip\par\noindent
   \hscodestyle
   \arrayrulewidth=\codeframewidth
   \tabular{@{}|p{\linewidth-2\arraycolsep-2\arrayrulewidth-2pt}|@{}}%
   \hline\framedhslinecorrect\\{-1.5ex}%
   \let\endoflinesave=\\
   \let\\=\@normalcr
   \(\pboxed}%
  {\endpboxed\)%
   \framedhslinecorrect\endoflinesave{.5ex}\hline
   \endtabular
   \parskip=\belowdisplayskip\par\noindent
   \ignorespacesafterend}

\newcommand{\framedhslinecorrect}[2]%
  {#1[#2]}

  {\(\def\column##1##2{}%
   \let\>\undefined\let\<\undefined\let\\\undefined
   \newcommand\>[1][]{}\newcommand\<[1][]{}\newcommand\\[1][]{}%
   \def\fromto##1##2##3{##3}%
   }{\) }%

\newenvironment{joincode}%
  {\let\orighscode=\hscode
   \let\origendhscode=\endhscode
   \def\endhscode{\def\hscode{\endgroup\def\@currenvir{hscode}\\}\begingroup}
   \orighscode\def\hscode{\endgroup\def\@currenvir{hscode}}}%
  {\origendhscode
   \global\let\hscode=\orighscode
   \global\let\endhscode=\origendhscode}%

\makeatother
\EndFmtInput
\def\doubleequals{\mathrel{\unitlength 0.01em
  \begin{picture}(78,40)
    \put(7,34){\line(1,0){25}} \put(45,34){\line(1,0){25}}
    \put(7,14){\line(1,0){25}} \put(45,14){\line(1,0){25}}
  \end{picture}}}

\DeclareMathAlphabet{\mathcal}{OMS}{cmsy}{m}{n}

\newtheorem{lemma}{Lemma}

\newcommand{\Rational}{\mathbb{Q}}
\newcommand{\Real}{\mathbb{R}}

\newcommand{\CalTime}{\mathcal{T}}
\newcommand{\ee}{\doteq}

\def\commentbegin{\quad\{\ }
\def\commentend{\}}

\providecommand{\DONE}[1]{}

\begin{document}

\jfpVolume{36}
\jfpArticle{3}
\jfpDOI{10.46298/jfp.17762}
\jfpYear{2026}
\received[Submitted]{August 2023}
\received[accepted]{March 2026}

\title{Types, equations, dimensions and the Pi theorem}

\author{Nicola Botta}
\orcid{0000-0002-8923-2734}
\affiliation{%
\institution{Potsdam Institute for Climate Impact Research}
\city{Potsdam}
\state{Brandenburg}
\country{Germany}
\institution{\ and Chalmers University of Technology and University of Gothenburg}
\city{Göteborg}
\country{Sweden}
\authoremail{botta@pik-potsdam.de}
}
\author{Patrik Jansson}
\orcid{0000-0003-3078-1437}
\affiliation{%
\institution{Chalmers University of Technology and University of Gothenburg}
\city{Göteborg}
\country{Sweden}
\authoremail{patrikj@chalmers.se}
}

\begin{abstract}
  The languages of mathematical physics and modelling are endowed with a
  rich ``grammar of dimensions'' that common abstractions of programming
  languages fail to represent. We propose a dependently typed
  domain-specific language (embedded in Idris) that captures this
  grammar. We apply it to formalize basic notions of dimensional
  analysis: those of dimension function, physical quantity, homomorphic
  measurement, the covariance principle and Buckingham's Pi theorem. We
  hope that the language makes mathematical physics more accessible to
  computer scientists and functional programming more palatable to
  modellers and physicists.
\end{abstract}

\maketitle


\setlength{\mathindent}{1em}
\newcommand{\fixlengths}{\setlength{\abovedisplayskip}{6pt plus 1pt minus 1pt}\setlength{\belowdisplayskip}{6pt plus 1pt minus 1pt}}
\renewcommand{\hscodestyle}{\small\fixlengths}
\fixlengths

\section{Introduction}
\label{section:intro}

\paragraph*{Motivation.}

\ The main motivation for this work comes from a failure. During more
than one decade, the authors have been advocating
mathematical specifications, type-driven analysis and functional
programming (FP) as methodologies to better understand, specify and
solve problems in climate impact research, climate policy advice and,
by large, global systems science 
\citep{botta+al2011a, ionescujansson:LIPIcs:2013:3899,
  2017_Botta_Jansson_Ionescu, esd-9-525-2018, ijb:ISoLA:2018,
  Botta2023MatterMost}.

Alas, after ten years of advocacy and intense collaborations, we have
hardly been able to convince any climate modeller of the usefulness of
FP, let alone convert them to ``thinking functionally'' with Haskell
\citep{bird2014thinking}, Agda~\citep{norell2007thesis},
Idris~\citep{idrisbook}, or Rocq~\citep{RocqProofAssistant}.
Have we just done a bad job or is this failure a symptom of a deeper
problem?

\paragraph*{There is no need for FP in mathematical physics, is there?}

\ Physicists and modellers are well trained in exploiting established
numerical libraries \citep{NAG, GSL, NumPy} and frameworks
\citep{OpenFOAM, AMReX_JOSS}
%
%
%
for approximating solutions of (ordinary, partial, stochastic)
differential equations efficiently, implementing large computer-based
models in FORTRAN, C, C++, Java, Python, or Julia and
testing model predictions against analytical solutions or observations.

In engineering and in many physical sciences, this methodology has led
to reliable computations and to computer-based modelling almost fully
replacing physical modelling: for many applications, running computer
programs is more flexible and much cheaper than running wind tunnels or
full-scale experiments.

But there are important research areas in which empirical validations
are beyond reach and the predictive capability of computer-based models
is poorly known and needs to be questioned.
In climate science but also in plasma physics, for example, it is
simply impossible (or just too dangerous or too expensive) to test the
correctness of computations empirically.
It is not possible to study the effectiveness of a policy designed to
reduce greenhouse gas (GHG) emissions without implementing that policy;
or, as argued by \citet{lucarini2004}, ``the usual Galilean scientific
validation criteria do not apply to climate science''. In much the same
way, plasma physicists and engineers cannot afford to damage hundreds of
experimental tokamak fusion reactors to assess and validate optimal
control options for such devices \citep{HOPPE2021108098,
  Pusztai2023BayesOptMMI}.

In these domains, scientists need methodologies that bring confidence
that their computations are correct well before such computations can
actually be applied to real world systems.
Formal specification is the key to both testing programs and
showing the \emph{absence} of errors in computations
\citep{ionescujansson:LIPIcs:2013:3899} and dependently typed FP
languages have reached enough expressive power to support formulating
very precise specifications.
And yet climate modellers and physicists have, by and large,
stayed away from FP languages. Why so?

\paragraph*{Educational gaps.}

\ It is probably fair to say that most physicists and modellers have never
heard of mathematical program specifications, not to mention FP and
dependently typed languages.
In much the same way, most computer scientists have hardly been exposed
to the language of mathematical physics, say, for example, that of
\citet{courant89, arnold1989mathematical, barenblatt1996scaling,
  Kuznetsov:1998:EAB:289919}.

Originally very close to elementary set theory and calculus, this
language has evolved over the last decades and fragmented into a
multitude of dialects or DSLs,
driven perhaps by the pervasive usage of imperative programming and
computer-based modelling.
Common traits of these dialects are the limited usage of currying and
higher order functions, the overloading of the equality sign, the lack
of referential transparency and explicit type information (although
mnemonic rules are often introduced for encoding such information like
in $x_{[0,T]}$ instead of \ensuremath{\Varid{x}\ \mathop{:}\ [\mskip1.5mu \mathrm{0},\Conid{T}\mskip1.5mu]\to \Real}) and the usage of
parentheses to denote both function application and function composition
as in $\dot{x} = f(x)$ instead of $\forall t, \ \dot{x}(t) = f(x(t))$ or, in
point-free notation, \ensuremath{\dot{x}\mathrel{=}\Varid{f}\mathbin{\circ}\Varid{x}}.
\DONE{The examples can be hard to read at this stage (but that may be OK with a forward pointer).}

These DSLs represent a major difficulty for computer scientists and
systematic efforts have been undertaken by one of the authors to make
them more accessible to computer science students \citep{Ionescu_2016,
  JanssonIonescuBernardyDSLsofMathBook2022, bernardy_jansson_2025_tensors} and
improve the dialogue between the computational sciences and the physical
sciences. Our paper is also a contribution to such dialogue.

We argue that the
DSLs of mathematical physics are endowed with a rich but hidden
``grammar of dimensions'' that (functional) programming languages have
failed to exploit or even recognize.
This grammar informs important notions of consistency and of correctness
which, in turn, are the bread and butter of program analysis, testing
and derivation.

From this perspective, it is not very surprising that physicists and
modellers have hardly been interested in FP.
Standard FP abstractions emphasize type annotations that do not matter
to physicists and modellers (for the climate scientist, all functions
are, bluntly speaking, of type $\mathbb{R}^m \to \mathbb{R}^n$ for some
natural numbers $m$ and $n$), while at the same time failing to highlight
differences that do matter like the one between a
\emph{length} and a \emph{time}.
We hope that this work will also help make FP a bit more palatable to
physicists and modellers.

\paragraph*{Outline.}

\ In \cref{section:equations} we briefly review the role of equations,
laws, types and \emph{dimensions} in computer science, mathematical
physics and modelling.

In \cref{section:dimensions} we discuss the ideas of dimension,
\emph{physical quantity}, and \emph{units of measurement}
informally. This is mainly meant to guide the computer scientist out of
her comfort zone but also to answer a question that should be of
interest also to readers who are familiar with modelling: ``what does it
mean for a parameter or for a variable to have a dimension?''

\Cref{section:pi} is a short account of similarity theory
\citep{Buckingham1914, Rayleigh1915, bridgman1922} and of Buckingham's
Pi theorem, mainly following
\citet[Section 1]{barenblatt1996scaling}. This is going to be new ground for most
computer scientists but, again, we hope to also provide a new angle to
modellers who are young and have therefore mainly been imprinted with
computer-based modelling.

In \cref{section:dsl1} we introduce a minimal domain specific language
for dimensionally consistent programming in Idris. We formalize the
notions of dimension function, physical quantity, homomorphic
measurement and dimensional (in)dependence.

In \cref{section:piexplained1}, 
we apply the DSL to formulate the
covariance principle (principle of relativity of measurements that is,
there is no privileged system of units of measurement) and the Pi
theorem in type theory.
These formulations are, on the one hand, a benchmark for the DSL
developed in \cref{section:dsl1}. On the other hand, they are a way to
understand covariance and the Pi theorem and to make these notions more
accessible to computer scientists. Perhaps not surprisingly, the theorem
as discussed in dimensional analysis textbooks is not implementable. We
discuss its non-constructive nature in \cref{subsection:piexplained1.2}
and present a methodology for ``applying'' the theorem and constructing
functions that provably respect the covariance principle in
\cref{subsection:piexplained1.3} .

In \cref{section:generalization} we discuss possible extensions and
generalizations of the DSL from \cref{section:dsl1} and give a meaning
to the equations and physical laws from \cref{section:equations}, while
\cref{section:conclusions} wraps up and discusses links between
dimensional analysis and more general relativity principles.

All sections of this manuscript are available as literate Idris code in our gitlab repository \citep{Pi2023}.
%
The code associated with \cref{section:dsl1,section:piexplained1} is also available in
non-literate Idris and Agda form in the \texttt{idris} and \texttt{agda}
folders of the same repository.

\paragraph*{Related work.}

\ Dimensional analysis (DA) \citep{bridgman1922,
  barenblatt1996scaling, alma991010162519703414,jonsson2014dimensional},
is closely connected with the theory
of physical similarity: under which conditions is it possible to
translate observations made on a scaled model of a physical system to
the system itself. 
This is captured in the fundamental principles of invariance and
relativity (of units of measurement), as described by
\citet{arnold1989mathematical}.

The theory of physical similarity was formulated well before the advent
of digital computers and programming languages \citep{Buckingham1914,
  Buckingham1915, Rayleigh1915, bridgman1922} and its formalizations
have been rare \citep{quade1961, hassler1, hassler2}.
With the advent of digital computers and massive numerical
simulations, physical modelling has been almost completely replaced by
computer-based modelling, mainly for economic reasons, and the theory
of physical similarity and DA are not any longer an integral component
of the education of physicists, modellers and data analysts%
\footnote{But Bridgman's work has been republished in 2007 by
  Kessinger Publishing and in 2018 by Forgotten Books and DA still
  plays a crucial role in the analysis of computer-based models in climate
  science \citep{esd-11-281-2020, esd-12-63-2021, esd-13-879-2022,
    cp-2023-30}.}.

The notions of invariance (with respect to a group of transformations) and
relativity (e.g., of units of measurement) in physics are similar to the
notions of parametricity and polymorphism in computer science and thus
it is perhaps not surprising that, as programming languages have gained
more and more expressive power, mathematicians and computer scientists
have started ``rediscovering'' DA, see \citet{10.1145/263699.263761,
  10.48550/arxiv.1107.4520, dimension-models}.
More recently the idea that dependent types can be applied to enforce
the dimensional consistency of expressions involving scalar physical
quantities has been generalized to expressions with vectors, tensors
and other derived quantities by \citet{doi:10.1142/9789811242380_0020}.
We return to these papers with a bit more details in
\cref{subsection:relatedwork}.

Dependent types can certainly be applied to enforce the dimensional
consistency of expressions and libraries for annotating values of
standard types with dimensional information are available in most
programming languages \citep{units, dimensional, Unitful, pint, astropy,
  Boost.Units} and since quite some time \citep{10.1093/comjnl/26.4.366}.

But dependently typed languages can do more. The type checker of Idris,
for example, can effectively assist the implementation of verified
programs by interactively resolving the types of holes, suggesting
tactics and recommending type consistent implementations. Similar
support is available in other languages based on
intensional type theory.

A DSL built on top of a dependently typed language that supports
expressing Buckingham's Pi theorem should in principle be able to assist
the interactive implementation of dimensionally consistent programs.
It should support the programmer formulating the question of how a
function that computes a force, say \ensuremath{\Conid{F}}, may depend on arguments \ensuremath{\Varid{m}} and
\ensuremath{\Varid{a}} representing mass and acceleration, leverage on the type system
of the host language and automatically derive \ensuremath{\Conid{F}\;\Varid{m}\;\Varid{a}\mathrel{=}\alpha\ensuremath{\cdot}\Varid{m}\ensuremath{\cdot}\Varid{a}} (for some \ensuremath{\alpha}).
\DONE{Add a constant factor? (that is what the Pi theorem would give)}
The work presented here is a first step in this direction. We are not
yet there and in \cref{section:conclusions} we discuss which steps are
left.

\section{Equations, physical laws and types}
\label{section:equations}

In the preface to the second edition of ``Programming from
specifications'' \citep{DBLP:books/daglib/0073499}, Carroll Morgan
starts with the observation that, in mathematics, $x^2 = 1$ is an
equation and that \ensuremath{\Varid{x}\mathrel{=}\mathrm{1}} and \ensuremath{\Varid{x}\mathrel{=}\mathbin{-}\mathrm{1}} are equations too.
He then goes on to point out that, because of the relationships between
these three equations (the implications \ensuremath{\Varid{x}\mathrel{=}\mathrm{1}} $\Rightarrow$ $x^2 = 1$ and
\ensuremath{\Varid{x}\mathrel{=}\mathbin{-}\mathrm{1}} $\Rightarrow$ $x^2 = 1$) and because \ensuremath{\Varid{x}\mathrel{=}\mathrm{1}} and \ensuremath{\Varid{x}\mathrel{=}\mathbin{-}\mathrm{1}}
define the value of \ensuremath{\Varid{x}} ``without further calculation'', these two
equations are called \emph{solutions} of $x^2 = 1$.

Thus equations in mathematics sometimes represent \emph{problems}. In
dependently typed languages, these problems can be formulated explicitly
and the resulting expressions can be checked for consistency.
For example, in Idris 
one can specify the problem of
finding a real number $x$ whose square is $1$ as
\begin{hscode}\SaveRestoreHook
\column{B}{@{}>{\hspre}l<{\hspost}@{}}%
\column{3}{@{}>{\hspre}l<{\hspost}@{}}%
\column{10}{@{}>{\hspre}c<{\hspost}@{}}%
\column{10E}{@{}l@{}}%
\column{13}{@{}>{\hspre}l<{\hspost}@{}}%
\column{E}{@{}>{\hspre}l<{\hspost}@{}}%
\>[3]{}\Varid{x}{}\<[10]%
\>[10]{}\ \mathop{:}\ {}\<[10E]%
\>[13]{}\Real{}\<[E]%
\\
\>[3]{}\Varid{xSpec}{}\<[10]%
\>[10]{}\ \mathop{:}\ {}\<[10E]%
\>[13]{}\Varid{x}\mathbin{\uparrow}\mathrm{2}\mathrel{=}\mathrm{1}{}\<[E]%
\ColumnHook
\end{hscode}\resethooks
\noindent
where
\begin{hscode}\SaveRestoreHook
\column{B}{@{}>{\hspre}l<{\hspost}@{}}%
\column{3}{@{}>{\hspre}l<{\hspost}@{}}%
\column{14}{@{}>{\hspre}l<{\hspost}@{}}%
\column{E}{@{}>{\hspre}l<{\hspost}@{}}%
\>[3]{}(\mathbin{\uparrow})\ \mathop{:}\ \Real\to \mathbb{N}\to \Real{}\<[E]%
\\
\>[3]{}\Varid{x}\mathbin{\uparrow}\Conid{Z}{}\<[14]%
\>[14]{}\mathrel{=}\mathrm{1}{}\<[E]%
\\
\>[3]{}\Varid{x}\mathbin{\uparrow}(\Conid{S}\;\Varid{n}){}\<[14]%
\>[14]{}\mathrel{=}\Varid{x}\ensuremath{\cdot}(\Varid{x}\mathbin{\uparrow}\Varid{n}){}\<[E]%
\ColumnHook
\end{hscode}\resethooks
\noindent
It is worth pointing out that 1) it is a context that is \emph{not}
immediately deducible from $x^2 = 1$ that determines the meaning of the
equality sign in this equation and 2) that it is the types of \ensuremath{\Varid{x}} and
the squaring function that make such context clear.
For example, with \ensuremath{\Varid{x}\ \mathop{:}\ \mathbb{N}} and \ensuremath{(\mathbin{\uparrow})\ \mathop{:}\ \mathbb{N}\to \mathbb{N}\to \mathbb{N}} there is
just one solution, and in \ensuremath{\Conid{Double}}, \ensuremath{\Varid{x}\mathbin{\uparrow}\mathrm{2}\mathrel{=}\mathrm{2}} has no (exact) solutions.

In this paper we will often use equations between functions.
In such equations, we use extensional equality implemented as follow

\begin{hscode}\SaveRestoreHook
\column{B}{@{}>{\hspre}l<{\hspost}@{}}%
\column{3}{@{}>{\hspre}l<{\hspost}@{}}%
\column{E}{@{}>{\hspre}l<{\hspost}@{}}%
\>[3]{}(\ee)\ \mathop{:}\ \{\mskip1.5mu \Conid{A},\Conid{B}\ \mathop{:}\ \Conid{Type}\mskip1.5mu\}\to (\Conid{A}\to \Conid{B})\to (\Conid{A}\to \Conid{B})\to \Conid{Type}{}\<[E]%
\\
\>[3]{}(\ee)\;\{\mskip1.5mu \Conid{A}\mskip1.5mu\}\;\Varid{f}\;\Varid{g}\mathrel{=}(\Varid{x}\ \mathop{:}\ \Conid{A})\to \Varid{f}\;\Varid{x}\mathrel{=}\Varid{g}\;\Varid{x}{}\<[E]%
\ColumnHook
\end{hscode}\resethooks
Because of the equivalence between logical propositions and types
\citep{DBLP:journals/cacm/Wadler15}, the type \ensuremath{\Varid{f}\;\ee\;\Varid{g}} means $\forall
x,$ \ensuremath{\Varid{f}\;\Varid{x}\mathrel{=}\Varid{g}\;\Varid{x}} and values of this type are proofs of this equality.
While it would be possible to work with setoid equality, or homotopy
type theory, we make the pragmatic choice to use extensional equality
(we explored the consequences in
\citep{botta_brede_jansson_richter_2021}).

\subsection{Generic differential equations}
\label{subsection:de}

In mathematical physics but also in the social sciences and in modelling
it is common to specify problems implicitly in terms of equations.

Typically, such specifications are \emph{generic} and come in the form
of systems of differential equations. In this context, generic means
that the problem equations are given in terms of functions which are
not defined explicitly. (Thus, one has a family of systems parameterised
  by a function. The idea is then to study how properties of these
  systems, for example that of having stationary solutions, depend on
  properties of the parameter.)
The focus is on the semantics and the syntax can be confusing. For
example, the ordinary differential Equation (1.4) at page 18 of
\citet{Kuznetsov:1998:EAB:289919}

\begin{equation}
\dot{x} = f(x)
\label{eq:ode0}
\end{equation}

\noindent
is said to define a continuous-time \emph{dynamical system} \ensuremath{(\CalTime,\Conid{X},\varphi)}. In this context, \ensuremath{\CalTime} is the \emph{time} set (a real
interval), \ensuremath{\Conid{X}} is the \emph{state space} of the system, \ensuremath{\Varid{x}} is a
function of type \ensuremath{\CalTime\to \Conid{X}}, \ensuremath{\dot{x}} (also of type \ensuremath{\CalTime\to \Conid{X}})
is the first derivative of \ensuremath{\Varid{x}}, and \ensuremath{\Varid{f}} is a function of type \ensuremath{\Conid{X}\to \Conid{X}}
smooth enough to ensure existence and uniqueness of solutions. Thus,
\cref{eq:ode0} contains a type error. The twist here is that the
equation is just an abbreviation for the specification
\begin{hscode}\SaveRestoreHook
\column{B}{@{}>{\hspre}l<{\hspost}@{}}%
\column{3}{@{}>{\hspre}l<{\hspost}@{}}%
\column{10}{@{}>{\hspre}c<{\hspost}@{}}%
\column{10E}{@{}l@{}}%
\column{13}{@{}>{\hspre}l<{\hspost}@{}}%
\column{E}{@{}>{\hspre}l<{\hspost}@{}}%
\>[3]{}\Varid{x}{}\<[10]%
\>[10]{}\ \mathop{:}\ {}\<[10E]%
\>[13]{}\CalTime\to \Conid{X}{}\<[E]%
\\
\>[3]{}\Varid{xSpec}{}\<[10]%
\>[10]{}\ \mathop{:}\ {}\<[10E]%
\>[13]{}\Varid{D}\;\Varid{x}\;\ee\;\Varid{f}\mathbin{\circ}\Varid{x}{}\<[E]%
\ColumnHook
\end{hscode}\resethooks
\noindent
where we adopt the notation of \citet{JanssonIonescuBernardyDSLsofMathBook2022} and use \ensuremath{\Varid{D}\;\Varid{x}}
to denote the derivative of \ensuremath{\Varid{x}}. When not otherwise stated, \ensuremath{\Varid{D}\;\Varid{x}} has
the type of \ensuremath{\Varid{x}}. When quoting textbooks verbatim, we also denote \ensuremath{\Varid{D}\;\Varid{x}}
by \ensuremath{\dot{x}} (and \ensuremath{\Varid{D}\;(\Varid{D}\;\Varid{x})} by \ensuremath{\ddot{x}}), $dx/dt$ or similar.
The third element of the dynamical system associated to
\cref{eq:ode0}, \ensuremath{\varphi}, is a function of type \ensuremath{\CalTime\to \Conid{X}\to \Conid{X}}. The idea
is that \ensuremath{\varphi\;\Varid{t}\;x_{0}} is the value of \emph{the} solution
of \cref{eq:ode0} for initial condition \ensuremath{x_{0}} at time \ensuremath{\Varid{t}}. Thus \ensuremath{\varphi}
does depend on \ensuremath{\Varid{f}} although this is not immediately visible from its
type.
\DONE{Perhaps not important here, but our intro would suggest that a
  solution would be a value of the same type as that of \ensuremath{\Varid{x}}, thus
  \ensuremath{\CalTime\to \Conid{X}}.  Then with the initial condition \ensuremath{\Varid{x}\;\mathrm{0}\mathrel{=}x_{0}} for a
  parameter \ensuremath{x_{0}} we get a family of unique solutions \ensuremath{\psi\ \mathop{:}\ \Conid{X}\to \CalTime\to \Conid{X}}. This is \ensuremath{\Varid{flip}\;\varphi}, but I would find the type of
  \ensuremath{\psi} more natural then that of \ensuremath{\varphi}.}

In mathematical physics, it is very common to use the same notation
for function application and function composition as in
\cref{eq:ode0}. For example, Newton's principle of
determinacy\footnote{Newton's principle of determinacy maintains that
  the initial state of a mechanical systems (the positions and the
  velocities of its points at an initial time) uniquely determines its
  motion, see \citep{arnold1989mathematical}, page 4.} is formalized in
Equation (1) of \citep{arnold1989mathematical} as

\begin{equation}
\ddot{\mathbf{x}} = F(\mathbf{x}, \dot{\mathbf{x}}, t)
\label{eq:newton0}
\end{equation}

\noindent
where
$\mathbf{x}, \dot{\mathbf{x}}, \ddot{\mathbf{x}} : \CalTime \to
\Real^N$ and $F : (\Real^N, \Real^N, \CalTime) \to \Real^N$. Again,
the equation is to be understood as an abbreviation for the
composition between $F$ and
the function taking $t$ to the triple $(\mathbf{x}(t), \dot{\mathbf{x}}(t), t)$.
The function $F$ only has access to a state triple at the current
instant in time, not to the full time evolution of the state.

Keeping in mind that $f(x)$ often just means \ensuremath{\Varid{f}\mathbin{\circ}\Varid{x}} and with a little
bit of knowledge of the specific domain, it is often not difficult to
understand the problem that equations represent. For example, in
bifurcation theory, a slightly more general form of \cref{eq:ode0}

\begin{equation}
\dot{x} = f(x, p)
\label{eq:ode3}
\end{equation}

\noindent
with parameter \ensuremath{\Varid{p}\ \mathop{:}\ \Conid{P}} and \ensuremath{\Varid{f}\ \mathop{:}\ (\Conid{X},\Conid{P})\to \Conid{X}} (again, with some abuse of
notation) is often put forward to discuss three different but closely
related problems:

\begin{itemize}

\item The problem of predicting how the system evolves in time from an
  initial condition \ensuremath{x_{0}\ \mathop{:}\ \Conid{X}} and for a given \ensuremath{\Varid{p}\ \mathop{:}\ \Conid{P}}.

\item The problem of finding \emph{stationary} points that is, values of
  \ensuremath{\Varid{x}\ \mathop{:}\ \Conid{X}} such that \(f(x,p) = 0\) for a given \ensuremath{\Varid{p}\ \mathop{:}\ \Conid{P}} and to study
  their \emph{stability}.

\item The problem of finding \emph{critical} parameters that is, values
  of \ensuremath{\Varid{p}\ \mathop{:}\ \Conid{P}} at which the set of stationary points $\{x \mid x : X,
  f(x, p) = 0\}$ (or a more general \emph{invariant} set associated with
  $f$) exhibits structural changes\footnote{In the context of climate
    research, such critical values are often called ``tipping
    points''.}.

\end{itemize}

\noindent
Mathematical models of physical systems often involve \emph{functional}
equations and systems of partial differential equations.
Understanding the problems associated with these equations from a FP
perspective requires some more domain-specific expertise.
But even these more advanced problems can often be understood by
applying ``type-driven'' analysis, see for example the discussion of the
Lagrangian in Chapter 3 ``Types in Mathematics'' \citep{JanssonIonescuBernardyDSLsofMathBook2022}.

\subsection{Physical laws, specific models}
\label{subsection:laws}

Perhaps surprisingly, understanding basic physical principles (and also
mathematical models of specific systems) in terms of well-typed
expressions is often tricky.
Consider, for example, Newton's second law as often stated in
elementary textbooks

\begin{equation}
F = m a
\label{eq:newton1}
\end{equation}

\noindent
or the relationship between pressure, density and temperature in an
ideal gas

\begin{equation}
p = \rho R T
\label{eq:gas0}
\end{equation}

\noindent
In these equations \ensuremath{\Conid{F}} denotes a \emph{force}, \ensuremath{\Varid{m}} a \emph{mass}, \ensuremath{\Varid{a}} an
\emph{acceleration}, \ensuremath{\Varid{p}} a \emph{pressure}, \ensuremath{\rho} a \emph{density}, \ensuremath{\Conid{T}}
a \emph{temperature} and \ensuremath{\Conid{R}} a gas-specific constant. But what are the
types of these quantities? What kind of equalities do these types imply?

In climate science, ``conceptual'' models of specific components of the
climate system often consist of simple, low-dimensional systems of
ordinary differential equations.
For example, Stommel's seminal 1961 paper starts by describing a simple
system consisting of a vessel of water with ``temperature \ensuremath{\Conid{T}} and
salinity \ensuremath{\Conid{S}} (in general variable in time) separated by porous walls
from an outside vessel whose temperature \ensuremath{T_e} and salinity \ensuremath{S_e} are
maintained at constant values'', see Fig. I at page 1 of
\citep{doi:10.3402/tellusa.v13i2.9491}.

The evolution of the temperature and of the salinity in the vessel are
then modeled by two uncoupled linear differential equations:


\begin{equation}
  \frac{dT}{dt} = c (T_e - T) \quad \mathrm{and} \quad \frac{dS}{dt} = d (S_e - S)
\label{eq:stommel0}
\end{equation}

\noindent
In this context, \ensuremath{\Conid{T}} and \ensuremath{\Conid{S}} represent functions of type \ensuremath{\Real\to \Real}
and \ensuremath{T_e}, \ensuremath{S_e}, the ``temperature transfer coefficient'' \ensuremath{\Varid{c}}, and the
``salinity transfer coefficient'' \ensuremath{\Varid{d}} are real numbers.
%
%
We can formulate the problem of computing solutions of
\cref{eq:stommel0} through the specification:

\begin{hscode}\SaveRestoreHook
\column{B}{@{}>{\hspre}l<{\hspost}@{}}%
\column{3}{@{}>{\hspre}l<{\hspost}@{}}%
\column{10}{@{}>{\hspre}c<{\hspost}@{}}%
\column{10E}{@{}l@{}}%
\column{13}{@{}>{\hspre}l<{\hspost}@{}}%
\column{18}{@{}>{\hspre}l<{\hspost}@{}}%
\column{65}{@{}>{\hspre}c<{\hspost}@{}}%
\column{65E}{@{}l@{}}%
\column{68}{@{}>{\hspre}l<{\hspost}@{}}%
\column{73}{@{}>{\hspre}l<{\hspost}@{}}%
\column{E}{@{}>{\hspre}l<{\hspost}@{}}%
\>[3]{}\Conid{TSpec}{}\<[10]%
\>[10]{}\ \mathop{:}\ {}\<[10E]%
\>[13]{}\Varid{D}\;\Conid{T}{}\<[18]%
\>[18]{}\;\ee\;\lambda \Varid{t}\to\Varid{c}\ensuremath{\cdot}(T_e\mathbin{-}\Conid{T}\;\Varid{t});\qquad\Conid{SSpec}{}\<[65]%
\>[65]{}\ \mathop{:}\ {}\<[65E]%
\>[68]{}\Varid{D}\;\Conid{S}{}\<[73]%
\>[73]{}\;\ee\;\lambda \Varid{t}\to\Varid{d}\ensuremath{\cdot}(S_e\mathbin{-}\Conid{S}\;\Varid{t}){}\<[E]%
\ColumnHook
\end{hscode}\resethooks
The $\lambda$-expression in \ensuremath{\Conid{TSpec}} denotes the function that maps \ensuremath{\Varid{t}}
to \ensuremath{\Varid{c}\ensuremath{\cdot}(T_e\mathbin{-}\Conid{T}\;\Varid{t})}. Thus (again, because of the equivalence between
types and logical propositions) any (total) \emph{definition} of \ensuremath{\Conid{TSpec}}
is (equivalent to) a proof that, $\forall t, \ dT(t)/dt = c (T_e -
T(t))$ and its \emph{declaration} specifies the task\footnote{\emph{Aufgabe},
\citep{Kolmogoroff1932}.} of providing one such proof.

The next step at the end of page 1 of Stommel's paper is to make
\cref{eq:stommel0} ``non-dimensional'' by introducing

\begin{equation}
  \tau = c \ t, \quad \delta = \frac{d}{c}, \quad y = T/T_e, \quad x = S/S_e
\label{eq:stommel1}
\end{equation}

\noindent
This yields


\begin{equation}
  \frac{dy}{d\tau} = 1 - y \quad \mathrm{and} \quad \frac{dx}{d\tau} = \delta (1 - x)
\label{eq:stommel2}
\end{equation}

\noindent
at the top of page 2. From there, Stommel then goes on discussing how a
\emph{density} of the vessel (a function of its temperature and salinity
and, thus, of \ensuremath{\Varid{x}} and \ensuremath{\Varid{y}}) evolves in time as the solution of
\cref{eq:stommel2} for arbitrary initial conditions \ensuremath{x_{0}}, \ensuremath{y_{0}}


\begin{equation}
  y(\tau) = 1 + (y_0 - 1)e^{-\tau} \quad \mathrm{and} \quad x(\tau) = 1 + (x_0 - 1)e^{-\delta\tau}
\label{eq:stommel3}
\end{equation}

\noindent
approaches 1 as \ensuremath{\tau} increases. Although Stommel's discussion
is in many ways interesting, we do not need to be concerned with it
here.
What we need to understand, however, are the types of \ensuremath{\tau}, \ensuremath{\delta},
\ensuremath{\Varid{x}} and \ensuremath{\Varid{y}} and how \cref{eq:stommel2} can be derived from
\cref{eq:stommel0} given \cref{eq:stommel1}. The first two equations of
\cref{eq:stommel1} are definitions of \ensuremath{\tau} and \ensuremath{\delta}:
\begin{hscode}\SaveRestoreHook
\column{B}{@{}>{\hspre}l<{\hspost}@{}}%
\column{3}{@{}>{\hspre}l<{\hspost}@{}}%
\column{10}{@{}>{\hspre}c<{\hspost}@{}}%
\column{10E}{@{}l@{}}%
\column{13}{@{}>{\hspre}l<{\hspost}@{}}%
\column{26}{@{}>{\hspre}l<{\hspost}@{}}%
\column{45}{@{}>{\hspre}l<{\hspost}@{}}%
\column{52}{@{}>{\hspre}c<{\hspost}@{}}%
\column{52E}{@{}l@{}}%
\column{55}{@{}>{\hspre}l<{\hspost}@{}}%
\column{E}{@{}>{\hspre}l<{\hspost}@{}}%
\>[3]{}\tau\ \mathop{:}\ \Real\to \Real;{}\<[26]%
\>[26]{}\qquad {}\<[45]%
\>[45]{}\delta{}\<[52]%
\>[52]{}\ \mathop{:}\ {}\<[52E]%
\>[55]{}\Real{}\<[E]%
\\
\>[3]{}\tau\;\Varid{t}{}\<[10]%
\>[10]{}\mathrel{=}{}\<[10E]%
\>[13]{}\Varid{c}\ensuremath{\cdot}\Varid{t};{}\<[45]%
\>[45]{}\delta{}\<[52]%
\>[52]{}\mathrel{=}{}\<[52E]%
\>[55]{}\Varid{d}\mathbin{/}\Varid{c}{}\<[E]%
\ColumnHook
\end{hscode}\resethooks
\DONE{It may be helpful to use different type names such that we get \ensuremath{\tau\ \mathop{:}\ \Conid{TTime}\to \Conid{TauTime}}. (A bit unfortunate that Tau is now the type of t instead of the type of tau.)}%
However, the last two equations are not \emph{explicit} definitions of
\ensuremath{\Varid{y}} and \ensuremath{\Varid{x}}. Instead, they are \emph{implicit} definitions,
corresponding to the specifications
\DONE{I think the type of y should be \ensuremath{\Conid{TauTime}\to \Real}, not \ensuremath{\Conid{TTime}\to \Real}. Only the composition \ensuremath{\Varid{y}\mathbin{\circ}\tau} has type \ensuremath{\Conid{TTime}\to \Real}.}
\begin{hscode}\SaveRestoreHook
\column{B}{@{}>{\hspre}l<{\hspost}@{}}%
\column{3}{@{}>{\hspre}l<{\hspost}@{}}%
\column{10}{@{}>{\hspre}c<{\hspost}@{}}%
\column{10E}{@{}l@{}}%
\column{13}{@{}>{\hspre}l<{\hspost}@{}}%
\column{43}{@{}>{\hspre}l<{\hspost}@{}}%
\column{59}{@{}>{\hspre}l<{\hspost}@{}}%
\column{66}{@{}>{\hspre}c<{\hspost}@{}}%
\column{66E}{@{}l@{}}%
\column{69}{@{}>{\hspre}l<{\hspost}@{}}%
\column{E}{@{}>{\hspre}l<{\hspost}@{}}%
\>[3]{}\Varid{y}{}\<[10]%
\>[10]{}\ \mathop{:}\ {}\<[10E]%
\>[13]{}\Real\to \Real;{}\<[59]%
\>[59]{}\Varid{x}{}\<[66]%
\>[66]{}\ \mathop{:}\ {}\<[66E]%
\>[69]{}\Real\to \Real{}\<[E]%
\\
\>[3]{}\Varid{ySpec}{}\<[10]%
\>[10]{}\ \mathop{:}\ {}\<[10E]%
\>[13]{}\Varid{y}\mathbin{\circ}\tau\;\ee\;\lambda \Varid{t}\to\Conid{T}\;\Varid{t}\mathbin{/}T_e;{}\<[43]%
\>[43]{}\qquad {}\<[59]%
\>[59]{}\Varid{xSpec}{}\<[66]%
\>[66]{}\ \mathop{:}\ {}\<[66E]%
\>[69]{}\Varid{x}\mathbin{\circ}\tau\;\ee\;\lambda \Varid{t}\to\Conid{S}\;\Varid{t}\mathbin{/}S_e{}\<[E]%
\ColumnHook
\end{hscode}\resethooks
and it is easy to see that the definitions
\DONE{Then it would be useful to have \ensuremath{\tau} as a variable name here. We need to think about what that would mean for the name of the conversion function (called \ensuremath{\tau} above) which clashes. Perhaps \ensuremath{\Varid{norm}} for normalize? or scale? or conv? Or live with the clash?}
\begin{hscode}\SaveRestoreHook
\column{B}{@{}>{\hspre}l<{\hspost}@{}}%
\column{3}{@{}>{\hspre}l<{\hspost}@{}}%
\column{36}{@{}>{\hspre}l<{\hspost}@{}}%
\column{56}{@{}>{\hspre}l<{\hspost}@{}}%
\column{E}{@{}>{\hspre}l<{\hspost}@{}}%
\>[3]{}\Varid{y}\;\sigma\mathrel{=}\Conid{T}\;(\sigma\mathbin{/}\Varid{c})\mathbin{/}T_e;{}\<[36]%
\>[36]{}\qquad {}\<[56]%
\>[56]{}\Varid{x}\;\sigma\mathrel{=}\Conid{S}\;(\sigma\mathbin{/}\Varid{c})\mathbin{/}S_e{}\<[E]%
\ColumnHook
\end{hscode}\resethooks
fulfil \ensuremath{\Varid{ySpec}} and \ensuremath{\Varid{xSpec}}:

\begin{hscode}\SaveRestoreHook
\column{B}{@{}>{\hspre}l<{\hspost}@{}}%
\column{3}{@{}>{\hspre}l<{\hspost}@{}}%
\column{12}{@{}>{\hspre}c<{\hspost}@{}}%
\column{12E}{@{}l@{}}%
\column{15}{@{}>{\hspre}l<{\hspost}@{}}%
\column{E}{@{}>{\hspre}l<{\hspost}@{}}%
\>[3]{}\Varid{ySpec}\;\Varid{t}{}\<[12]%
\>[12]{}\mathrel{=}{}\<[12E]%
\>[15]{}((\Varid{y}\mathbin{\circ}\tau)\;\Varid{t})=\hspace{-3pt}\{\; \Conid{Refl}\;\}\hspace{-3pt}=(\Conid{T}\;(\Varid{c}\ensuremath{\cdot}\Varid{t}\mathbin{/}\Varid{c})\mathbin{/}T_e)=\hspace{-3pt}\{\; \mathbf{?h_0}\;\}\hspace{-3pt}=(\Conid{T}\;\Varid{t}\mathbin{/}T_e)\;\Conid{QED}{}\<[E]%
\ColumnHook
\end{hscode}\resethooks
and similarly for \ensuremath{\Varid{xSpec}}.
Filling the \ensuremath{\mathbf{?h_0}} hole in the last step requires invoking
congruence and a proof of \ensuremath{\Varid{c}\ensuremath{\cdot}\Varid{t}\mathbin{/}\Varid{c}\mathrel{=}\Varid{t}}.
Here, the types of \ensuremath{\Varid{c}} and \ensuremath{\Varid{t}} are aliases for double-precision
floating-point numbers for which such an equality needs to be
postulated.
The differential equations (\ref{eq:stommel2}) for \ensuremath{\Varid{y}} and \ensuremath{\Varid{x}} follow
then from the definitions of \ensuremath{\tau} and \ensuremath{\delta}, from the
specifications \ensuremath{\Conid{TSpec}}, \ensuremath{\Varid{ySpec}}, \ensuremath{\Conid{SSpec}}, \ensuremath{\Varid{xSpec}} and from the rules
for derivation.
\newpage\noindent
Informally:
\begin{hscode}\SaveRestoreHook
\column{B}{@{}>{\hspre}l<{\hspost}@{}}%
\column{3}{@{}>{\hspre}l<{\hspost}@{}}%
\column{74}{@{}>{\hspre}l<{\hspost}@{}}%
\column{E}{@{}>{\hspre}l<{\hspost}@{}}%
\>[3]{}\Varid{y}\mathbin{\circ}\tau\;\ee\;\lambda \Varid{t}\to\Conid{T}\;\Varid{t}\mathbin{/}T_e{}\<[74]%
\>[74]{}\Rightarrow \mbox{\onelinecomment  congruence}{}\<[E]%
\\[\blanklineskip]%
\>[3]{}\Varid{D}\;(\Varid{y}\mathbin{\circ}\tau)\;\ee\;\Varid{D}\;(\lambda \Varid{t}\to\Conid{T}\;\Varid{t}\mathbin{/}T_e){}\<[74]%
\>[74]{}\mathrel{=}\mbox{\onelinecomment  D-composition, \dots}{}\<[E]%
\\[\blanklineskip]%
\>[3]{}\lambda \Varid{t}\to(\Varid{D}\;\Varid{y})\;(\tau\;\Varid{t})\ensuremath{\cdot}(\Varid{D}\;\tau)\;\Varid{t}\;\ee\;\lambda \Varid{t}\to(\Varid{D}\;\Conid{T})\;\Varid{t}\mathbin{/}T_e\qquad{}\<[74]%
\>[74]{}\mathrel{=}\mbox{\onelinecomment  \ensuremath{(\Varid{D}\;\tau)\;\Varid{t}\mathrel{=}(\Varid{const}\;\Varid{c})\;\Varid{t}\mathrel{=}\Varid{c}}, \ensuremath{\Conid{TSpec}}}{}\<[E]%
\\[\blanklineskip]%
\>[3]{}\lambda \Varid{t}\to(\Varid{D}\;\Varid{y})\;(\tau\;\Varid{t})\ensuremath{\cdot}\Varid{c}\;\ee\;\lambda \Varid{t}\to\Varid{c}\ensuremath{\cdot}(T_e\mathbin{-}\Conid{T}\;\Varid{t})\mathbin{/}T_e{}\<[74]%
\>[74]{}\mathrel{=}\mbox{\onelinecomment  arithmetics}{}\<[E]%
\\[\blanklineskip]%
\>[3]{}\lambda \Varid{t}\to(\Varid{D}\;\Varid{y})\;(\tau\;\Varid{t})\;\ee\;\lambda \Varid{t}\to\mathrm{1}\mathbin{-}\Conid{T}\;(\Varid{c}\ensuremath{\cdot}\Varid{t}\mathbin{/}\Varid{c})\mathbin{/}T_e{}\<[74]%
\>[74]{}\mathrel{=}\mbox{\onelinecomment  \ensuremath{\Varid{ySpec}}, def. composition}{}\<[E]%
\\[\blanklineskip]%
\>[3]{}\lambda \Varid{t}\to((\Varid{D}\;\Varid{y})\mathbin{\circ}\tau)\;\Varid{t}\;\ee\;\lambda \Varid{t}\to\mathrm{1}\mathbin{-}\Varid{y}\;(\tau\;\Varid{t}){}\<[74]%
\>[74]{}\mathrel{=}\mbox{\onelinecomment  \ensuremath{\beta}-reduction}{}\<[E]%
\\[\blanklineskip]%
\>[3]{}(\Varid{D}\;\Varid{y})\mathbin{\circ}\tau\;\ee\;\lambda \Varid{t}\to\mathrm{1}\mathbin{-}(\Varid{y}\mathbin{\circ}\tau)\;\Varid{t}{}\<[74]%
\>[74]{}\mathrel{=}\mbox{\onelinecomment  factoring on the RHS}{}\<[E]%
\\[\blanklineskip]%
\>[3]{}(\Varid{D}\;\Varid{y})\mathbin{\circ}\tau\;\ee\;(\lambda \sigma\to\mathrm{1}\mathbin{-}\Varid{y}\;\sigma)\mathbin{\circ}\tau{}\<[74]%
\>[74]{}\mathrel{=}\mbox{\onelinecomment  cancel composition (inv. \ensuremath{\tau})}{}\<[E]%
\\[\blanklineskip]%
\>[3]{}\Varid{D}\;\Varid{y}\;\ee\;\lambda \sigma\to\mathrm{1}\mathbin{-}\Varid{y}\;\sigma{}\<[E]%
\ColumnHook
\end{hscode}\resethooks
\DONE{I think two steps are missing:}
\DONE{Possibly a last step with 1 standing for const 1}
\noindent
We can turn this into a valid Idris proof but the point here is that the
computation looks significantly more contrived than the straightforward
(again, informal) derivation
\begin{hscode}\SaveRestoreHook
\column{B}{@{}>{\hspre}l<{\hspost}@{}}%
\column{3}{@{}>{\hspre}l<{\hspost}@{}}%
\column{57}{@{}>{\hspre}l<{\hspost}@{}}%
\column{71}{@{}>{\hspre}l<{\hspost}@{}}%
\column{E}{@{}>{\hspre}l<{\hspost}@{}}%
\>[3]{}\Varid{y}\mathrel{=}\Conid{T}\mathbin{/}T_e{}\<[71]%
\>[71]{}\Rightarrow \mbox{\onelinecomment  chain rule}{}\<[E]%
\\[\blanklineskip]%
\>[3]{}\frac{\Varid{dy}}{d\tau}\phantom{x} \frac{d\tau}{\Varid{dt}}\mathrel{=}\frac{\Varid{dT}}{\Varid{dt}}\phantom{x} \frac{\mathrm{1}}{T_e}{}\<[71]%
\>[71]{}\mathrel{=}\mbox{\onelinecomment  def. of \ensuremath{\tau}, \cref{eq:stommel2}}{}\<[E]%
\\[\blanklineskip]%
\>[3]{}\frac{\Varid{dy}}{d\tau}\phantom{x} \Varid{c}\mathrel{=}\Varid{c}\;(T_e\mathbin{-}\Conid{T})\mathbin{/}T_e{}\<[57]%
\>[57]{}\qquad{}\<[71]%
\>[71]{}\mathrel{=}\mbox{\onelinecomment  def. of \ensuremath{\Varid{y}}}{}\<[E]%
\\[\blanklineskip]%
\>[3]{}\frac{\Varid{dy}}{d\tau}\mathrel{=}\mathrm{1}\mathbin{-}\Varid{y}{}\<[E]%
\ColumnHook
\end{hscode}\resethooks
that is taken for granted and not even mentioned in Stommel's
paper.
This complication is a consequence of having insisted on explicit
types and on consistent typing rules in a language that lacks important
abstractions.
Specifically, it lacks rules for lifting real-number operations (like
multiplication,
subtraction, etc.) to operations
between real numbers and functions with matching co-domain.

Such complications may partly explain why modellers and mathematicians
have not found much added value in dependent types and functional
languages: even for derivations as simple as those of the Stommel model,
these languages add a formal layer that seems to unnecessarily obfuscate
computations that are nearly self-evident.

It is not difficult to make the Idris (or Agda, Rocq, etc.) type checker
digest definitions like \cref{eq:stommel1} and simplify derivations like
those of \ensuremath{\Varid{ySpec}} and \ensuremath{\Varid{xSpec}} but, unfortunately, there are more bad
news: the ``dimensionality'' of the expressions involved in Stommel's
derivation is not visible in their types!
Stommel shows that by ``making the equations non-dimensional'' the
number of parameters of \cref{eq:stommel0} has been reduced to just one:
the ``non-dimensional'' ratio~\ensuremath{\delta}.

This is an important result. It implies that the solution of
\cref{eq:stommel2} for a given value of \ensuremath{\delta} allows one to recover a
whole family of solutions of \cref{eq:stommel0}, namely, all those
corresponding to values of \ensuremath{\Varid{c}} and \ensuremath{\Varid{d}} such that $d / c = \delta$.
Conversely, it demonstrates that the problem defined by
\cref{eq:stommel0} is \emph{inflated}.

In mathematical modelling but also in data analysis, inflated problems
are ubiquitous. Recognizing that the solution of a problem can be
reduced to the solution of a much simpler problem is often crucial to
avoid the ``curse of dimensionality'', e.g. when solving large systems
of partial differential equations or in training deep neural networks.

But how can one recognize that a problem is inflated? What does it mean
to say that \ensuremath{\delta} is \emph{non-dimensional}? And why are
\cref{eq:stommel0} dimensional and \cref{eq:stommel2}
non-dimensional? The types of the expressions involved do not
provide us with any clue.

Like in the case of Newton's second law and of the equation for ideal
gases, we have to do with a \emph{grammar} of properties (being a force,
being a mass, being non-dimensional) that we do not grasp.
As a consequence, our types are not well suited to that grammar: \ensuremath{\Conid{T}} and
\ensuremath{\Varid{y}}, \ensuremath{\Conid{S}} and \ensuremath{\Varid{x}}, \ensuremath{\Varid{m}} and \ensuremath{\rho} have different properties and these
properties are important in the domain of discourse. Yet, they all have
the same types!
\DONE{In a natural (DSLsofMath-inspired) ``encoding'' of the problem, we would have separate types for t, T, S, and then let the ``non-dimensional'' versions by just REALs. But that is perhaps the point here. It is a bit hidden behind other notational differences, though.}

If we want to take mathematical physics and modelling seriously (and if
we want experts in these domains to take FP seriously) we have to encode
that grammar through suitable types.
We discuss a minimal DSL of dimensional quantities in
\cref{section:dsl1}. Before getting there, however, we need to build a
better understanding of the questions raised in this section. In the
next one, we start with the question of what it means for a physical
quantity to \emph{have a dimension}.

\section{Dimensions, physical quantities, units of measurement}
\label{section:dimensions}

In the last section we have discussed simple examples of equations and
implicit problem specifications in the context of mathematical physics
and modelling.
We have seen that the variables that appear in equations like Newton's
second law \cref{eq:newton1} or in the ideal gas law are endowed
with properties, like being a force or a temperature, that we do not
know how to represent through the type system.

In the case of the Stommel model, we have encountered variables that
were said to ``have a dimension'', for example \ensuremath{\Varid{c}}. Other expressions
were said to be ``non-dimensional''. But what does it mean for \ensuremath{\Varid{c}} to
have a dimension?  In a nutshell, it means two things:

\begin{enumerate}
  \item That \ensuremath{\Varid{c}} represents a \emph{physical quantity} that can be
    measured in a system of \emph{units} of measurement of a given
    \emph{class}.
  \item That the numerical measure of \ensuremath{\Varid{c}} will change in a specific,
    predictable way if the system of units is changed (within the same
    class).
\end{enumerate}

\noindent
An example will help illustrate the idea. Consider the sheet of paper
on which this article is printed. Assume its width to be 20 centimetres
and its height to be 30 centimetres.

If we measure \emph{lengths} in centimetres, the measures of width and
height will be 20 and 30, respectively. In metres, these measures will
instead be 0.2 and 0.3. A change in the units of measurement of lengths
has resulted in a change in the measures of the width and of the height
of the paper: we say that the width and the height have a dimension or,
equivalently, that they are dimensional quantities.

By contrast, the ratio between the height and the width of the paper is
3/2 no matter whether we measure lengths in centimetres, metres or in
other units: we say that the ratio is a non-dimensional quantity.

Notice that the distinction between dimensional and non-dimensional
quantities crucially relies on the (implicit) assumption of measuring
\emph{both} the height and the width of the paper (more generally, all
lengths) with the same units.


\paragraph*{The dimension judgment.} \
In strongly typed languages like Idris and Agda, the judgment \ensuremath{\Varid{e}\ \mathop{:}\ \Varid{t}}
means that expression \ensuremath{\Varid{e}} has type \ensuremath{\Varid{t}}.
In the physical sciences, the judgment \ensuremath{[\mskip1.5mu \Varid{e}\mskip1.5mu]\mathrel{=}\Varid{d}} means that expression
\ensuremath{\Varid{e}} (representing a physical quantity) has dimension \ensuremath{\Varid{d}}. At this point,
we do not know how to formalize this judgment in type theory (we
discuss how to do so in \cref{section:dsl1}) but the idea
is that \ensuremath{\Varid{d}} is a function of type $\Real_{+}^n \to \Real_{+}$ where the
number \ensuremath{\Varid{n}\ \mathop{:}\ \mathbb{N}} is domain-specific and $\Real_{+}$ denotes the set of
positive real numbers.

For example, in mechanics \ensuremath{\Varid{n}\mathrel{=}\mathrm{3}}. In this domain, the judgment \ensuremath{[\mskip1.5mu \Varid{e}\mskip1.5mu]\mathrel{=}\lambda (\Conid{L},\Conid{T},\Conid{M})\Rightarrow \Conid{L}\ensuremath{\cdot}T^{-1}} means that \ensuremath{\Varid{e}} is a quantity that can be measured
in a system of units of measurements, for example SI (international) or
CGS (centimetre-gram-second), for \emph{lengths}, \emph{times} and
\emph{masses}, and that the measure of \ensuremath{\Varid{e}} \emph{increases} by a
factor \ensuremath{\Conid{L}\ensuremath{\cdot}T^{-1}} when the units for lengths, times and masses are
\emph{decreased} by factors \ensuremath{\Conid{L}}, \ensuremath{\Conid{T}} and \ensuremath{\Conid{M}}, respectively. Another way
to express the same judgment is to say that ``\ensuremath{\Varid{e}} is a velocity'' or
that ``\ensuremath{\Varid{e}} has the dimension of a velocity''. The notation was
originally introduced by Maxwell, see \citep[Section 1.1.3]{barenblatt1996scaling},
and in DA (dimensional analysis) it is common to write
\ensuremath{[\mskip1.5mu \Varid{e}\mskip1.5mu]\mathrel{=}\Conid{L}\;T^{-1}} as an abbreviation for \ensuremath{[\mskip1.5mu \Varid{e}\mskip1.5mu]\mathrel{=}\lambda (\Conid{L},\Conid{T},\Conid{M})\Rightarrow \Conid{L}\ensuremath{\cdot}T^{-1}}.

In mechanics, physical quantities can alternatively be measured in a system of
units for \emph{lengths}, \emph{times} and \emph{forces}. This defines a
different \emph{class} of units in the same domain (thus \ensuremath{\Varid{n}\mathrel{=}\mathrm{3}}). In plain geometry \ensuremath{\Varid{n}\mathrel{=}\mathrm{1}} and in classical mechanics with heat transfer \ensuremath{\Varid{n}\mathrel{=}\mathrm{4}}: beside units
for lengths, times and forces, in this domain we also need units for
\emph{temperatures}.

%

For the reader who finds all this very confusing and suspiciously far
away from the clear and deep waters of type theory: it is!
As we have seen in \cref{section:equations}, the standard arsenal
of functional programming abstractions is not yet ready to encode the
grammar of dimensions that informs the language of mathematical physics
and modelling.

If we want to develop DSLs that are palatable to mathematicians,
physicists and modellers, we need to spend some time wading in shallow
and muddy waters. The ideas summarized in this section are discussed
more authoritatively and to a greater extent in the introduction and in
Section 1 of \citep{barenblatt1996scaling}.

We will give a precise specification of the higher order function \ensuremath{[\mskip1.5mu \phantom{x} \mskip1.5mu]}
and formalize the notion of physical quantity in type theory in
\cref{section:dsl1}.
To get there, however, we need to first get an idea of the problems
addressed by the theory of \emph{similarity} and by Buckingham's Pi
theorem. This is done in the next two sections. We conclude this one
with three remarks:

\paragraph*{Remark 1: Equations relate quantities through measurements.} \
Equations like Newton's second principle
\cref{eq:newton1} or the ideal gas law \cref{eq:gas0} represent
relationships between physical quantities (like force, mass, and acceleration).
The idea is that these equations summarize empirical facts about
measurements of these quantities (or put forward axioms about such measurements).
%
These facts (or axioms) about measurements are understood to hold under
a number of assumptions, typically implicit. A crucial one is that all
measurements are done in the same class and system of units.

For example, \cref{eq:gas0} maintains that measurements of the
pressure \ensuremath{\Varid{p}} of an ideal gas are proportional to the product of
measurements of density \ensuremath{\rho} and measurements of temperature \ensuremath{\Conid{T}},
the factor of proportionality being a gas-specific constant \ensuremath{\Conid{R}}.
We come back to the idea that equations like Newton's second principle
represent relationships between physical quantities and we present a more
consistent interpretation of such equations in
\cref{subsection:covariance}.

\paragraph*{Remark 2: Context matters.} \ The result of measurements
(and thus the type of the variables entering the equations) depends on
a context that is not visible in the equations themselves.
For example, when the measurements of pressure, density and temperature
are pertinent to a homogeneous gas, \cref{eq:gas0} can be
interpreted as the specification
\begin{hscode}\SaveRestoreHook
\column{B}{@{}>{\hspre}l<{\hspost}@{}}%
\column{3}{@{}>{\hspre}l<{\hspost}@{}}%
\column{18}{@{}>{\hspre}c<{\hspost}@{}}%
\column{18E}{@{}l@{}}%
\column{21}{@{}>{\hspre}l<{\hspost}@{}}%
\column{E}{@{}>{\hspre}l<{\hspost}@{}}%
\>[3]{}\Varid{pSpec}{}\<[18]%
\>[18]{}\ \mathop{:}\ {}\<[18E]%
\>[21]{}\mu\;\Varid{p}\mathrel{=}\mu\;\rho\ensuremath{\cdot}\mu\;\Conid{R}\ensuremath{\cdot}\mu\;\Conid{T}{}\<[E]%
\ColumnHook
\end{hscode}\resethooks
where \ensuremath{\mu} is the measure function (which will talk more about later).
In a context in which \ensuremath{\Varid{p}}, \ensuremath{\rho} and \ensuremath{\Conid{T}} represent the pressure, the
density and the temperature of a gas in local thermodynamical
equilibrium, the same equation can be interpreted as the specification
\begin{hscode}\SaveRestoreHook
\column{B}{@{}>{\hspre}l<{\hspost}@{}}%
\column{3}{@{}>{\hspre}l<{\hspost}@{}}%
\column{18}{@{}>{\hspre}c<{\hspost}@{}}%
\column{18E}{@{}l@{}}%
\column{21}{@{}>{\hspre}l<{\hspost}@{}}%
\column{E}{@{}>{\hspre}l<{\hspost}@{}}%
\>[3]{}\Varid{pSpec}{}\<[18]%
\>[18]{}\ \mathop{:}\ {}\<[18E]%
\>[21]{}\mu\;\Varid{p}\ee\lambda (\Varid{x},\Varid{t})\Rightarrow \mu\;(\rho\;(\Varid{x},\Varid{t}))\ensuremath{\cdot}\mu\;\Conid{R}\ensuremath{\cdot}\mu\;(\Conid{T}\;(\Varid{x},\Varid{t})){}\<[E]%
\ColumnHook
\end{hscode}\resethooks
This is typically the case when a symbol that represents the pressure
appears in the right hand side of a system of partial differential
equations like the Euler or the Navier-Stokes equations of fluid
mechanics \citep{chorin2000mathematical}.
In yet another context, \ensuremath{\Varid{p}}, \ensuremath{\rho} and \ensuremath{\Conid{T}} could represent
probability density functions or other higher order functions.
But what could be the types of \ensuremath{\Varid{p}}, \ensuremath{\rho}, \ensuremath{\Conid{R}} and \ensuremath{\Conid{T}} in these
contexts?  We answer this question in \cref{subsection:quantities}.

\paragraph*{Remark 3: Dimensions, functions, and derivatives.} \
When \ensuremath{\Varid{p}} is a function taking inputs from, e.g.,
\ensuremath{(\Real,\CalTime)}, the judgment ``\ensuremath{\Varid{p}} is a pressure'' (or ``\ensuremath{\Varid{p}} has the
dimension of a pressure'' or, \ensuremath{[\mskip1.5mu \Varid{p}\mskip1.5mu]\mathrel{=}\Conid{M}\;L^{-1}\;T^{-2}}) is just an
abbreviation for ``\ensuremath{\forall\;(\Varid{x},\Varid{t})\ \mathop{:}\ (\Real,\CalTime)}, \ \ensuremath{\Varid{p}\;(\Varid{x},\Varid{t})} is a
pressure''.
If \ensuremath{\Varid{p}} is differentiable with respect to both space and time and if the
space and time coordinates are dimensional (that is, \ensuremath{\forall\;(\Varid{x},\Varid{t})\ \mathop{:}\ (\Real,\CalTime)}, \ensuremath{[\mskip1.5mu \Varid{x}\mskip1.5mu]\mathrel{=}\Conid{L}} and \ensuremath{[\mskip1.5mu \Varid{t}\mskip1.5mu]\mathrel{=}\Conid{T}}) then the partial derivatives of
\ensuremath{\Varid{p}} with respect to space and time have the dimensions \ensuremath{\Conid{M}\;L^{-2}\;T^{-2}}
and \ensuremath{\Conid{M}\;L^{-1}\;T^{-3}}, respectively.
More generally, if \ensuremath{\Varid{p}} is a function from a dimensional space-time set
into real numbers and \ensuremath{\Varid{p}} has dimension \ensuremath{\Varid{d}}, the partial derivatives of
\ensuremath{\Varid{p}} with respect to time and space are functions of the same type as \ensuremath{\Varid{p}}
but with dimensions \ensuremath{\Varid{d}\ensuremath{\cdot}T^{-1}} and \ensuremath{\Varid{d}\ensuremath{\cdot}L^{-1}}. Again, these are shortcuts
for \ensuremath{\lambda (\Conid{L},\Conid{T},\Conid{M})\Rightarrow \Varid{d}\;(\Conid{L},\Conid{T},\Conid{M})\ensuremath{\cdot}T^{-1}} and \ensuremath{\lambda (\Conid{L},\Conid{T},\Conid{M})\Rightarrow \Varid{d}\;(\Conid{L},\Conid{T},\Conid{M})\ensuremath{\cdot}L^{-1}}, respectively.

As already mentioned in \cref{section:equations}, the last
specification for \ensuremath{\Varid{p}} could be written more concisely as \ensuremath{\Varid{p}\ee\rho\ensuremath{\cdot}\Conid{R}\ensuremath{\cdot}\Conid{T}} by introducing canonical abstractions for functions that return
numerical values.
To the best of our knowledge, no standard Idris library provides such
abstractions although, as discussed in the introduction, there have been
proposals for making types in FP languages more aware of dimensions and
physical quantities, for example by \citet{IdrisQuantities}.

\section{Similarity theory and the Pi theorem}
\label{section:pi}

The notion of dimension function informally introduced in the last
section is closely related with a fundamental principle in physical
sciences and with a very pragmatic question in physical modelling.

We start with the pragmatic question. At the turn of the 20th century, no computers
were available for approximating numerical solutions of mathematical
models of physical systems, typically in the form of partial
differential equations.
Thus, models of physical systems, say of a ship cruising in shallow
waters or of an airplane, were themselves physical systems, for
convenience often at a reduced \emph{scale}.
This raised the obvious question of the conditions under which careful
measurements made on the scaled model could be ``scaled up'' to the real
system and how this scaling up should be done.

For example, if the drag on a 1:50 model of a ship cruising in a water
channel was found to be \ensuremath{\Varid{x}} Newton, what would be the drag of the
real ship? And under which conditions is it possible to give a definite
answer to this question?

At first glance, the problem seems to be one of engineering. But the
theory that was developed to answer this question, similarity theory or
dimensional analysis (DA), turned out to be a logical consequence of a
fundamental principle: ``physical laws do not depend on arbitrarily
chosen basic units of measurement'', see \citep[section 0.1]{barenblatt1996scaling}.
This is, in turn, an instance of Galileo's principle of
relativity, see \citep[page 3]{arnold1989mathematical}.

The core results of DA can be summarized in Buckingham's Pi theorem and
this theorem is also an answer to the question(s) of model similarity
raised above.
We do not need to be concerned with these answers and with specific
applications of the Pi theorem in the physical sciences here.
But we need to understand the notions that are at core of this theorem
in order to develop a DSL that is suitable for mathematical physics and
for modelling.

We introduce Buckingham's Pi theorem as it is stated in
\citep{barenblatt1996scaling}. This formulation is consistent with
textbook presentations and raises a number of questions.
We flag these questions here and then tackle them from a functional
programming perspective in \cref{section:dsl1} where we build a minimal
DSL for dimensionally consistent programming.

\subsection{Buckingham's Pi theorem}
\label{subsection:pi}

The theorem is stated at page 42 of \citep{barenblatt1996scaling}:

\begin{quotation}
A physical relationship between some dimensional (generally speaking)
quantity and several dimensional governing parameters can be rewritten
as a relationship between some dimensionless parameter and several
dimensionless products of the governing parameters; the number of
dimensionless products is equal to the total number of governing
parameters minus the number of governing parameters with independent
dimensions.
\end{quotation}

\noindent
This formulation comes at the end of a derivation that starts at page 39
by positing a ``physical relationship'' between a ``dimensional
quantity'' $a$ and $k + m$ ``dimensional governing parameters''
$a_1,\dots,a_k$ and $b_1,\dots,b_m$
\begin{equation}
a = f(a_1,\dots,a_k,b_1,\dots,b_m)
\label{eq:pih0}
\end{equation}
such that $a_1,\dots,a_k$ ``have independent dimensions, while the
dimensions of parameters $b_1,\dots,b_m$ can be expressed as products of
powers of the dimensions of the parameters $a_1,\dots,a_k$'':
\begin{equation}
[b_i] = [a_1]^{p_{i1}} \dots [a_k]^{p_{ik}} \quad i = 1,\dots,m
\label{eq:pih1}
\end{equation}
With these premises, the conclusions are then 1) that ``the dimension of
$a$ must be expressible in terms of the dimensions of the governing
parameters $a_1,\dots,a_k$'':
\begin{equation}
[a] = [a_1]^{p_1} \dots [a_k]^{p_k}
\label{eq:pi0}
\end{equation}
and 2) that the function $f$ ``possesses the property of generalized
homogeneity or symmetry, i.e., it can be written in terms of a function
of a smaller number of variables, and is of the following special
form'':
\begin{equation}
  f(a_1, \dots, a_k, b_1, \dots, b_m) = a_1^{p_1} \dots a_k^{p_k} \ \Phi(\Pi_1, \dots, \Pi_m)
\label{eq:pi1}
\end{equation}
where $\Pi_i = b_i / (a_1^{p_{i1}} \dots a_k^{p_{ik}})$ for $i = 1, \dots,
m$. On page 42ff, the author comments that the term ``physical
relationship'' for $f$ ``is used to emphasize that it should obey the
covariance principle'' and, further, that the $\Pi$-theorem is
``completely obvious at an intuitive level'' and that ``it is clear that
physical laws should not depend on the choice of units'' and that this
``was realized long ago, and concepts from dimensional analysis were in
use long before the $\Pi$-theorem had been explicitly recognized,
formulated and proved formally'' among others, by ``Galileo, Newton,
Fourier, Maxwell, Reynolds and Rayleigh''.
Indeed, one of the most successful applications of the theorem was
Reynolds' scaling law for fluid flows in pipes in 1883, well before
Buckingham's seminal papers \citep{Buckingham1914, Buckingham1915}.

Here we do not discuss applications of the $\Pi$-theorem, but its
relevance for data analysis, parameter identification, and sensitivity
analysis is obvious: the computational cost of these algorithms is
typically exponential in the number of parameters. The theorem allows
cost reductions that are exponential in the number of parameters with
independent dimensions. In the case of $f$, for example, the theorem
allows the cost to be reduced from $N^{k + m}$ to $N^m$ where $N$ denotes
a sampling size.
In data-based modelling and machine learning,
this can make the difference between being able to solve a problem in
principle and being able to solve it in practice.

The bottom line is that, even though DA and the $\Pi$-theorem were
formulated at a time in which computer based modelling was not available
and were mainly motivated by the questions of model similarity mentioned
above, they remain highly relevant today.

For us, the challenge is to understand how to apply dependent types to
1) systematically check the dimensional consistency of expressions and
2) assist DA.

In \cref{section:dimensions}, we have seen that what in mathematical
physics and modelling are called physical quantities are equipped with
a dimension function.
In analogy with the judgment \ensuremath{\Varid{e}\ \mathop{:}\ \Varid{t}} (or, \ensuremath{\Varid{typeOf}\;\Varid{e}\mathrel{=}\Varid{t}}), we have
informally introduced the judgment \ensuremath{[\mskip1.5mu \Varid{e}\mskip1.5mu]\mathrel{=}\Varid{d}} to denote that expression
\ensuremath{\Varid{e}} has dimension function \ensuremath{\Varid{d}}. With the $\Pi$-theorem, we have seen
that physical quantities may have ``independent dimensions'' or be
dimensionally dependent.
In \cref{eq:pih1,eq:pi0} we have encountered (systems of) equations
between dimension functions whose solutions play a crucial role in the
$\Pi$-theorem \cref{eq:pi1}.
In the next section we come back to the notion of dimension function of
a physical quantity \ensuremath{[\mskip1.5mu \Varid{x}\mskip1.5mu]\ \mathop{:}\ \Real_{+}^n\to \Real_{+}}, discuss its
relationship with units of measurement, and argue that \ensuremath{[\mskip1.5mu \Varid{x}\mskip1.5mu]} is a
power-law monomial. This property is at the core of the dependently
typed formalization of DA presented in
\cref{section:dsl1,section:piexplained1}.


\section{A minimal DSL for dimensionally consistent programming}
\label{section:dsl1}

In this section we introduce a minimal domain-specific language for
dimensionally consistent programming in Idris and in the context of
classical mechanics.
We formalize the notions of dimension function, physical quantity,
measurement and units of measurement from \cref{section:dimensions} and
the notion of dimensional (in)dependence which are at the core of the Pi
theorem from \cref{section:pi}.

The DSL supports expressing dimensional judgments and dimensionally
consistent programming in the domain of classical mechanics. It can be
straightforwardly extended to higher dimensional domains (for example,
thermodynamics) or restricted to subdomains of mechanics like kinematics
or non-fractal geometry.

At the core of the DSL is the idea that any physical quantity can be
associated with a dimension function. We follow a concrete, minimalist
approach and encode dimension functions in terms of vectors of integers.

\subsection{Dimension function}
\label{subsection:df}

In \cref{section:dimensions}, we said that the dimension function
of a physical quantity \ensuremath{\Varid{x}}, \ensuremath{[\mskip1.5mu \Varid{x}\mskip1.5mu]\ \mathop{:}\ \Real_{+}^n\to \Real_{+}} encodes the idea
that \ensuremath{\Varid{x}} can be measured in a system of \ensuremath{\Varid{n}} fundamental units of
measurement and that the number \ensuremath{[\mskip1.5mu \Varid{x}\mskip1.5mu]\;(L_1,\dots,L_n)} denotes the factor
by which the measure of \ensuremath{\Varid{x}} increases (is multiplied) when the \ensuremath{\Varid{n}} units
are decreased (divided) by \ensuremath{L_1,\dots,L_n}.

We can give a precise meaning to this idea by denoting the measure of
\ensuremath{\Varid{x}} in the units of measurement \ensuremath{u_1,\dots,u_n} by \ensuremath{\mu\;(u_1,\dots,u_n)\;\Varid{x}}.
We will sometimes write \ensuremath{\Varid{u}} for the tuple \ensuremath{(u_1,\dots,u_n)} and shorten
the notation to \ensuremath{\mu_u\;\Varid{x}} for \ensuremath{\mu\;\Varid{u}\;\Varid{x}}.
For the time being, we posit that \ensuremath{\mu_u\;\Varid{x}\ \mathop{:}\ \Real} but defer the
specification of the types of \ensuremath{\Varid{x}} and \ensuremath{\Varid{u}} to
\cref{subsection:quantities,subsection:units}.
The reader should keep in mind that \ensuremath{\Varid{x}} could represent a velocity or a
stress tensor and the result of measuring \ensuremath{\Varid{x}} could be a vector or a
tensor of real numbers.

With these premises, we can make precise the notion of dimension
function through (for any non-zero \ensuremath{u_1}, \dots, \ensuremath{u_n}):
%
%
\begin{equation}
\text{\ensuremath{[\mskip1.5mu \Varid{x}\mskip1.5mu]\;(L_1,\dots,L_n)}} = \frac{\text{\ensuremath{\mu\;(u_1\mathbin{/}L_1,\dots,u_n\mathbin{/}L_n)\;\Varid{x}}}}{\text{\ensuremath{\mu\;(u_1,\dots,u_n)\;\Varid{x}}}}
\label{eq:df0}
\end{equation}
The specification suggests that the dimension function of \ensuremath{\Varid{x}} does not
depend on the units of measurement. It formalizes the principle of
covariance (or the relativity of measurements): that there is no
privileged system of units of measurement or, in other words, that all
systems are equally good:
\begin{equation}
  \frac{\text{\ensuremath{\mu\;(u_1\mathbin{/}L_1,\dots,u_n\mathbin{/}L_n)\;\Varid{x}}}}{\text{\ensuremath{\mu\;(u_1,\dots,u_n)\;\Varid{x}}}}
  =
  \frac{\text{\ensuremath{\mu\;(u_1^{\prime}\mathbin{/}L_1,\dots,u_n^{\prime}\mathbin{/}L_n)\;\Varid{x}}}}{\text{\ensuremath{\mu\;(u_1^{\prime},\dots,u_n^{\prime})\;\Varid{x}}}}
\label{eq:cov0}
\end{equation}
for any physical quantity \ensuremath{\Varid{x}}, systems of units
\ensuremath{\Varid{u}} and \ensuremath{\Varid{u'}}
and scaling factors \ensuremath{L_1,\dots,L_n}. It is easy to see that the
principle \ref{eq:cov0} implies that the dimension function fulfils
\begin{equation}
  \text{\ensuremath{[\mskip1.5mu \Varid{x}\mskip1.5mu]\;(L_1\mathbin{/}L_1^{\prime},\dots,L_n\mathbin{/}L_n^{\prime})}}
  =
  \frac{\text{\ensuremath{[\mskip1.5mu \Varid{x}\mskip1.5mu]\;(L_1,\dots,L_n)}}}{\text{\ensuremath{[\mskip1.5mu \Varid{x}\mskip1.5mu]\;(L_1^{\prime},\dots,L_n^{\prime})}}}
\label{eq:fun0}
\end{equation}
by equational reasoning
\begin{hscode}\SaveRestoreHook
\column{B}{@{}>{\hspre}l<{\hspost}@{}}%
\column{3}{@{}>{\hspre}l<{\hspost}@{}}%
\column{5}{@{}>{\hspre}l<{\hspost}@{}}%
\column{E}{@{}>{\hspre}l<{\hspost}@{}}%
\>[3]{}[\mskip1.5mu \Varid{x}\mskip1.5mu]\;(L_1,\dots,L_n)\mathbin{/}[\mskip1.5mu \Varid{x}\mskip1.5mu]\;(L_1^{\prime},\dots,L_n^{\prime}){}\<[E]%
\\[\blanklineskip]%
\>[3]{}\hsindent{2}{}\<[5]%
\>[5]{}\mathrel{=}\mbox{\onelinecomment  use \cref{eq:df0}: def. of \ensuremath{[\mskip1.5mu \Varid{x}\mskip1.5mu]} for units \ensuremath{(u_1,\dots,u_n)}}{}\<[E]%
\\[\blanklineskip]%
\>[3]{}\mu\;(u_1\mathbin{/}L_1,\dots,u_n\mathbin{/}L_n)\;\Varid{x}\mathbin{/}\mu\;(u_1\mathbin{/}L_1^{\prime},\dots,u_n\mathbin{/}L_n^{\prime})\;\Varid{x}{}\<[E]%
\\[\blanklineskip]%
\>[3]{}\hsindent{2}{}\<[5]%
\>[5]{}\mathrel{=}\mbox{\onelinecomment  let \ensuremath{u_1^{\prime}\mathrel{=}u_1\mathbin{/}L_1^{\prime}}, \dots, \ensuremath{u_n^{\prime}\mathrel{=}u_n\mathbin{/}L_n^{\prime}}}{}\<[E]%
\\[\blanklineskip]%
\>[3]{}\mu\;(u_1^{\prime}\mathbin{/}(L_1\mathbin{/}L_1^{\prime}),\dots,u_n^{\prime}\mathbin{/}(L_n\mathbin{/}L_n^{\prime}))\;\Varid{x}\mathbin{/}\mu\;(u_1^{\prime},\dots,u_n^{\prime})\;\Varid{x}{}\<[E]%
\\[\blanklineskip]%
\>[3]{}\hsindent{2}{}\<[5]%
\>[5]{}\mathrel{=}\mbox{\onelinecomment  use \cref{eq:df0}: def. of \ensuremath{[\mskip1.5mu \Varid{x}\mskip1.5mu]} for units \ensuremath{(u_1^{\prime},\dots,u_n^{\prime})}}{}\<[E]%
\\[\blanklineskip]%
\>[3]{}[\mskip1.5mu \Varid{x}\mskip1.5mu]\;(L_1\mathbin{/}L_1^{\prime},\dots,L_n\mathbin{/}L_n^{\prime}){}\<[E]%
\ColumnHook
\end{hscode}\resethooks
\noindent
From \cref{eq:fun0} it follows that
\begin{equation}
\text{\ensuremath{[\mskip1.5mu \Varid{x}\mskip1.5mu]\;(\mathrm{1},\dots,\mathrm{1})\mathrel{=}\mathrm{1}}}
\label{eq:fun1}
\end{equation}
and that dimension functions have the form of power-law monomials
\begin{equation}
\text{\ensuremath{[\mskip1.5mu \Varid{x}\mskip1.5mu]\;(L_1,\dots,L_n)}} = L_1^{\ensuremath{d_1\;\Varid{x}}} \ensuremath{\ensuremath{\cdot}} \dots \ensuremath{\ensuremath{\cdot}} L_n^{d_n \; x}
\label{eq:fun2}
\end{equation}
The derivation of this power-law form is not straightforward and
involves solving the functional equation \cref{eq:fun0}, see
\citep[section 1.1.4]{barenblatt1996scaling}.
The exponents $d_1 \; x$, \dots $d_n \; x$ are sometimes called (perhaps
confusingly) the ``dimensions'' of \ensuremath{\Varid{x}} and \ensuremath{\Varid{x}} is said to ``have
dimensions'' $d_1 \; x$, \dots $d_n \; x$. Their value can be obtained by
recalling the specification \cref{eq:df0}.

For concreteness, consider the case of mechanics already discussed in
\cref{section:dimensions}. Here \ensuremath{\Varid{n}\mathrel{=}\mathrm{3}} and, in a system of units
of measurements for lengths, times and masses, the scaling factors
$L_1$, $L_2$ and $L_3$ are denoted by \ensuremath{\Conid{L}}, \ensuremath{\Conid{T}} and \ensuremath{\Conid{M}}. For consistency,
we denote the exponents $d_1 \; x$, $d_2 \; x$ and $d_3 \; x$ by $d_L \; x$,
$d_T \; x$ and $d_M \; x$, respectively.

Thus, in mechanics, \cref{eq:df0} tells us that $L^{d_L \; x} \ensuremath{\ensuremath{\cdot}}
T^{d_T \; x} \ensuremath{\ensuremath{\cdot}} M^{d_M \; x}$ is the factor by which the measure
of \ensuremath{\Varid{x}} gets multiplied when the units of measurement for lengths, times, and masses
are divided by \ensuremath{\Conid{L}}, \ensuremath{\Conid{T}} and \ensuremath{\Conid{M}}.
Therefore, when \ensuremath{\Varid{x}} represents a length (for example, the distance
between two points or a space coordinate) we have $d_L \ x = 1$ and
$d_T \ x = d_M \ x = 0$. Similarly, when \ensuremath{\Varid{x}} represents a mass we have $d_M
\ x = 1$ and $d_L \ x = d_T \ x = 0$ and when \ensuremath{\Varid{x}} represents a time we
have $d_T \ x = 1$ and $d_L \ x = d_M \ x = 0$. And when \ensuremath{\Varid{x}} represents
a velocity (a distance divided by a time), the factor by which
the measure of \ensuremath{\Varid{x}} gets multiplied when the units of measurement for
lengths, times, and masses are divided by \ensuremath{\Conid{L}}, \ensuremath{\Conid{T}} and \ensuremath{\Conid{M}} shall be $L /
T$ and thus $d_L \ x = 1$, $d_T \ x = -1$, and $d_M \ x = 0$ .

These judgments are a consequence of the notion of (direct or indirect)
measurement as \emph{counting}: when we say that the length of a pen is
20 centimetres we mean that we have to add 20 times the length of a
centimetre to obtain the length of that pen.

The above analysis suggests that, in classical mechanics, the type of
dimension functions is isomorphic to \ensuremath{\mathbb{Z}^3}, see also page 3 of
\citet{doi:10.1142/9789811242380_0020}. Thus, in this domain, we can
represent a dimension function with a vector of integers of length 3:

\begin{joincode}%
\begin{hscode}\SaveRestoreHook
\column{B}{@{}>{\hspre}l<{\hspost}@{}}%
\column{3}{@{}>{\hspre}l<{\hspost}@{}}%
\column{5}{@{}>{\hspre}l<{\hspost}@{}}%
\column{E}{@{}>{\hspre}l<{\hspost}@{}}%
\>[B]{}\mathbf{namespace}\;\Conid{Mechanics}\;{}\<[E]%
\\[\blanklineskip]%
\>[B]{}\hsindent{3}{}\<[3]%
\>[3]{}\mathbf{namespace}\;\Conid{LTM}{}\<[E]%
\\[\blanklineskip]%
\>[3]{}\hsindent{2}{}\<[5]%
\>[5]{}\mathbf{data}\;\Conid{Units}\mathrel{=}\Conid{SI}\mid \Conid{CGS}{}\<[E]%
\\[\blanklineskip]%
\>[3]{}\hsindent{2}{}\<[5]%
\>[5]{}\Conid{D}\ \mathop{:}\ \Conid{Type}{}\<[E]%
\\
\>[3]{}\hsindent{2}{}\<[5]%
\>[5]{}\Conid{D}\mathrel{=}\Conid{Vec}\;\mathrm{3}\;\mathbb{Z}{}\<[E]%
\ColumnHook
\end{hscode}\resethooks
\end{joincode}
\noindent
Here, we have embedded the datatypes \ensuremath{\Conid{Units}} (with just two codes for
the \ensuremath{\Conid{SI}} and \ensuremath{\Conid{CGS}} systems of units for simplicity) and \ensuremath{\Conid{D}} in the
namespaces \ensuremath{\Conid{Mechanics}} and \ensuremath{\Conid{LTM}}, the latter representing the class of
units for lengths, times and masses, see \cref{section:dimensions}.
Following our analysis, we can model the syntax of dimensional
expressions in terms of a \ensuremath{\Conid{DimLess}} vector for dimensionless quantities,
of \ensuremath{\Conid{Length}}, \ensuremath{\Conid{Time}} and \ensuremath{\Conid{Mass}} vectors for lengths, times and masses
(the dimensions associated with the fundamental units of measurement in
\ensuremath{\Conid{LTM}})

\begin{hscode}\SaveRestoreHook
\column{B}{@{}>{\hspre}l<{\hspost}@{}}%
\column{5}{@{}>{\hspre}l<{\hspost}@{}}%
\column{14}{@{}>{\hspre}c<{\hspost}@{}}%
\column{14E}{@{}l@{}}%
\column{17}{@{}>{\hspre}l<{\hspost}@{}}%
\column{39}{@{}>{\hspre}l<{\hspost}@{}}%
\column{47}{@{}>{\hspre}c<{\hspost}@{}}%
\column{47E}{@{}l@{}}%
\column{50}{@{}>{\hspre}l<{\hspost}@{}}%
\column{72}{@{}>{\hspre}l<{\hspost}@{}}%
\column{78}{@{}>{\hspre}c<{\hspost}@{}}%
\column{78E}{@{}l@{}}%
\column{81}{@{}>{\hspre}l<{\hspost}@{}}%
\column{103}{@{}>{\hspre}l<{\hspost}@{}}%
\column{109}{@{}>{\hspre}c<{\hspost}@{}}%
\column{109E}{@{}l@{}}%
\column{112}{@{}>{\hspre}l<{\hspost}@{}}%
\column{E}{@{}>{\hspre}l<{\hspost}@{}}%
\>[5]{}\Conid{DimLess}{}\<[14]%
\>[14]{}\ \mathop{:}\ {}\<[14E]%
\>[17]{}\Conid{D};{}\<[39]%
\>[39]{}\Conid{Length}{}\<[47]%
\>[47]{}\ \mathop{:}\ {}\<[47E]%
\>[50]{}\Conid{D};{}\<[72]%
\>[72]{}\Conid{Time}{}\<[78]%
\>[78]{}\ \mathop{:}\ {}\<[78E]%
\>[81]{}\Conid{D};{}\<[103]%
\>[103]{}\Conid{Mass}{}\<[109]%
\>[109]{}\ \mathop{:}\ {}\<[109E]%
\>[112]{}\Conid{D}{}\<[E]%
\\
\>[5]{}\Conid{DimLess}{}\<[14]%
\>[14]{}\mathrel{=}{}\<[14E]%
\>[17]{}[\mskip1.5mu \mathrm{0},\mathrm{0},\mathrm{0}\mskip1.5mu];\quad{}\<[39]%
\>[39]{}\Conid{Length}{}\<[47]%
\>[47]{}\mathrel{=}{}\<[47E]%
\>[50]{}[\mskip1.5mu \mathrm{1},\mathrm{0},\mathrm{0}\mskip1.5mu];\quad{}\<[72]%
\>[72]{}\Conid{Time}{}\<[78]%
\>[78]{}\mathrel{=}{}\<[78E]%
\>[81]{}[\mskip1.5mu \mathrm{0},\mathrm{1},\mathrm{0}\mskip1.5mu];\quad{}\<[103]%
\>[103]{}\Conid{Mass}{}\<[109]%
\>[109]{}\mathrel{=}{}\<[109E]%
\>[112]{}[\mskip1.5mu \mathrm{0},\mathrm{0},\mathrm{1}\mskip1.5mu]{}\<[E]%
\ColumnHook
\end{hscode}\resethooks
and of two combinators \ensuremath{\Conid{Times}} and \ensuremath{\Conid{Over}}:
\begin{hscode}\SaveRestoreHook
\column{B}{@{}>{\hspre}l<{\hspost}@{}}%
\column{5}{@{}>{\hspre}l<{\hspost}@{}}%
\column{14}{@{}>{\hspre}c<{\hspost}@{}}%
\column{14E}{@{}l@{}}%
\column{17}{@{}>{\hspre}l<{\hspost}@{}}%
\column{31}{@{}>{\hspre}l<{\hspost}@{}}%
\column{44}{@{}>{\hspre}l<{\hspost}@{}}%
\column{53}{@{}>{\hspre}c<{\hspost}@{}}%
\column{53E}{@{}l@{}}%
\column{56}{@{}>{\hspre}l<{\hspost}@{}}%
\column{E}{@{}>{\hspre}l<{\hspost}@{}}%
\>[5]{}\Conid{Times}{}\<[14]%
\>[14]{}\ \mathop{:}\ {}\<[14E]%
\>[17]{}\Conid{D}\to \Conid{D}\to \Conid{D};{}\<[31]%
\>[31]{}\quad{}\<[44]%
\>[44]{}\Conid{Over}{}\<[53]%
\>[53]{}\ \mathop{:}\ {}\<[53E]%
\>[56]{}\Conid{D}\to \Conid{D}\to \Conid{D}{}\<[E]%
\\
\>[5]{}\Conid{Times}{}\<[14]%
\>[14]{}\mathrel{=}{}\<[14E]%
\>[17]{}(\mathbin{+});{}\<[44]%
\>[44]{}\Conid{Over}{}\<[53]%
\>[53]{}\mathrel{=}{}\<[53E]%
\>[56]{}(\mathbin{-}){}\<[E]%
\ColumnHook
\end{hscode}\resethooks
where \ensuremath{(\mathbin{+})} and \ensuremath{(\mathbin{-})} are the canonical addition and subtraction between
vectors of integer numbers.
These correspond to the idea that the dimensions of derived units of
measurement (for example, for velocities or for energies) are obtained
by multiplying or by dividing the dimensions of the fundamental units:

\begin{hscode}\SaveRestoreHook
\column{B}{@{}>{\hspre}l<{\hspost}@{}}%
\column{5}{@{}>{\hspre}l<{\hspost}@{}}%
\column{53}{@{}>{\hspre}l<{\hspost}@{}}%
\column{E}{@{}>{\hspre}l<{\hspost}@{}}%
\>[5]{}\Conid{Velocity}\ \mathop{:}\ \Conid{D};{}\<[53]%
\>[53]{}\Conid{Acceleration}\ \mathop{:}\ \Conid{D}{}\<[E]%
\\
\>[5]{}\Conid{Velocity}\mathrel{=}\Conid{Length}\mathbin{`\Conid{Over}`}\Conid{Time};\quad{}\<[53]%
\>[53]{}\Conid{Acceleration}\mathrel{=}\Conid{Velocity}\mathbin{`\Conid{Over}`}\Conid{Time}{}\<[E]%
\\[\blanklineskip]%
\>[5]{}\Conid{Force}\ \mathop{:}\ \Conid{D};{}\<[53]%
\>[53]{}\Conid{Work}\ \mathop{:}\ \Conid{D}{}\<[E]%
\\
\>[5]{}\Conid{Force}\mathrel{=}\Conid{Mass}\mathbin{`\Conid{Times}`}\Conid{Acceleration};\quad{}\<[53]%
\>[53]{}\Conid{Work}\mathrel{=}\Conid{Force}\mathbin{`\Conid{Times}`}\Conid{Length}{}\<[E]%
\\[\blanklineskip]%
\>[5]{}\Conid{Energy}\ \mathop{:}\ \Conid{D}{}\<[E]%
\\
\>[5]{}\Conid{Energy}\mathrel{=}\Conid{Mass}\mathbin{`\Conid{Times}`}(\Conid{Velocity}\mathbin{`\Conid{Times}`}\Conid{Velocity}){}\<[E]%
\ColumnHook
\end{hscode}\resethooks
One can easily check that energy and mechanical work are equivalent, as
one would expect
\begin{hscode}\SaveRestoreHook
\column{B}{@{}>{\hspre}l<{\hspost}@{}}%
\column{5}{@{}>{\hspre}l<{\hspost}@{}}%
\column{E}{@{}>{\hspre}l<{\hspost}@{}}%
\>[5]{}\Varid{check}_{1 }\ \mathop{:}\ \Conid{Energy}\mathrel{=}\Conid{Work}{}\<[E]%
\\
\>[5]{}\Varid{check}_{1 }\mathrel{=}\Conid{Refl}{}\<[E]%
\ColumnHook
\end{hscode}\resethooks
that force and energy are different notions
\begin{hscode}\SaveRestoreHook
\column{B}{@{}>{\hspre}l<{\hspost}@{}}%
\column{5}{@{}>{\hspre}l<{\hspost}@{}}%
\column{E}{@{}>{\hspre}l<{\hspost}@{}}%
\>[5]{}\Varid{check}_{2 }\ \mathop{:}\ \Conid{Not}\;(\Conid{Force}\mathrel{=}\Conid{Energy}){}\<[E]%
\\
\>[5]{}\Varid{check}_{2 }\;\Conid{Refl}\;\Varid{impossible}{}\<[E]%
\ColumnHook
\end{hscode}\resethooks
and, we can compute the dimension functions of \ensuremath{\Conid{D}}-values:
\begin{hscode}\SaveRestoreHook
\column{B}{@{}>{\hspre}l<{\hspost}@{}}%
\column{5}{@{}>{\hspre}l<{\hspost}@{}}%
\column{E}{@{}>{\hspre}l<{\hspost}@{}}%
\>[5]{}\Varid{df}\ \mathop{:}\ \Conid{D}\to \Real_{+}^3\to \Real_{+}{}\<[E]%
\\
\>[5]{}\Varid{df}\;\Varid{d}\;\Varid{ls}\mathrel{=}\Varid{prodReal}\;(\Varid{zipWith}\;\Varid{ipow}\;\Varid{ls}\;\Varid{d}){}\<[E]%
\ColumnHook
\end{hscode}\resethooks
In the definition of \ensuremath{\Varid{df}}, \ensuremath{\Varid{ipow}} is the function that computes the
integer power of a real number and \ensuremath{\Varid{prodReal}\mathrel{=}\Varid{foldr}\;(\lambda \Varid{m}\to \mathbb{R} )\;(\ensuremath{\cdot})\;\mathrm{1.0}}.
The dimension function satisfies \cref{eq:fun2} by construction. One can
see that it also satisfies \cref{eq:fun0} (this is \ensuremath{\Varid{dfLemma1}} with
\ensuremath{\Varid{ls}\mathrel{=}(L_1\mathbin{/}L_1^{\prime},\dots,L_n\mathbin{/}L_n^{\prime})}, \ensuremath{\Varid{ls'}\mathrel{=}(L_1^{\prime},\dots,L_n^{\prime})} and with the
componentwise multiplication \ensuremath{\Varid{ls}\ensuremath{\cdot}\Varid{ls'}}), \cref{eq:fun1} that is, \ensuremath{\Varid{dfLemma2}}
\begin{joincode}%
\begin{hscode}\SaveRestoreHook
\column{B}{@{}>{\hspre}l<{\hspost}@{}}%
\column{5}{@{}>{\hspre}l<{\hspost}@{}}%
\column{56}{@{}>{\hspre}l<{\hspost}@{}}%
\column{E}{@{}>{\hspre}l<{\hspost}@{}}%
\>[5]{}\Varid{dfLemma1}\ \mathop{:}\ (\Varid{d}\ \mathop{:}\ \Conid{D})\to (\Varid{ls},\Varid{ls'}\ \mathop{:}\ \Real_{+}^3)\to {}\<[56]%
\>[56]{}\Varid{df}\;\Varid{d}\;\Varid{ls}\ensuremath{\cdot}\Varid{df}\;\Varid{d}\;\Varid{ls'}\mathrel{=}\Varid{df}\;\Varid{d}\;(\Varid{ls}\ensuremath{\cdot}\Varid{ls'}){}\<[E]%
\ColumnHook
\end{hscode}\resethooks
\begin{hscode}\SaveRestoreHook
\column{B}{@{}>{\hspre}l<{\hspost}@{}}%
\column{5}{@{}>{\hspre}l<{\hspost}@{}}%
\column{E}{@{}>{\hspre}l<{\hspost}@{}}%
\>[5]{}\Varid{dfLemma2}\ \mathop{:}\ (\Varid{d}\ \mathop{:}\ \Conid{D})\to \Varid{df}\;\Varid{d}\;\Varid{one3}\mathrel{=}\mathrm{1.0}{}\<[E]%
\ColumnHook
\end{hscode}\resethooks
\end{joincode}
and the following properties
\begin{joincode}%
\begin{hscode}\SaveRestoreHook
\column{B}{@{}>{\hspre}l<{\hspost}@{}}%
\column{5}{@{}>{\hspre}l<{\hspost}@{}}%
\column{56}{@{}>{\hspre}l<{\hspost}@{}}%
\column{E}{@{}>{\hspre}l<{\hspost}@{}}%
\>[5]{}\Varid{dfLemma3}\ \mathop{:}\ \{\mskip1.5mu \Varid{d}\ \mathop{:}\ \Conid{D}\mskip1.5mu\}\to \{\mskip1.5mu \Varid{u},\Varid{v},\Varid{w}\ \mathop{:}\ \Conid{Units}\mskip1.5mu\}\to {}\<[56]%
\>[56]{}\Varid{df}\;\Varid{d}\;(\Varid{fs}\;\Varid{u}\;\Varid{v})\ensuremath{\cdot}\Varid{df}\;\Varid{d}\;(\Varid{fs}\;\Varid{v}\;\Varid{w})\mathrel{=}\Varid{df}\;\Varid{d}\;(\Varid{fs}\;\Varid{u}\;\Varid{w}){}\<[E]%
\ColumnHook
\end{hscode}\resethooks
\begin{hscode}\SaveRestoreHook
\column{B}{@{}>{\hspre}l<{\hspost}@{}}%
\column{5}{@{}>{\hspre}l<{\hspost}@{}}%
\column{E}{@{}>{\hspre}l<{\hspost}@{}}%
\>[5]{}\Varid{dfDimLessLemma}\ \mathop{:}\ (\Varid{ls}\ \mathop{:}\ \Real_{+}^3)\to \Varid{df}\;\Conid{DimLess}\;\Varid{ls}\mathrel{=}\mathrm{1.0}{}\<[E]%
\ColumnHook
\end{hscode}\resethooks
\begin{hscode}\SaveRestoreHook
\column{B}{@{}>{\hspre}l<{\hspost}@{}}%
\column{5}{@{}>{\hspre}l<{\hspost}@{}}%
\column{17}{@{}>{\hspre}c<{\hspost}@{}}%
\column{17E}{@{}l@{}}%
\column{20}{@{}>{\hspre}l<{\hspost}@{}}%
\column{56}{@{}>{\hspre}l<{\hspost}@{}}%
\column{72}{@{}>{\hspre}l<{\hspost}@{}}%
\column{82}{@{}>{\hspre}l<{\hspost}@{}}%
\column{92}{@{}>{\hspre}c<{\hspost}@{}}%
\column{92E}{@{}l@{}}%
\column{95}{@{}>{\hspre}l<{\hspost}@{}}%
\column{E}{@{}>{\hspre}l<{\hspost}@{}}%
\>[5]{}\Varid{dfHomTimes}{}\<[17]%
\>[17]{}\ \mathop{:}\ {}\<[17E]%
\>[20]{}\{\mskip1.5mu d_1,d_2\ \mathop{:}\ \Conid{D}\mskip1.5mu\}\to \{\mskip1.5mu \Varid{ls}\ \mathop{:}\ \Real_{+}^3\mskip1.5mu\}\to {}\<[56]%
\>[56]{}\Varid{df}\;(d_1\mathbin{`\Conid{Times}`}{}\<[72]%
\>[72]{}d_2)\;\Varid{ls}\mathrel{=}{}\<[82]%
\>[82]{}\Varid{df}\;d_1\;\Varid{ls}{}\<[92]%
\>[92]{}\ensuremath{\cdot}{}\<[92E]%
\>[95]{}\Varid{df}\;d_2\;\Varid{ls}{}\<[E]%
\ColumnHook
\end{hscode}\resethooks
\begin{hscode}\SaveRestoreHook
\column{B}{@{}>{\hspre}l<{\hspost}@{}}%
\column{5}{@{}>{\hspre}l<{\hspost}@{}}%
\column{17}{@{}>{\hspre}c<{\hspost}@{}}%
\column{17E}{@{}l@{}}%
\column{20}{@{}>{\hspre}l<{\hspost}@{}}%
\column{56}{@{}>{\hspre}l<{\hspost}@{}}%
\column{72}{@{}>{\hspre}l<{\hspost}@{}}%
\column{82}{@{}>{\hspre}l<{\hspost}@{}}%
\column{92}{@{}>{\hspre}c<{\hspost}@{}}%
\column{92E}{@{}l@{}}%
\column{95}{@{}>{\hspre}l<{\hspost}@{}}%
\column{E}{@{}>{\hspre}l<{\hspost}@{}}%
\>[5]{}\Varid{dfHomOver}{}\<[17]%
\>[17]{}\ \mathop{:}\ {}\<[17E]%
\>[20]{}\{\mskip1.5mu d_1,d_2\ \mathop{:}\ \Conid{D}\mskip1.5mu\}\to \{\mskip1.5mu \Varid{ls}\ \mathop{:}\ \Real_{+}^3\mskip1.5mu\}\to {}\<[56]%
\>[56]{}\Varid{df}\;(d_1\mathbin{`\Conid{Over}`}{}\<[72]%
\>[72]{}d_2)\;\Varid{ls}\mathrel{=}{}\<[82]%
\>[82]{}\Varid{df}\;d_1\;\Varid{ls}{}\<[92]%
\>[92]{}\mathbin{/}{}\<[92E]%
\>[95]{}\Varid{df}\;d_2\;\Varid{ls}{}\<[E]%
\ColumnHook
\end{hscode}\resethooks
\end{joincode}
These follow from elementary properties of real number and integer power arithmetic (see the Idris and Agda code in the repository \citep{Pi2023})
%
%
and are crucial for ensuring that measurements of physical quantities
fulfil the covariance principle and for the results to be discussed in
\cref{subsection:piexplained1.1}.

Notice that for both \ensuremath{\Conid{D}} and \ensuremath{\Conid{Units}} we use datatypes of codes (just
syntactic expressions) and that \ensuremath{\Varid{df}} here and \ensuremath{\Varid{fs}} later
(\cref{subsection:units}) are used to translate these codes to their
intended semantics.
The use of integers for the exponents of the dimension function makes
dimensional judgments decidable, which would not hold for real
numbered exponents, or functions.

We come back to the choice of types in \cref{section:generalization}
where we put forward specifications for data types that implement
dimension functions in terms of type classes. For the rest of this
section, we stick to the example of mechanics and to the
representation of dimension functions in terms of vectors of three
integer exponents.

\subsection{Physical quantities and homomorphic measurement}
\label{subsection:quantities}

We are now ready to formalize the notion of physical quantity informally
introduced in \cref{section:dimensions}. There, we posited that a
parameter (e.g., the parameter \ensuremath{\Varid{c}} of the Stommel model discussed in
\cref{subsection:laws}) represents a \emph{dimensional} physical
quantity when 1) that parameter can be measured in a system of units of
a given class and 2) one can define another system of units in the same
class that gives a different measurement for the parameter.
By contrast, measurements of \emph{dimensionless} physical quantities
do not change when the units of measurement are re-scaled.

This suggests that a (dimensional or dimensionless) physical quantity
can be represented by a value of type
\begin{hscode}\SaveRestoreHook
\column{B}{@{}>{\hspre}l<{\hspost}@{}}%
\column{5}{@{}>{\hspre}l<{\hspost}@{}}%
\column{7}{@{}>{\hspre}l<{\hspost}@{}}%
\column{E}{@{}>{\hspre}l<{\hspost}@{}}%
\>[5]{}\mathbf{data}\;\Conid{Q}\ \mathop{:}\ \Conid{D}\to \Conid{Type}\;\mathbf{where}{}\<[E]%
\\
\>[5]{}\hsindent{2}{}\<[7]%
\>[7]{}\Conid{Val}\ \mathop{:}\ \{\mskip1.5mu \Varid{d}\ \mathop{:}\ \Conid{D}\mskip1.5mu\}\to (\Varid{u}\ \mathop{:}\ \Conid{Units})\to \Real\to \Conid{Q}\;\Varid{d}{}\<[E]%
\ColumnHook
\end{hscode}\resethooks
\noindent
effectively annotating \ensuremath{\Real} values with different dimensions and
systems of units, for example
\begin{hscode}\SaveRestoreHook
\column{B}{@{}>{\hspre}l<{\hspost}@{}}%
\column{5}{@{}>{\hspre}l<{\hspost}@{}}%
\column{8}{@{}>{\hspre}c<{\hspost}@{}}%
\column{8E}{@{}l@{}}%
\column{11}{@{}>{\hspre}l<{\hspost}@{}}%
\column{22}{@{}>{\hspre}l<{\hspost}@{}}%
\column{35}{@{}>{\hspre}l<{\hspost}@{}}%
\column{38}{@{}>{\hspre}c<{\hspost}@{}}%
\column{38E}{@{}l@{}}%
\column{41}{@{}>{\hspre}l<{\hspost}@{}}%
\column{53}{@{}>{\hspre}l<{\hspost}@{}}%
\column{66}{@{}>{\hspre}l<{\hspost}@{}}%
\column{69}{@{}>{\hspre}c<{\hspost}@{}}%
\column{69E}{@{}l@{}}%
\column{72}{@{}>{\hspre}l<{\hspost}@{}}%
\column{E}{@{}>{\hspre}l<{\hspost}@{}}%
\>[5]{}\Varid{x}{}\<[8]%
\>[8]{}\ \mathop{:}\ {}\<[8E]%
\>[11]{}\Conid{Q}\;\Conid{Length};{}\<[22]%
\>[22]{}\quad{}\<[35]%
\>[35]{}\Varid{t}{}\<[38]%
\>[38]{}\ \mathop{:}\ {}\<[38E]%
\>[41]{}\Conid{Q}\;\Conid{Time};{}\<[53]%
\>[53]{}\quad{}\<[66]%
\>[66]{}\Varid{m}{}\<[69]%
\>[69]{}\ \mathop{:}\ {}\<[69E]%
\>[72]{}\Conid{Q}\;\Conid{Mass}{}\<[E]%
\\
\>[5]{}\Varid{x}{}\<[8]%
\>[8]{}\mathrel{=}{}\<[8E]%
\>[11]{}\Conid{Val}\;\Conid{SI}\;\mathrm{3};{}\<[22]%
\>[22]{}\quad{}\<[35]%
\>[35]{}\Varid{t}{}\<[38]%
\>[38]{}\mathrel{=}{}\<[38E]%
\>[41]{}\Conid{Val}\;\Conid{CGS}\;\mathrm{1};{}\<[53]%
\>[53]{}\quad{}\<[66]%
\>[66]{}\Varid{m}{}\<[69]%
\>[69]{}\mathrel{=}{}\<[69E]%
\>[72]{}\Conid{Val}\;\Conid{SI}\;\mathrm{2}{}\<[E]%
\ColumnHook
\end{hscode}\resethooks
\noindent
What kind of combinators are required for physical quantities? As a
minimum, one wants to be able to project the \ensuremath{\Conid{D}}-value of a physical
quantity
\begin{hscode}\SaveRestoreHook
\column{B}{@{}>{\hspre}l<{\hspost}@{}}%
\column{5}{@{}>{\hspre}l<{\hspost}@{}}%
\column{E}{@{}>{\hspre}l<{\hspost}@{}}%
\>[5]{}\Varid{dim}\ \mathop{:}\ \{\mskip1.5mu \Varid{d}\ \mathop{:}\ \Conid{D}\mskip1.5mu\}\to \Conid{Q}\;\Varid{d}\to \Conid{D}{}\<[E]%
\\
\>[5]{}\Varid{dim}\;\{\mskip1.5mu \Varid{d}\mskip1.5mu\}\;\anonymous \mathrel{=}\Varid{d}{}\<[E]%
\ColumnHook
\end{hscode}\resethooks
and thus compute its dimension function \ensuremath{\Varid{df}\mathbin{\circ}\Varid{dim}}. Crucially, we need
a way to measure physical quantities in different units of measurement
\begin{hscode}\SaveRestoreHook
\column{B}{@{}>{\hspre}l<{\hspost}@{}}%
\column{5}{@{}>{\hspre}l<{\hspost}@{}}%
\column{E}{@{}>{\hspre}l<{\hspost}@{}}%
\>[5]{}\mu\ \mathop{:}\ \{\mskip1.5mu \Varid{d}\ \mathop{:}\ \Conid{D}\mskip1.5mu\}\to \Conid{Units}\to \Conid{Q}\;\Varid{d}\to \Real{}\<[E]%
\\
\>[5]{}\mu\;\{\mskip1.5mu \Varid{d}\mskip1.5mu\}\;\Varid{u'}\;(\Conid{Val}\;\Varid{u}\;\Varid{x})\mathrel{=}\Varid{x}\ensuremath{\cdot}\Varid{df}\;\Varid{d}\;(\Varid{fs}\;\Varid{u}\;\Varid{u'}){}\<[E]%
\ColumnHook
\end{hscode}\resethooks
In the definition of \ensuremath{\mu}, \ensuremath{\Varid{fs}} is the table that returns the scaling
factors between units of measurement. It fulfills
\begin{hscode}\SaveRestoreHook
\column{B}{@{}>{\hspre}l<{\hspost}@{}}%
\column{5}{@{}>{\hspre}l<{\hspost}@{}}%
\column{15}{@{}>{\hspre}c<{\hspost}@{}}%
\column{15E}{@{}l@{}}%
\column{18}{@{}>{\hspre}l<{\hspost}@{}}%
\column{37}{@{}>{\hspre}c<{\hspost}@{}}%
\column{37E}{@{}l@{}}%
\column{41}{@{}>{\hspre}l<{\hspost}@{}}%
\column{E}{@{}>{\hspre}l<{\hspost}@{}}%
\>[5]{}\Varid{fsLemma}_1{}\<[15]%
\>[15]{}\ \mathop{:}\ {}\<[15E]%
\>[18]{}(\Varid{u}\ \mathop{:}\ \Conid{Units}){}\<[37]%
\>[37]{}\to {}\<[37E]%
\>[41]{}\Varid{fs}\;\Varid{u}\;\Varid{u}\mathrel{=}[\mskip1.5mu \mathrm{1},\mathrm{1},\mathrm{1}\mskip1.5mu]{}\<[E]%
\\
\>[5]{}\Varid{fsLemma}_2{}\<[15]%
\>[15]{}\ \mathop{:}\ {}\<[15E]%
\>[18]{}(\Varid{u},\Varid{v}\ \mathop{:}\ \Conid{Units}){}\<[37]%
\>[37]{}\to {}\<[37E]%
\>[41]{}\Varid{fs}\;\Varid{u}\;\Varid{v}\mathrel{=}\Varid{invRealPos3}\;(\Varid{fs}\;\Varid{v}\;\Varid{u}){}\<[E]%
\\
\>[5]{}\Varid{fsLemma}_3{}\<[15]%
\>[15]{}\ \mathop{:}\ {}\<[15E]%
\>[18]{}(\Varid{u},\Varid{v},\Varid{w}\ \mathop{:}\ \Conid{Units}){}\<[37]%
\>[37]{}\to {}\<[37E]%
\>[41]{}(\Varid{fs}\;\Varid{u}\;\Varid{v})\ensuremath{\cdot}(\Varid{fs}\;\Varid{v}\;\Varid{w})\mathrel{=}\Varid{fs}\;\Varid{u}\;\Varid{w}{}\<[E]%
\ColumnHook
\end{hscode}\resethooks
by construction. Remember that, as discussed in
\cref{section:dimensions}, measurements of dimensionless physical
quantities are to be the same in all units of measurement. In other
words, \ensuremath{\mu} has to fulfil
\begin{hscode}\SaveRestoreHook
\column{B}{@{}>{\hspre}l<{\hspost}@{}}%
\column{5}{@{}>{\hspre}l<{\hspost}@{}}%
\column{E}{@{}>{\hspre}l<{\hspost}@{}}%
\>[5]{}\Varid{measDimLessLemma}\ \mathop{:}\ \{\mskip1.5mu \Varid{u},\Varid{u'}\ \mathop{:}\ \Conid{Units}\mskip1.5mu\}\to (\Varid{q}\ \mathop{:}\ \Conid{Q}\;\Conid{DimLess})\to \mu_{\Varid{u}}\;\Varid{q}\mathrel{=}\mu_{\Varid{u'}}\;\Varid{q}{}\<[E]%
\ColumnHook
\end{hscode}\resethooks
This lemma is a straightforward consequence of \cref{eq:fun1} that is
\ensuremath{\Varid{dfDimLessLemma}}. Further, it is useful to define elementary binary
operations \ensuremath{(\mathbin{+})}, \ensuremath{(\ensuremath{\cdot})}, \ensuremath{(\ensuremath{\triangleleft})}:
\begin{hscode}\SaveRestoreHook
\column{B}{@{}>{\hspre}l<{\hspost}@{}}%
\column{5}{@{}>{\hspre}l<{\hspost}@{}}%
\column{10}{@{}>{\hspre}c<{\hspost}@{}}%
\column{10E}{@{}l@{}}%
\column{13}{@{}>{\hspre}l<{\hspost}@{}}%
\column{E}{@{}>{\hspre}l<{\hspost}@{}}%
\>[5]{}(\mathbin{+}){}\<[10]%
\>[10]{}\ \mathop{:}\ {}\<[10E]%
\>[13]{}\{\mskip1.5mu \Varid{d}\ \mathop{:}\ \Conid{D}\mskip1.5mu\}\to \Conid{Q}\;\Varid{d}\to \Conid{Q}\;\Varid{d}\to \Conid{Q}\;\Varid{d}{}\<[E]%
\\
\>[5]{}q_1\mathbin{+}q_2\mathrel{=}\Conid{Val}\;\Conid{SI}\;(\mu_{\Conid{SI}}\;q_1\mathbin{+}\mu_{\Conid{SI}}\;q_2){}\<[E]%
\\[\blanklineskip]%
\>[5]{}(\ensuremath{\cdot}){}\<[10]%
\>[10]{}\ \mathop{:}\ {}\<[10E]%
\>[13]{}\{\mskip1.5mu d_1,d_2\ \mathop{:}\ \Conid{D}\mskip1.5mu\}\to \Conid{Q}\;d_1\to \Conid{Q}\;d_2\to \Conid{Q}\;(d_1\mathbin{`\Conid{Times}`}d_2){}\<[E]%
\\
\>[5]{}q_1\ensuremath{\cdot}q_2\mathrel{=}\Conid{Val}\;\Conid{SI}\;(\mu_{\Conid{SI}}\;q_1\ensuremath{\cdot}\mu_{\Conid{SI}}\;q_2){}\<[E]%
\\[\blanklineskip]%
\>[5]{}(\ensuremath{\triangleleft})\ \mathop{:}\ \{\mskip1.5mu \Varid{d}\ \mathop{:}\ \Conid{D}\mskip1.5mu\}\to \Real\to \Conid{Q}\;\Varid{d}\to \Conid{Q}\;\Varid{d}{}\<[E]%
\\
\>[5]{}\Varid{x}\ensuremath{\triangleleft}(\Conid{Val}\;\Varid{u}\;\Varid{y})\mathrel{=}\Conid{Val}\;\Varid{u}\;(\Varid{x}\ensuremath{\cdot}\Varid{y}){}\<[E]%
\ColumnHook
\end{hscode}\resethooks
with similar definitions for subtraction \ensuremath{(\mathbin{-})}, division \ensuremath{(\mathbin{/})} and right
scaling \ensuremath{(\ensuremath{\triangleright})} of physical quantities.
Notice that addition (subtraction) is only defined for quantities of the
same dimensions. This helps avoiding ``adding apples and oranges'' in
expressions involving physical quantities. When such additions make
sense, as in the example at page 6 of \citep{barenblatt1996scaling}, the
computations can be implemented by pattern matching, as done in the
definition of \ensuremath{\mu}.

Remember that the core result of the Pi theorem is that physical laws
can be expressed through products of powers of physical variables, see
\cref{eq:pi1}. In order to formalize this result in
\cref{subsection:piexplained1.1}, we need two more combinators
\begin{hscode}\SaveRestoreHook
\column{B}{@{}>{\hspre}l<{\hspost}@{}}%
\column{5}{@{}>{\hspre}l<{\hspost}@{}}%
\column{10}{@{}>{\hspre}c<{\hspost}@{}}%
\column{10E}{@{}l@{}}%
\column{13}{@{}>{\hspre}l<{\hspost}@{}}%
\column{49}{@{}>{\hspre}l<{\hspost}@{}}%
\column{E}{@{}>{\hspre}l<{\hspost}@{}}%
\>[5]{}\Varid{pow}{}\<[10]%
\>[10]{}\ \mathop{:}\ {}\<[10E]%
\>[13]{}\{\mskip1.5mu \Varid{d}\ \mathop{:}\ \Conid{D}\mskip1.5mu\}\to \Conid{Q}\;\Varid{d}\to (\Varid{n}\ \mathop{:}\ \mathbb{Z})\to {}\<[49]%
\>[49]{}\Conid{Q}\;(\Varid{d}\mathbin{`\Conid{Pow}`}\Varid{n}){}\<[E]%
\\
\>[5]{}\Varid{pow}\;(\Conid{Val}\;\Varid{u}\;\Varid{x})\;\Varid{n}\mathrel{=}\Conid{Val}\;\Varid{u}\;(\Varid{ipow}\;\Varid{x}\;\Varid{n}){}\<[E]%
\ColumnHook
\end{hscode}\resethooks
\begin{hscode}\SaveRestoreHook
\column{B}{@{}>{\hspre}l<{\hspost}@{}}%
\column{5}{@{}>{\hspre}l<{\hspost}@{}}%
\column{15}{@{}>{\hspre}l<{\hspost}@{}}%
\column{26}{@{}>{\hspre}l<{\hspost}@{}}%
\column{37}{@{}>{\hspre}c<{\hspost}@{}}%
\column{37E}{@{}l@{}}%
\column{40}{@{}>{\hspre}l<{\hspost}@{}}%
\column{E}{@{}>{\hspre}l<{\hspost}@{}}%
\>[5]{}\Varid{prodPows}\ \mathop{:}\ \{\mskip1.5mu \Varid{n}\ \mathop{:}\ \mathbb{N}\mskip1.5mu\}\to \{\mskip1.5mu \Varid{ds}\ \mathop{:}\ \Conid{Vec}\;\Varid{n}\;\Conid{D}\mskip1.5mu\}\to \Conid{Vec}_{\scalebox{.5}{Q}}\;\Varid{n}\;\Varid{ds}\to (\Varid{ps}\ \mathop{:}\ \Conid{Vec}\;\Varid{n}\;\mathbb{Z})\to \Conid{Q}\;(\Conid{ProdPows}\;\Varid{ds}\;\Varid{ps}){}\<[E]%
\\
\>[5]{}\Varid{prodPows}\;{}\<[15]%
\>[15]{}\Conid{Nil}\;{}\<[26]%
\>[26]{}\Conid{Nil}{}\<[37]%
\>[37]{}\mathrel{=}{}\<[37E]%
\>[40]{}\Conid{Val}\;\Conid{SI}\;\mathrm{1.0}{}\<[E]%
\\
\>[5]{}\Varid{prodPows}\;{}\<[15]%
\>[15]{}(\Varid{q}\mathbin{::}\Varid{qs})\;{}\<[26]%
\>[26]{}(\Varid{p}\mathbin{::}\Varid{ps}){}\<[37]%
\>[37]{}\mathrel{=}{}\<[37E]%
\>[40]{}(\Varid{pow}\;\Varid{q}\;\Varid{p})\ensuremath{\cdot}(\Varid{prodPows}\;\Varid{qs}\;\Varid{ps}){}\<[E]%
\ColumnHook
\end{hscode}\resethooks
In the return types of \ensuremath{\Varid{pow}} and \ensuremath{\Varid{prodPows}} we have applied integer
powers of \ensuremath{\Conid{D}}-values \ensuremath{\Conid{Pow}} and its generalization \ensuremath{\Conid{ProdPows}} to products
of integer powers. \ensuremath{\Conid{Pow}} fulfils the specification
\begin{hscode}\SaveRestoreHook
\column{B}{@{}>{\hspre}l<{\hspost}@{}}%
\column{5}{@{}>{\hspre}l<{\hspost}@{}}%
\column{16}{@{}>{\hspre}l<{\hspost}@{}}%
\column{20}{@{}>{\hspre}c<{\hspost}@{}}%
\column{20E}{@{}l@{}}%
\column{23}{@{}>{\hspre}l<{\hspost}@{}}%
\column{E}{@{}>{\hspre}l<{\hspost}@{}}%
\>[5]{}\Conid{Pow}\;\Varid{d}\;{}\<[16]%
\>[16]{}\mathrm{0}{}\<[20]%
\>[20]{}\mathrel{=}{}\<[20E]%
\>[23]{}\Conid{DimLess}{}\<[E]%
\\
\>[5]{}\Conid{Pow}\;\Varid{d}\;(\Varid{n}\mathbin{+}\mathrm{1}){}\<[20]%
\>[20]{}\mathrel{=}{}\<[20E]%
\>[23]{}\Conid{Pow}\;\Varid{d}\;\Varid{n}\mathbin{`\Conid{Times}`}\Varid{d}{}\<[E]%
\\
\>[5]{}\Conid{Pow}\;\Varid{d}\;(\Varid{n}\mathbin{-}\mathrm{1}){}\<[20]%
\>[20]{}\mathrel{=}{}\<[20E]%
\>[23]{}\Conid{Pow}\;\Varid{d}\;\Varid{n}\mathbin{`\Conid{Over}`}\Varid{d}{}\<[E]%
\ColumnHook
\end{hscode}\resethooks
and \ensuremath{\Conid{ProdPows}} is a fold after a zip:
\begin{hscode}\SaveRestoreHook
\column{B}{@{}>{\hspre}l<{\hspost}@{}}%
\column{5}{@{}>{\hspre}l<{\hspost}@{}}%
\column{15}{@{}>{\hspre}l<{\hspost}@{}}%
\column{26}{@{}>{\hspre}l<{\hspost}@{}}%
\column{37}{@{}>{\hspre}c<{\hspost}@{}}%
\column{37E}{@{}l@{}}%
\column{40}{@{}>{\hspre}l<{\hspost}@{}}%
\column{E}{@{}>{\hspre}l<{\hspost}@{}}%
\>[5]{}\Conid{ProdPows}\ \mathop{:}\ \{\mskip1.5mu \Varid{n}\ \mathop{:}\ \mathbb{N}\mskip1.5mu\}\to \Conid{Vec}\;\Varid{n}\;\Conid{D}\to \Conid{Vec}\;\Varid{n}\;\mathbb{Z}\to \Conid{D}{}\<[E]%
\\
\>[5]{}\Conid{ProdPows}\;{}\<[15]%
\>[15]{}\Conid{Nil}\;{}\<[26]%
\>[26]{}\Conid{Nil}{}\<[37]%
\>[37]{}\mathrel{=}{}\<[37E]%
\>[40]{}\Conid{DimLess}{}\<[E]%
\\
\>[5]{}\Conid{ProdPows}\;{}\<[15]%
\>[15]{}(\Varid{d}\mathbin{::}\Varid{ds})\;{}\<[26]%
\>[26]{}(\Varid{p}\mathbin{::}\Varid{ps}){}\<[37]%
\>[37]{}\mathrel{=}{}\<[37E]%
\>[40]{}\Conid{Pow}\;\Varid{d}\;\Varid{p}\mathbin{`\Conid{Times}`}\Conid{ProdPows}\;\Varid{ds}\;\Varid{ps}{}\<[E]%
\ColumnHook
\end{hscode}\resethooks
Note that the definition of \ensuremath{\Varid{prodPows}} has the same structure as the
definition of its type (index) \ensuremath{\Conid{ProdPows}}.
This is an example of type-driven development~\citep{idrisbook}.

Because of the definition of \ensuremath{\Conid{DimLess}}, \ensuremath{\Conid{Times}} and \ensuremath{\Conid{Over}}
from \cref{subsection:df}, \ensuremath{\Conid{Pow}} and \ensuremath{\Conid{ProdPows}} also fulfil
\begin{hscode}\SaveRestoreHook
\column{B}{@{}>{\hspre}l<{\hspost}@{}}%
\column{5}{@{}>{\hspre}l<{\hspost}@{}}%
\column{E}{@{}>{\hspre}l<{\hspost}@{}}%
\>[5]{}\Conid{Pow}\;\Varid{d}\;\Varid{n}\mathrel{=}\Varid{n}\triangleleft \,\Varid{d}{}\<[E]%
\\
\>[5]{}\Conid{ProdPows}\;\Varid{ds}\;\Varid{ps}\mathrel{=}\Varid{foldr}\;(\mathbin{+})\;\Conid{DimLess}\;(\Varid{zipWith}\;(\triangleleft \,)\;\Varid{ps}\;\Varid{ds}){}\<[E]%
\ColumnHook
\end{hscode}\resethooks
where \ensuremath{(\triangleleft \,)} is generic scaling for vectors of numerical types:
\begin{hscode}\SaveRestoreHook
\column{B}{@{}>{\hspre}l<{\hspost}@{}}%
\column{5}{@{}>{\hspre}l<{\hspost}@{}}%
\column{8}{@{}>{\hspre}l<{\hspost}@{}}%
\column{11}{@{}>{\hspre}c<{\hspost}@{}}%
\column{11E}{@{}l@{}}%
\column{14}{@{}>{\hspre}l<{\hspost}@{}}%
\column{16}{@{}>{\hspre}l<{\hspost}@{}}%
\column{E}{@{}>{\hspre}l<{\hspost}@{}}%
\>[5]{}(\triangleleft \,){}\<[11]%
\>[11]{}\ \mathop{:}\ {}\<[11E]%
\>[14]{}\{\mskip1.5mu \Varid{n}\ \mathop{:}\ \mathbb{N}\mskip1.5mu\}\to \{\mskip1.5mu \Conid{T}\ \mathop{:}\ \Conid{Type}\mskip1.5mu\}\to \Conid{Num}\;\Conid{T}\Rightarrow\Conid{T}\to \Conid{Vec}\;\Varid{n}\;\Conid{T}\to \Conid{Vec}\;\Varid{n}\;\Conid{T}{}\<[E]%
\\
\>[5]{}\Varid{x}{}\<[8]%
\>[8]{}\triangleleft \,\Varid{v}\mathrel{=}{}\<[16]%
\>[16]{}\Varid{map}\;(\Varid{x}\ensuremath{\cdot})\;\Varid{v}{}\<[E]%
\ColumnHook
\end{hscode}\resethooks
Finally, the data type \ensuremath{\Conid{Vec}_{\scalebox{.5}{Q}}} in the signature of \ensuremath{\Varid{prodPows}} represents
vectors of physical quantities of potentially different dimensions:
\ensuremath{\Conid{Vec}_{\scalebox{.5}{Q}}\;\Varid{n}\;\Varid{ds}\mathrel{=}\Conid{All}\;\Conid{Q}\;\Varid{ds}}.

\paragraph*{Physical quantities and ``phantom'' types.} \ We have
introduced physical quantities through the ``phantom\footnote{See
\url{https://wiki.haskell.org/Phantom_type}.}'' data type \ensuremath{\Conid{Q}\ \mathop{:}\ \Conid{D}\to \Conid{Type}}. The type parameter \ensuremath{\Conid{D}} allows one to \emph{specify} physical
quantities of given dimensions as in \ensuremath{\Varid{x}\ \mathop{:}\ \Conid{Q}\;\Conid{Length}}. When
\emph{defining} \ensuremath{\Varid{x}} one has to provide its measure in a system of units
as in \ensuremath{\Varid{x}\mathrel{=}\Conid{Val}\;\Conid{SI}\;\mathrm{3}}.
An alternative design could be to make also the units of measurement visible
in the type of physical quantities:
\begin{hscode}\SaveRestoreHook
\column{B}{@{}>{\hspre}l<{\hspost}@{}}%
\column{5}{@{}>{\hspre}l<{\hspost}@{}}%
\column{7}{@{}>{\hspre}l<{\hspost}@{}}%
\column{E}{@{}>{\hspre}l<{\hspost}@{}}%
\>[5]{}\mathbf{data}\;Q_1\ \mathop{:}\ \Conid{D}\to \Conid{Units}\to \Conid{Type}\;\mathbf{where}{}\<[E]%
\\
\>[5]{}\hsindent{2}{}\<[7]%
\>[7]{}\Conid{Val}_1\ \mathop{:}\ \{\mskip1.5mu \Varid{d}\ \mathop{:}\ \Conid{D}\mskip1.5mu\}\to (\Varid{u}\ \mathop{:}\ \Conid{Units})\to \Real\to Q_1\;\Varid{d}\;\Varid{u}{}\<[E]%
\ColumnHook
\end{hscode}\resethooks
This would allow one to avoid the introduction of an apparently
distinguished system of units in the definition of the elementary binary
operations, \ensuremath{\Conid{SI}} in our implementation. With \ensuremath{Q_1},
addition would read
\begin{hscode}\SaveRestoreHook
\column{B}{@{}>{\hspre}l<{\hspost}@{}}%
\column{5}{@{}>{\hspre}l<{\hspost}@{}}%
\column{11}{@{}>{\hspre}c<{\hspost}@{}}%
\column{11E}{@{}l@{}}%
\column{14}{@{}>{\hspre}l<{\hspost}@{}}%
\column{E}{@{}>{\hspre}l<{\hspost}@{}}%
\>[5]{}(\mathbin{+}){}\<[11]%
\>[11]{}\ \mathop{:}\ {}\<[11E]%
\>[14]{}\{\mskip1.5mu \Varid{d}\ \mathop{:}\ \Conid{D}\mskip1.5mu\}\to \{\mskip1.5mu \Varid{u},\Varid{v},\Varid{w}\ \mathop{:}\ \Conid{Units}\mskip1.5mu\}\to Q_1\;\Varid{d}\;\Varid{u}\to Q_1\;\Varid{d}\;\Varid{v}\to Q_1\;\Varid{d}\;\Varid{w}{}\<[E]%
\\
\>[5]{}(\mathbin{+})\;q_1\;q_2\mathrel{=}\Conid{Val}_1\;\Varid{w}\;(\mu\;\Varid{w}\;q_1\mathbin{+}\mu\;\Varid{w}\;q_2){}\<[E]%
\ColumnHook
\end{hscode}\resethooks
A practical drawback of that approach is the proliferation of implicit
parameters. More importantly, exposing the units of measurement in the
type of physical quantities is conceptually unsatisfactory: whether \ensuremath{\Varid{x}}
has been defined to be a length of one meter or 1.093613 yards does not
really matter and we argue that this choice shall not be visible in its
type.

\paragraph*{The covariance principle for elementary operations.} \
Perhaps not surprisingly, \ensuremath{\mu_u} is a homomorphism from \ensuremath{(\mathbin{+})}, \ensuremath{(\ensuremath{\cdot})},
\ensuremath{(\ensuremath{\triangleleft})}, \ensuremath{(\mathbin{-})}, \ensuremath{(\mathbin{/})} and \ensuremath{(\ensuremath{\triangleright})} between physical quantities and the
corresponding binary operations in \ensuremath{\Real}. For example
\begin{hscode}\SaveRestoreHook
\column{B}{@{}>{\hspre}l<{\hspost}@{}}%
\column{5}{@{}>{\hspre}l<{\hspost}@{}}%
\column{7}{@{}>{\hspre}l<{\hspost}@{}}%
\column{12}{@{}>{\hspre}l<{\hspost}@{}}%
\column{14}{@{}>{\hspre}l<{\hspost}@{}}%
\column{16}{@{}>{\hspre}l<{\hspost}@{}}%
\column{19}{@{}>{\hspre}c<{\hspost}@{}}%
\column{19E}{@{}l@{}}%
\column{22}{@{}>{\hspre}l<{\hspost}@{}}%
\column{E}{@{}>{\hspre}l<{\hspost}@{}}%
\>[5]{}\mu{}\Conid{HomMult}{}\<[19]%
\>[19]{}\ \mathop{:}\ {}\<[19E]%
\>[22]{}(\Varid{u}\ \mathop{:}\ \Conid{Units})\to \{\mskip1.5mu d_1,d_2\ \mathop{:}\ \Conid{D}\mskip1.5mu\}\to {}\<[E]%
\\
\>[22]{}(q_1\ \mathop{:}\ \Conid{Q}\;d_1)\to (q_2\ \mathop{:}\ \Conid{Q}\;d_2)\to \mu_{\Varid{u}}\;(q_1\ensuremath{\cdot}q_2)\mathrel{=}\mu_{\Varid{u}}\;q_1\ensuremath{\cdot}\mu_{\Varid{u}}\;q_2{}\<[E]%
\\[\blanklineskip]%
\>[5]{}\mu{}\Conid{HomMult}\;\Varid{u}\;\{\mskip1.5mu d_1\mskip1.5mu\}\;\{\mskip1.5mu d_2\mskip1.5mu\}\;(\Conid{Val}\;u_1\;\Varid{x}_{1})\;(\Conid{Val}\;u_2\;\Varid{x}_{2})\mathrel{=}{}\<[E]%
\\
\>[5]{}\hsindent{2}{}\<[7]%
\>[7]{}\mathbf{let}\;{}\<[12]%
\>[12]{}\mbox{\onelinecomment  Factors to convert from local units to SI}{}\<[E]%
\\
\>[12]{}\Varid{f}_{1}{}\<[16]%
\>[16]{}\mathrel{=}\Varid{df}\;d_1\;(\Varid{fs}\;u_1\;\Conid{SI}){}\<[E]%
\\
\>[12]{}\Varid{f}_{2}{}\<[16]%
\>[16]{}\mathrel{=}\Varid{df}\;d_2\;(\Varid{fs}\;u_2\;\Conid{SI}){}\<[E]%
\\
\>[12]{}\mbox{\onelinecomment  Factors to convert from SI to target unit u}{}\<[E]%
\\
\>[12]{}\Varid{g}_{1}{}\<[16]%
\>[16]{}\mathrel{=}\Varid{df}\;d_1\;(\Varid{fs}\;\Conid{SI}\;\Varid{u}){}\<[E]%
\\
\>[12]{}\Varid{g}_{2}{}\<[16]%
\>[16]{}\mathrel{=}\Varid{df}\;d_2\;(\Varid{fs}\;\Conid{SI}\;\Varid{u}){}\<[E]%
\\
\>[12]{}\Varid{g}_{12}\mathrel{=}\Varid{df}\;(d_1\mathbin{`\Conid{Times}`}d_2)\;(\Varid{fs}\;\Conid{SI}\;\Varid{u}){}\<[E]%
\\
\>[5]{}\hsindent{2}{}\<[7]%
\>[7]{}\mathbf{in}\;{}\<[12]%
\>[12]{}\mu_{\Varid{u}}\;(\Conid{Val}\;u_1\;\Varid{x}_{1}\ensuremath{\cdot}\Conid{Val}\;u_2\;\Varid{x}_{2}){}\<[E]%
\\
\>[12]{}\hsindent{2}{}\<[14]%
\>[14]{}=\hspace{-3pt}\{\; \Conid{Refl}\;\}\hspace{-3pt}={}\<[E]%
\\
\>[12]{}(((\Varid{x}_{1}\ensuremath{\cdot}\Varid{f}_{1})\ensuremath{\cdot}(\Varid{x}_{2}\ensuremath{\cdot}\Varid{f}_{2}))\ensuremath{\cdot}\Varid{g}_{12}){}\<[E]%
\\
\>[12]{}\hsindent{2}{}\<[14]%
\>[14]{}=\hspace{-3pt}\{\; \Varid{cong}\;\{\mskip1.5mu \Varid{f}\mathrel{=}\lambda \Varid{y}\Rightarrow((\Varid{x}_{1}\ensuremath{\cdot}\Varid{f}_{1})\ensuremath{\cdot}(\Varid{x}_{2}\ensuremath{\cdot}\Varid{f}_{2}))\ensuremath{\cdot}\Varid{y}\mskip1.5mu\}\;\Varid{dfHomTimes}\;\}\hspace{-3pt}={}\<[E]%
\\
\>[12]{}(((\Varid{x}_{1}\ensuremath{\cdot}\Varid{f}_{1})\ensuremath{\cdot}(\Varid{x}_{2}\ensuremath{\cdot}\Varid{f}_{2}))\ensuremath{\cdot}(\Varid{g}_{1}\ensuremath{\cdot}\Varid{g}_{2})){}\<[E]%
\\
\>[12]{}\hsindent{2}{}\<[14]%
\>[14]{}=\hspace{-3pt}\{\; \Varid{lemma}\mathbb{R}_4\;\Varid{x}_{1}\;\Varid{f}_{1}\;\Varid{x}_{2}\;\Varid{f}_{2}\;\Varid{g}_{1}\;\Varid{g}_{2}\;\}\hspace{-3pt}={}\<[E]%
\\
\>[12]{}((\Varid{x}_{1}\ensuremath{\cdot}(\Varid{f}_{1}\ensuremath{\cdot}\Varid{g}_{1}))\ensuremath{\cdot}(\Varid{x}_{2}\ensuremath{\cdot}(\Varid{f}_{2}\ensuremath{\cdot}\Varid{g}_{2}))){}\<[E]%
\\
\>[12]{}\hsindent{2}{}\<[14]%
\>[14]{}=\hspace{-3pt}\{\; \Varid{cong}\;\{\mskip1.5mu \Varid{f}\mathrel{=}\lambda \Varid{y}\Rightarrow(\Varid{x}_{1}\ensuremath{\cdot}\Varid{y})\ensuremath{\cdot}(\Varid{x}_{2}\ensuremath{\cdot}(\Varid{f}_{2}\ensuremath{\cdot}\Varid{g}_{2}))\mskip1.5mu\}\;\Varid{dfLemma3}\;\}\hspace{-3pt}={}\<[E]%
\\
\>[12]{}((\Varid{x}_{1}\ensuremath{\cdot}\Varid{df}\;d_1\;(\Varid{fs}\;u_1\;\Varid{u}))\ensuremath{\cdot}(\Varid{x}_{2}\ensuremath{\cdot}(\Varid{f}_{2}\ensuremath{\cdot}\Varid{g}_{2}))){}\<[E]%
\\
\>[12]{}\hsindent{2}{}\<[14]%
\>[14]{}=\hspace{-3pt}\{\; \Varid{cong}\;\{\mskip1.5mu \Varid{f}\mathrel{=}\lambda \Varid{y}\Rightarrow(\Varid{x}_{1}\ensuremath{\cdot}\Varid{df}\;d_1\;(\Varid{fs}\;u_1\;\Varid{u}))\ensuremath{\cdot}(\Varid{x}_{2}\ensuremath{\cdot}\Varid{y})\mskip1.5mu\}\;\Varid{dfLemma3}\;\}\hspace{-3pt}={}\<[E]%
\\
\>[12]{}((\Varid{x}_{1}\ensuremath{\cdot}\Varid{df}\;d_1\;(\Varid{fs}\;u_1\;\Varid{u}))\ensuremath{\cdot}(\Varid{x}_{2}\ensuremath{\cdot}\Varid{df}\;d_2\;(\Varid{fs}\;u_2\;\Varid{u}))){}\<[E]%
\\
\>[12]{}\hsindent{2}{}\<[14]%
\>[14]{}=\hspace{-3pt}\{\; \Conid{Refl}\;\}\hspace{-3pt}={}\<[E]%
\\
\>[12]{}(\mu_{\Varid{u}}\;(\Conid{Val}\;u_1\;\Varid{x}_{1})\ensuremath{\cdot}\mu_{\Varid{u}}\;(\Conid{Val}\;u_2\;\Varid{x}_{2}))\;{}\<[E]%
\\
\>[12]{}\hsindent{2}{}\<[14]%
\>[14]{}\Conid{QED}{}\<[E]%
\ColumnHook
\end{hscode}\resethooks

\noindent
The proofs require postulating commutativity and associativity of \ensuremath{(\ensuremath{\cdot})},
distributivity of \ensuremath{(\ensuremath{\cdot})} over \ensuremath{(\mathbin{+})} and \ensuremath{(\mathbin{-})} in \ensuremath{\Real} (packaged up in
\ensuremath{\Varid{lemma}\mathbb{R}_4}) and, crucially, the properties of dimension functions
\ensuremath{\Varid{dfLemma3}}, \ensuremath{\Varid{dfHomTimes}} and \ensuremath{\Varid{dfHomTimes}} discussed in
\cref{subsection:df}.

The fact that \ensuremath{\mu_u} is a homomorphism between elementary binary
operations on physical quantities and the corresponding operations in
\ensuremath{\Real} is a special form of the covariance principle. We discuss a
general form of this principle in \cref{section:piexplained1}.

\subsection{Dimensional judgments, dimensional consistent programming}
\label{subsection:judgments}

The infrastructure introduced so far allows one to define new physical
quantities from existing ones (remember that \ensuremath{\Varid{x}} and \ensuremath{\Varid{t}} were declared to be a
length and a time in \cref{subsection:quantities})
\begin{hscode}\SaveRestoreHook
\column{B}{@{}>{\hspre}l<{\hspost}@{}}%
\column{5}{@{}>{\hspre}l<{\hspost}@{}}%
\column{8}{@{}>{\hspre}c<{\hspost}@{}}%
\column{8E}{@{}l@{}}%
\column{11}{@{}>{\hspre}l<{\hspost}@{}}%
\column{E}{@{}>{\hspre}l<{\hspost}@{}}%
\>[5]{}\Varid{v}{}\<[8]%
\>[8]{}\ \mathop{:}\ {}\<[8E]%
\>[11]{}\Conid{Q}\;\Conid{Velocity}{}\<[E]%
\\
\>[5]{}\Varid{v}{}\<[8]%
\>[8]{}\mathrel{=}{}\<[8E]%
\>[11]{}(\mathrm{2}\ensuremath{\triangleleft}\Varid{x})\mathbin{/}\Varid{t}{}\<[E]%
\ColumnHook
\end{hscode}\resethooks
and implement dimensional judgments for verified programming like
\begin{hscode}\SaveRestoreHook
\column{B}{@{}>{\hspre}l<{\hspost}@{}}%
\column{5}{@{}>{\hspre}l<{\hspost}@{}}%
\column{E}{@{}>{\hspre}l<{\hspost}@{}}%
\>[5]{}\Varid{check}_{3 }\ \mathop{:}\ \Varid{dim}\;(\Varid{x}\mathbin{/}(\Varid{t}\ensuremath{\cdot}\Varid{t}))\mathrel{=}\Conid{Acceleration}{}\<[E]%
\\
\>[5]{}\Varid{check}_{3 }\mathrel{=}\Conid{Refl}{}\<[E]%
\ColumnHook
\end{hscode}\resethooks
Here, a value of type \ensuremath{\Varid{dim}\;(\Varid{x}\mathbin{/}(\Varid{t}\ensuremath{\cdot}\Varid{t}))\mathrel{=}\Conid{Acceleration}} serves as
proof that \ensuremath{\Varid{x}\mathbin{/}(\Varid{t}\ensuremath{\cdot}\Varid{t})} is an acceleration. We have seen
many examples of this kind of dimensional judgments in previous
sections: \ensuremath{\Varid{f}} is a force, \ensuremath{\Varid{m}} is a mass, \ensuremath{\Conid{T}} is a temperature, \ensuremath{\Varid{p}} is
a pressure, etc. We can express these judgments through the idiom
\begin{hscode}\SaveRestoreHook
\column{B}{@{}>{\hspre}l<{\hspost}@{}}%
\column{5}{@{}>{\hspre}l<{\hspost}@{}}%
\column{E}{@{}>{\hspre}l<{\hspost}@{}}%
\>[5]{}\Conid{Is}\ \mathop{:}\ \{\mskip1.5mu \Varid{d}\ \mathop{:}\ \Conid{D}\mskip1.5mu\}\to \Conid{Q}\;\Varid{d}\to \Conid{D}\to \Conid{Type}{}\<[E]%
\\
\>[5]{}\Conid{Is}\;\Varid{q}\;\Varid{d}\mathrel{=}(\Varid{dim}\;\Varid{q}\mathrel{=}\Varid{d}){}\<[E]%
\ColumnHook
\end{hscode}\resethooks
and write \ensuremath{\Conid{Is}\;\Varid{q}\;\Varid{d}} instead of \ensuremath{\Varid{dim}\;\Varid{q}\mathrel{=}\Varid{d}}:
\begin{hscode}\SaveRestoreHook
\column{B}{@{}>{\hspre}l<{\hspost}@{}}%
\column{5}{@{}>{\hspre}l<{\hspost}@{}}%
\column{E}{@{}>{\hspre}l<{\hspost}@{}}%
\>[5]{}\Varid{check}_{4 }\ \mathop{:}\ \Conid{Is}\;(\Varid{m}\ensuremath{\cdot}\Varid{x}\mathbin{/}(\Varid{t}\ensuremath{\cdot}\Varid{t}))\;\Conid{Force}{}\<[E]%
\\
\>[5]{}\Varid{check}_{4 }\mathrel{=}\Conid{Refl}{}\<[E]%
\ColumnHook
\end{hscode}\resethooks
Similarly, we can assess whether a physical quantity is dimensionless or
not
\begin{hscode}\SaveRestoreHook
\column{B}{@{}>{\hspre}l<{\hspost}@{}}%
\column{5}{@{}>{\hspre}l<{\hspost}@{}}%
\column{E}{@{}>{\hspre}l<{\hspost}@{}}%
\>[5]{}\Conid{IsDimLess}\ \mathop{:}\ \{\mskip1.5mu \Varid{d}\ \mathop{:}\ \Conid{D}\mskip1.5mu\}\to \Conid{Q}\;\Varid{d}\to \Conid{Type}{}\<[E]%
\\
\>[5]{}\Conid{IsDimLess}\;\Varid{q}\mathrel{=}(\Varid{dim}\;\Varid{q}\mathrel{=}\Conid{DimLess}){}\<[E]%
\\[\blanklineskip]%
\>[5]{}\Varid{check}_{5 }\ \mathop{:}\ \Conid{IsDimLess}\;((\Varid{x}\mathbin{+}\Varid{x})\mathbin{/}\Varid{x}){}\<[E]%
\\
\>[5]{}\Varid{check}_{5 }\mathrel{=}\Conid{Refl}{}\<[E]%
\ColumnHook
\end{hscode}\resethooks
As one would expect, dimensionless quantities are invariant under
re-scaling of the units of measurement
\begin{hscode}\SaveRestoreHook
\column{B}{@{}>{\hspre}l<{\hspost}@{}}%
\column{E}{@{}>{\hspre}l<{\hspost}@{}}%
\>[B]{}\lambda\sqcap\!>\;\mu\;\Conid{SI}\;((\Varid{x}\mathbin{+}\Varid{x})\mathbin{/}\Varid{x}){}\<[E]%
\\
\>[B]{}\mathrm{2.0}\ \mathop{:}\ \Conid{Double}{}\<[E]%
\\[\blanklineskip]%
\>[B]{}\lambda\sqcap\!>\;\mu\;\Conid{CGS}\;((\Varid{x}\mathbin{+}\Varid{x})\mathbin{/}\Varid{x}){}\<[E]%
\\
\>[B]{}\mathrm{2.0}\ \mathop{:}\ \Conid{Double}{}\<[E]%
\ColumnHook
\end{hscode}\resethooks
and the dimension function fulfils the specification (\ref{eq:df0}): by
the definition of \ensuremath{\mu}, measurements only depend on the scaling
factors between the units of measurement, not on the units themselves.

Notice, however, that a direct encoding of the perhaps most discussed
application of dimensional analysis does not work out. If we try to compute
the period \ensuremath{\tau} of small-angle oscillation of a simple pendulum of
length \ensuremath{\Varid{l}} in a gravitational field of strength \ensuremath{\Varid{g}}
\begin{hscode}\SaveRestoreHook
\column{B}{@{}>{\hspre}l<{\hspost}@{}}%
\column{5}{@{}>{\hspre}l<{\hspost}@{}}%
\column{8}{@{}>{\hspre}c<{\hspost}@{}}%
\column{8E}{@{}l@{}}%
\column{11}{@{}>{\hspre}l<{\hspost}@{}}%
\column{24}{@{}>{\hspre}l<{\hspost}@{}}%
\column{37}{@{}>{\hspre}l<{\hspost}@{}}%
\column{40}{@{}>{\hspre}c<{\hspost}@{}}%
\column{40E}{@{}l@{}}%
\column{43}{@{}>{\hspre}l<{\hspost}@{}}%
\column{60}{@{}>{\hspre}l<{\hspost}@{}}%
\column{73}{@{}>{\hspre}l<{\hspost}@{}}%
\column{77}{@{}>{\hspre}c<{\hspost}@{}}%
\column{77E}{@{}l@{}}%
\column{80}{@{}>{\hspre}l<{\hspost}@{}}%
\column{E}{@{}>{\hspre}l<{\hspost}@{}}%
\>[5]{}\Varid{l}{}\<[8]%
\>[8]{}\ \mathop{:}\ {}\<[8E]%
\>[11]{}\Conid{Q}\;\Conid{Length};{}\<[24]%
\>[24]{}\quad{}\<[37]%
\>[37]{}\Varid{g}{}\<[40]%
\>[40]{}\ \mathop{:}\ {}\<[40E]%
\>[43]{}\Conid{Q}\;\Conid{Acceleration};{}\<[60]%
\>[60]{}\quad{}\<[73]%
\>[73]{}\pi{}\<[77]%
\>[77]{}\ \mathop{:}\ {}\<[77E]%
\>[80]{}\Conid{Q}\;\Conid{DimLess}{}\<[E]%
\\
\>[5]{}\Varid{l}{}\<[8]%
\>[8]{}\mathrel{=}{}\<[8E]%
\>[11]{}\Conid{Val}\;\Conid{SI}\;\mathrm{0.5};{}\<[24]%
\>[24]{}\quad{}\<[37]%
\>[37]{}\Varid{g}{}\<[40]%
\>[40]{}\mathrel{=}{}\<[40E]%
\>[43]{}\Conid{Val}\;\Conid{SI}\;\mathrm{9.81};{}\<[60]%
\>[60]{}\quad{}\<[73]%
\>[73]{}\pi{}\<[77]%
\>[77]{}\mathrel{=}{}\<[77E]%
\>[80]{}\Conid{Val}\;\Conid{SI}\;\mathrm{3.14}{}\<[E]%
\ColumnHook
\end{hscode}\resethooks
as
\begin{hscode}\SaveRestoreHook
\column{B}{@{}>{\hspre}l<{\hspost}@{}}%
\column{5}{@{}>{\hspre}l<{\hspost}@{}}%
\column{10}{@{}>{\hspre}c<{\hspost}@{}}%
\column{10E}{@{}l@{}}%
\column{13}{@{}>{\hspre}l<{\hspost}@{}}%
\column{E}{@{}>{\hspre}l<{\hspost}@{}}%
\>[5]{}\tau{}\<[10]%
\>[10]{}\ \mathop{:}\ {}\<[10E]%
\>[13]{}\Conid{Q}\;\Conid{Time}{}\<[E]%
\\
\>[5]{}\tau{}\<[10]%
\>[10]{}\mathrel{=}{}\<[10E]%
\>[13]{}\mathrm{2}\ensuremath{\triangleleft}\pi\ensuremath{\cdot}\Varid{sqrt}\;(\Varid{l}\mathbin{/}\Varid{g}){}\<[E]%
\ColumnHook
\end{hscode}\resethooks
we get an error. This is because we have represented dimension functions
through vectors of \emph{integer} rather than \emph{rational} numbers
and therefore we cannot define a \ensuremath{\Varid{sqrt}} function for arbitrary physical
variables.

Representing dimension functions through integer vectors is what makes
our DSL ``minimal''. We motivate this choice in \cref{subsection:dind}
and discuss possible generalizations in \cref{section:generalization}.

%
\subsection{Units of measurement}
\label{subsection:units}

We have introduced the notion of units of measurement through a datatype
\ensuremath{\Conid{Units}} with just two data constructors: \ensuremath{\Conid{SI}} and \ensuremath{\Conid{CGS}}. This is also
the approach followed by \citet{barenblatt1996scaling} and reflects the
idea that units of measurement are just annotations.
There is nothing wrong with this approach and \ensuremath{\Conid{Units}} can readily be
extended to accommodate more systems of units by enumeration.

Notice, however, that this requires defining \ensuremath{\Varid{fs}}, the table that
returns the scaling factors between units of measurement, for all
possible pairs of data constructors. For the definition of \ensuremath{\Conid{Units}} above,
for example, \ensuremath{\Varid{fs}} was defined as:
\begin{hscode}\SaveRestoreHook
\column{B}{@{}>{\hspre}l<{\hspost}@{}}%
\column{5}{@{}>{\hspre}l<{\hspost}@{}}%
\column{13}{@{}>{\hspre}l<{\hspost}@{}}%
\column{18}{@{}>{\hspre}l<{\hspost}@{}}%
\column{28}{@{}>{\hspre}l<{\hspost}@{}}%
\column{32}{@{}>{\hspre}l<{\hspost}@{}}%
\column{39}{@{}>{\hspre}c<{\hspost}@{}}%
\column{39E}{@{}l@{}}%
\column{E}{@{}>{\hspre}l<{\hspost}@{}}%
\>[5]{}\Varid{fs}\ \mathop{:}\ \Conid{Units}\to \Conid{Units}\to \Real_{+}^3{}\<[E]%
\\
\>[5]{}\Varid{fs}\;\Conid{SI}\;{}\<[13]%
\>[13]{}\Conid{SI}{}\<[18]%
\>[18]{}\mathrel{=}[\mskip1.5mu \mathrm{1},{}\<[28]%
\>[28]{}\mathrm{1},{}\<[32]%
\>[32]{}\mathrm{1}{}\<[39]%
\>[39]{}\mskip1.5mu]{}\<[39E]%
\\
\>[5]{}\Varid{fs}\;\Conid{SI}\;{}\<[13]%
\>[13]{}\Conid{CGS}{}\<[18]%
\>[18]{}\mathrel{=}[\mskip1.5mu \mathrm{100},{}\<[28]%
\>[28]{}\mathrm{1},{}\<[32]%
\>[32]{}\mathrm{1000}{}\<[39]%
\>[39]{}\mskip1.5mu]{}\<[39E]%
\\
\>[5]{}\Varid{fs}\;\Conid{CGS}\;{}\<[13]%
\>[13]{}\Conid{SI}{}\<[18]%
\>[18]{}\mathrel{=}[\mskip1.5mu \mathrm{0.01},{}\<[28]%
\>[28]{}\mathrm{1},{}\<[32]%
\>[32]{}\mathrm{0.001}{}\<[39]%
\>[39]{}\mskip1.5mu]{}\<[39E]%
\\
\>[5]{}\Varid{fs}\;\Conid{CGS}\;{}\<[13]%
\>[13]{}\Conid{CGS}{}\<[18]%
\>[18]{}\mathrel{=}[\mskip1.5mu \mathrm{1},{}\<[28]%
\>[28]{}\mathrm{1},{}\<[32]%
\>[32]{}\mathrm{1}{}\<[39]%
\>[39]{}\mskip1.5mu]{}\<[39E]%
\ColumnHook
\end{hscode}\resethooks
The table fulfils \ensuremath{\Varid{fsLemma}_1}, \ensuremath{\Varid{fsLemma}_2} and \ensuremath{\Varid{fsLemma}_3} by construction
which can be exploited to reduce the amount of boiler-plate code in the
definition of \ensuremath{\Varid{fs}}.

Still, defining \ensuremath{\Conid{Units}} through enumeration is perhaps not completely
satisfactory from a conceptual point of view: first, it does not
explicitly encode the idea that units of measurement are distinguished
properties of \emph{reference} physical quantities.
For example, the standard metre is defined as the length of the path
traveled by light in vacuum during a time interval of 1/299792458 of a
second\footnote{See \url{https://en.wikipedia.org/wiki/Metre}.}.

Second, a datatype with only two (or even a countable number of)
inhabitants is perhaps a bit too small to encode a principle (that there
is no privileged system of units of measurement) that must hold for an
uncountable number of systems of units.

One can avoid these shortcomings by introducing a data constructor for
systems of units that takes as arguments strictly positive but otherwise
arbitrary reference lengths, times and masses. For example
\begin{hscode}\SaveRestoreHook
\column{B}{@{}>{\hspre}l<{\hspost}@{}}%
\column{5}{@{}>{\hspre}l<{\hspost}@{}}%
\column{7}{@{}>{\hspre}l<{\hspost}@{}}%
\column{9}{@{}>{\hspre}l<{\hspost}@{}}%
\column{14}{@{}>{\hspre}c<{\hspost}@{}}%
\column{14E}{@{}l@{}}%
\column{17}{@{}>{\hspre}l<{\hspost}@{}}%
\column{21}{@{}>{\hspre}l<{\hspost}@{}}%
\column{35}{@{}>{\hspre}c<{\hspost}@{}}%
\column{35E}{@{}l@{}}%
\column{39}{@{}>{\hspre}l<{\hspost}@{}}%
\column{48}{@{}>{\hspre}l<{\hspost}@{}}%
\column{66}{@{}>{\hspre}l<{\hspost}@{}}%
\column{E}{@{}>{\hspre}l<{\hspost}@{}}%
\>[5]{}\Varid{mutual}{}\<[E]%
\\[\blanklineskip]%
\>[5]{}\hsindent{2}{}\<[7]%
\>[7]{}\mathbf{data}\;\Conid{Units}\ \mathop{:}\ \Conid{Type}\;\mathbf{where}{}\<[E]%
\\
\>[7]{}\hsindent{2}{}\<[9]%
\>[9]{}\Conid{SI}{}\<[14]%
\>[14]{}\ \mathop{:}\ {}\<[14E]%
\>[17]{}\Conid{Units}{}\<[E]%
\\
\>[7]{}\hsindent{2}{}\<[9]%
\>[9]{}\Conid{Ref}\ \mathop{:}\ {}\<[17]%
\>[17]{}(\Varid{l}{}\<[21]%
\>[21]{}\ \mathop{:}\ \Conid{Q}\;\Conid{Length}){}\<[35]%
\>[35]{}\to {}\<[35E]%
\>[39]{}\{\mskip1.5mu \Varid{auto}\;\Varid{p}{}\<[48]%
\>[48]{}\ \mathop{:}\ (\mathrm{0.0}\mathbin{<}\Varid{val}\;\Varid{l}){}\<[66]%
\>[66]{}\mathrel{=}\Conid{True}\mskip1.5mu\}\to {}\<[E]%
\\
\>[17]{}(\Varid{t}{}\<[21]%
\>[21]{}\ \mathop{:}\ \Conid{Q}\;\Conid{Time}){}\<[35]%
\>[35]{}\to {}\<[35E]%
\>[39]{}\{\mskip1.5mu \Varid{auto}\;\Varid{q}{}\<[48]%
\>[48]{}\ \mathop{:}\ (\mathrm{0.0}\mathbin{<}\Varid{val}\;\Varid{t}){}\<[66]%
\>[66]{}\mathrel{=}\Conid{True}\mskip1.5mu\}\to {}\<[E]%
\\
\>[17]{}(\Varid{m}{}\<[21]%
\>[21]{}\ \mathop{:}\ \Conid{Q}\;\Conid{Mass}){}\<[35]%
\>[35]{}\to {}\<[35E]%
\>[39]{}\{\mskip1.5mu \Varid{auto}\;\Varid{r}{}\<[48]%
\>[48]{}\ \mathop{:}\ (\mathrm{0.0}\mathbin{<}\Varid{val}\;\Varid{m}){}\<[66]%
\>[66]{}\mathrel{=}\Conid{True}\mskip1.5mu\}\to \Conid{Units}{}\<[E]%
\\[\blanklineskip]%
\>[5]{}\hsindent{2}{}\<[7]%
\>[7]{}\mathbf{data}\;\Conid{Q}\ \mathop{:}\ \Conid{D}\to \Conid{Type}\;\mathbf{where}{}\<[E]%
\\
\>[7]{}\hsindent{2}{}\<[9]%
\>[9]{}\Conid{Val}\ \mathop{:}\ \{\mskip1.5mu \Varid{d}\ \mathop{:}\ \Conid{D}\mskip1.5mu\}\to (\Varid{u}\ \mathop{:}\ \Conid{Units})\to \Real\to \Conid{Q}\;\Varid{d}{}\<[E]%
\\[\blanklineskip]%
\>[5]{}\hsindent{2}{}\<[7]%
\>[7]{}\Varid{val}\ \mathop{:}\ \{\mskip1.5mu \Varid{d}\ \mathop{:}\ \Conid{D}\mskip1.5mu\}\to \Conid{Q}\;\Varid{d}\to \Real{}\<[E]%
\\
\>[5]{}\hsindent{2}{}\<[7]%
\>[7]{}\Varid{val}\;(\Conid{Val}\;\Varid{u}\;\Varid{x})\mathrel{=}\Varid{x}{}\<[E]%
\ColumnHook
\end{hscode}\resethooks
As one would expect from the notion of measurement as counting, negative
values and the neutral elements of \ensuremath{(\mathbin{+})} for lengths, times and masses
are not suitable reference quantities: in building an arbitrary system
of units in the LTM class with the \ensuremath{\Conid{Ref}} constructor, these conditions
are checked with the ``internal'' \ensuremath{\Varid{val}} helper.

Perhaps not surprisingly, the approach requires a mutual definition of
units of measurement and physical quantities but otherwise presents no
difficulties. The table that returns the scaling factors between units
of measurement \ensuremath{\Varid{fs}} and the measure function \ensuremath{\mu} also have to be
defined mutually:
\begin{hscode}\SaveRestoreHook
\column{B}{@{}>{\hspre}l<{\hspost}@{}}%
\column{7}{@{}>{\hspre}l<{\hspost}@{}}%
\column{14}{@{}>{\hspre}l<{\hspost}@{}}%
\column{30}{@{}>{\hspre}c<{\hspost}@{}}%
\column{30E}{@{}l@{}}%
\column{33}{@{}>{\hspre}l<{\hspost}@{}}%
\column{49}{@{}>{\hspre}c<{\hspost}@{}}%
\column{49E}{@{}l@{}}%
\column{52}{@{}>{\hspre}l<{\hspost}@{}}%
\column{60}{@{}>{\hspre}l<{\hspost}@{}}%
\column{80}{@{}>{\hspre}l<{\hspost}@{}}%
\column{101}{@{}>{\hspre}l<{\hspost}@{}}%
\column{E}{@{}>{\hspre}l<{\hspost}@{}}%
\>[7]{}\Varid{partial}\;\Varid{fs}\ \mathop{:}\ \Conid{Units}\to \Conid{Units}\to \Real_{+}^3{}\<[E]%
\\
\>[7]{}\Varid{fs}\;{}\<[14]%
\>[14]{}\Conid{SI}\;{}\<[33]%
\>[33]{}\Conid{SI}{}\<[49]%
\>[49]{}\mathrel{=}{}\<[49E]%
\>[52]{}[\mskip1.5mu \mathrm{1},\mathrm{1},\mathrm{1}\mskip1.5mu]{}\<[E]%
\\
\>[7]{}\Varid{fs}\;{}\<[14]%
\>[14]{}\Conid{SI}\;{}\<[30]%
\>[30]{}({}\<[30E]%
\>[33]{}\Conid{Ref}\;\Varid{l}\;\Varid{t}\;\Varid{m}){}\<[49]%
\>[49]{}\mathrel{=}{}\<[49E]%
\>[52]{}[\mskip1.5mu \mathrm{1.0}\mathbin{/}(\mu\;\Conid{SI}\;\Varid{l}),\mathrm{1.0}\mathbin{/}(\mu\;\Conid{SI}\;\Varid{t}),\mathrm{1.0}\mathbin{/}(\mu\;\Conid{SI}\;\Varid{m})\mskip1.5mu]{}\<[E]%
\\
\>[7]{}\Varid{fs}\;({}\<[14]%
\>[14]{}\Conid{Ref}\;\Varid{l}\;\Varid{t}\;\Varid{m})\;{}\<[33]%
\>[33]{}\Conid{SI}{}\<[49]%
\>[49]{}\mathrel{=}{}\<[49E]%
\>[52]{}[\mskip1.5mu {}\<[60]%
\>[60]{}\mu\;\Conid{SI}\;\Varid{l},{}\<[80]%
\>[80]{}\mu\;\Conid{SI}\;\Varid{t},{}\<[101]%
\>[101]{}\mu\;\Conid{SI}\;\Varid{m}\mskip1.5mu]{}\<[E]%
\\
\>[7]{}\Varid{fs}\;({}\<[14]%
\>[14]{}\Conid{Ref}\;l_1\;\Varid{t}_{1}\;\Varid{m}_{1})\;{}\<[30]%
\>[30]{}({}\<[30E]%
\>[33]{}\Conid{Ref}\;l_2\;\Varid{t}_{2}\;\Varid{m}_{2}){}\<[49]%
\>[49]{}\mathrel{=}{}\<[49E]%
\>[52]{}[\mskip1.5mu \mu\;\Conid{SI}\;l_1\mathbin{/}\mu\;\Conid{SI}\;l_2,\mu\;\Conid{SI}\;\Varid{t}_{1}\mathbin{/}\mu\;\Conid{SI}\;\Varid{t}_{2},\mu\;\Conid{SI}\;\Varid{m}_{1}\mathbin{/}\mu\;\Conid{SI}\;\Varid{m}_{2}\mskip1.5mu]{}\<[E]%
\\[\blanklineskip]%
\>[7]{}\Varid{partial}\;\mu\ \mathop{:}\ \{\mskip1.5mu \Varid{d}\ \mathop{:}\ \Conid{D}\mskip1.5mu\}\to \Conid{Units}\to \Conid{Q}\;\Varid{d}\to \Real{}\<[E]%
\\
\>[7]{}\mu\;\{\mskip1.5mu \Varid{d}\mskip1.5mu\}\;\Varid{u'}\;(\Conid{Val}\;\Varid{u}\;\Varid{x})\mathrel{=}\Varid{x}\ensuremath{\cdot}\Varid{df}\;\Varid{d}\;(\Varid{fs}\;\Varid{u}\;\Varid{u'}){}\<[E]%
\ColumnHook
\end{hscode}\resethooks
The mutual definition prevents the (necessarily conservative) Idris
termination checker to establish the totality of \ensuremath{\Varid{fs}} and \ensuremath{\mu} which
is why these functions are marked as partial.

Apart from these technicalities, the implementation of \ensuremath{\Varid{dim}} and the
definition of new systems of units and of physical quantities in terms
of arbitrary reference lengths, times and masses are straightforward,
see the literate Idris code that generates this document \citep{Pi2023}.

A variation on the mutual approach sketched above could be to avoid
introducing \ensuremath{\Conid{SI}} as an outstanding system of units altogether and
instead equip the data type of physical quantities with three reference
constructors for a length, a time and a mass. For the rest of this and
of the next section, we stick to the minimal DSL with just two codes for
systems of units.

\subsection{Dimensional (in)dependence}
\label{subsection:dind}

\textbf{Dependence.}
Remember that the Pi theorem is about two properties of a generic
``physical relationship'' \ensuremath{\Varid{f}} between a ``dimensional quantity'' $a$
and $k + m$ ``dimensional governing parameters'' $a_1,\dots,a_k$ and
$b_1,\dots,b_m$.

One property (conclusion of the theorem) is that ``the dimension of $a$
can be expressed as a product of the powers of the dimensions of the
parameters $a_1,\dots,a_k$'' as formulated in \cref{eq:pi0}.
We have just seen examples of physical quantities whose dimension
functions can be expressed as products of powers of the dimension
functions of other physical quantities:
\begin{equation*}
  [l]     = [g] [\tau]^{2}, \quad
  [g]     = [l] [\tau]^{-2}, \quad \mathrm{or} \quad
  [\tau]  = [l]^{1/2} [g]^{-1/2}
\end{equation*}
In the \ensuremath{\Conid{D}}-language, we can express and assert the first
equality with
\begin{hscode}\SaveRestoreHook
\column{B}{@{}>{\hspre}l<{\hspost}@{}}%
\column{5}{@{}>{\hspre}l<{\hspost}@{}}%
\column{E}{@{}>{\hspre}l<{\hspost}@{}}%
\>[5]{}\Varid{check}_{6 }\ \mathop{:}\ \Conid{Is}\;\Varid{l}\;(\Varid{dim}\;\Varid{g}\mathbin{`\Conid{Times}`}(\Varid{dim}\;\tau\mathbin{`\Conid{Times}`}\Varid{dim}\;\tau)){}\<[E]%
\\
\>[5]{}\Varid{check}_{6 }\mathrel{=}\Conid{Refl}{}\<[E]%
\ColumnHook
\end{hscode}\resethooks
or, equivalently
\begin{hscode}\SaveRestoreHook
\column{B}{@{}>{\hspre}l<{\hspost}@{}}%
\column{5}{@{}>{\hspre}l<{\hspost}@{}}%
\column{E}{@{}>{\hspre}l<{\hspost}@{}}%
\>[5]{}\Varid{check}_{7 }\ \mathop{:}\ \Conid{Is}\;\Varid{l}\;(\Varid{dim}\;\Varid{g}\mathbin{`\Conid{Times}`}\Conid{Pow}\;(\Varid{dim}\;\tau)\;\mathrm{2}){}\<[E]%
\\
\>[5]{}\Varid{check}_{7 }\mathrel{=}\Conid{Refl}{}\<[E]%
\ColumnHook
\end{hscode}\resethooks
and similarly for the second equality.

\noindent
As already mentioned, formulating $[\tau] = [l]^{1/2} [g]^{-1/2}$ would
require fractional exponents and extending our representation of
dimension functions to vectors of rational numbers.
In other words: the exponents in \cref{eq:pih1} (and those in
\cref{eq:pi0}) are, in general, rational numbers and
representing dimension functions in terms of vectors of rational numbers
is perhaps the most natural setting for formulating the Pi theorem in
type theory.

The drawback of this approach is that it requires implementing rational
numbers in type theory. This is not a problem in principle, but integers
have a simpler algebraic structure (initial ring) and more efficient
implementations.
Also, formulating the Pi theorem in terms of rational exponents does not seem
strictly necessary: if dimensional analysis with rational exponents
allows one to deduce that the period $\tau$ of a simple pendulum scales
with $(l/g)^{\frac{1}{2}}$, integer-based dimensional analysis should be
enough to deduce that $\tau^2$ scales with $l/g$! We explore this
possibility in the next section.

\paragraph*{Formalizing dependence.} \ To this end, let's
formalize the notion that \ensuremath{\Varid{d}\ \mathop{:}\ \Conid{D}} is dependent on \ensuremath{\Varid{ds}\ \mathop{:}\ \Conid{Vec}\;\Varid{k}\;\Conid{D}} iff a
non-zero integer power of \ensuremath{\Varid{d}} can be expressed as a product of integer
powers of \ensuremath{\Varid{ds}}:
\begin{hscode}\SaveRestoreHook
\column{B}{@{}>{\hspre}l<{\hspost}@{}}%
\column{5}{@{}>{\hspre}l<{\hspost}@{}}%
\column{12}{@{}>{\hspre}c<{\hspost}@{}}%
\column{12E}{@{}l@{}}%
\column{15}{@{}>{\hspre}l<{\hspost}@{}}%
\column{30}{@{}>{\hspre}l<{\hspost}@{}}%
\column{E}{@{}>{\hspre}l<{\hspost}@{}}%
\>[5]{}\Conid{IsDep}{}\<[12]%
\>[12]{}\ \mathop{:}\ {}\<[12E]%
\>[15]{}\{\mskip1.5mu \Varid{k}\ \mathop{:}\ \mathbb{N}\mskip1.5mu\}\to (\Varid{d}\ \mathop{:}\ \Conid{D})\to (\Varid{ds}\ \mathop{:}\ \Conid{Vec}\;\Varid{k}\;\Conid{D})\to \Conid{Type}{}\<[E]%
\\
\>[5]{}\Conid{IsDep}\;\{\mskip1.5mu \Varid{k}\mskip1.5mu\}\;\Varid{d}\;\Varid{ds}\mathrel{=}\Conid{Exists}\;{}\<[30]%
\>[30]{}(\mathbb{Z},\Conid{Vec}\;\Varid{k}\;\mathbb{Z})\;(\lambda (\Varid{p},\Varid{ps})\to(\Conid{Not}\;(\Varid{p}\mathrel{=}\mathrm{0}),\Conid{Pow}\;\Varid{d}\;\Varid{p}\mathrel{=}\Conid{ProdPows}\;\Varid{ds}\;\Varid{ps})){}\<[E]%
\ColumnHook
\end{hscode}\resethooks
Therefore, in the \ensuremath{\Conid{D}}-language, dependence between dimension functions
boils down to linear dependence between their representations, as one
would expect.
By extension, we say that a physical quantity \ensuremath{\Varid{q}} is \emph{dimensionally
dependent} on a vector of physical quantities \ensuremath{\Varid{qs}} if \ensuremath{\Varid{dim}\;\Varid{q}} is
dependent on the dimensions of \ensuremath{\Varid{qs}}:
\begin{hscode}\SaveRestoreHook
\column{B}{@{}>{\hspre}l<{\hspost}@{}}%
\column{5}{@{}>{\hspre}l<{\hspost}@{}}%
\column{15}{@{}>{\hspre}c<{\hspost}@{}}%
\column{15E}{@{}l@{}}%
\column{18}{@{}>{\hspre}l<{\hspost}@{}}%
\column{29}{@{}>{\hspre}c<{\hspost}@{}}%
\column{29E}{@{}l@{}}%
\column{32}{@{}>{\hspre}l<{\hspost}@{}}%
\column{E}{@{}>{\hspre}l<{\hspost}@{}}%
\>[5]{}\Conid{IsDimDep}{}\<[15]%
\>[15]{}\ \mathop{:}\ {}\<[15E]%
\>[18]{}\{\mskip1.5mu \Varid{d}\ \mathop{:}\ \Conid{D}\mskip1.5mu\}\to \{\mskip1.5mu \Varid{k}\ \mathop{:}\ \mathbb{N}\mskip1.5mu\}\to \{\mskip1.5mu \Varid{ds}\ \mathop{:}\ \Conid{Vec}\;\Varid{k}\;\Conid{D}\mskip1.5mu\}\to \Conid{Q}\;\Varid{d}\to \Conid{Vec}_{\scalebox{.5}{Q}}\;\Varid{k}\;\Varid{ds}\to \Conid{Type}{}\<[E]%
\\
\>[5]{}\Conid{IsDimDep}\;\{\mskip1.5mu \Varid{d}\mskip1.5mu\}\;\{\mskip1.5mu \Varid{ds}\mskip1.5mu\}\;\Varid{q}\;\Varid{qs}{}\<[29]%
\>[29]{}\mathrel{=}{}\<[29E]%
\>[32]{}\Conid{IsDep}\;\Varid{d}\;\Varid{ds}{}\<[E]%
\ColumnHook
\end{hscode}\resethooks
Notice that \ensuremath{\Conid{IsDimDep}\;\Varid{a}\;\Varid{as}} is an existential type. In order to assess
that \ensuremath{\Varid{a}} is dimensionally dependent on \ensuremath{\Varid{as}}, one has to provide suitable
integer exponents and an equality proof. For the simple pendulum, for
example:
\begin{hscode}\SaveRestoreHook
\column{B}{@{}>{\hspre}l<{\hspost}@{}}%
\column{5}{@{}>{\hspre}l<{\hspost}@{}}%
\column{E}{@{}>{\hspre}l<{\hspost}@{}}%
\>[5]{}\Varid{check}_{8 }\ \mathop{:}\ \Conid{IsDimDep}\;\tau\;[\mskip1.5mu \Varid{l},\Varid{g}\mskip1.5mu]{}\<[E]%
\\
\>[5]{}\Varid{check}_{8 }\mathrel{=}\Conid{Evidence}\;(\mathrm{2},[\mskip1.5mu \mathrm{1},\mathbin{-}\mathrm{1}\mskip1.5mu])\;(\Varid{not2eq0},\Conid{Refl}){}\<[E]%
\ColumnHook
\end{hscode}\resethooks
where \ensuremath{\Varid{not2eq0}} is a proof that 2 is not equal to 0. This evidence
(\ensuremath{\Conid{Evidence}} is the data constructor of \ensuremath{\Conid{Exists}}, a data type
representing dependent pairs first used in the definition of \ensuremath{\Conid{IsDep}}) is
just another way of asserting the equality
\begin{hscode}\SaveRestoreHook
\column{B}{@{}>{\hspre}l<{\hspost}@{}}%
\column{5}{@{}>{\hspre}l<{\hspost}@{}}%
\column{E}{@{}>{\hspre}l<{\hspost}@{}}%
\>[5]{}\Varid{check}_{9 }\ \mathop{:}\ \Conid{Pow}\;(\Varid{dim}\;\tau)\;\mathrm{2}\mathrel{=}\Conid{Pow}\;(\Varid{dim}\;\Varid{l})\;\mathrm{1}\mathbin{`\Conid{Times}`}\Conid{Pow}\;(\Varid{dim}\;\Varid{g})\;(\mathbin{-}\mathrm{1}){}\<[E]%
\\
\>[5]{}\Varid{check}_{9 }\mathrel{=}\Conid{Refl}{}\<[E]%
\ColumnHook
\end{hscode}\resethooks
For the simple pendulum example, it allows one to deduce that, under
quite general assumptions\footnote{The assumptions are
that the period of oscillation of the pendulum only depends on its mass,
its length, the acceleration of gravity and the initial angle, see for
example the introduction of \citep{barenblatt1996scaling}.}, the period
of oscillation $\tau$ is proportional to the square root of $l/g$.

\paragraph*{Forming dimensionless products.} \ Given a physical quantity
which is dimensionally dependent on other physical quantities, one can
make it dimensionless like the ``$\Pi$'' quantities of the Pi theorem:
\begin{hscode}\SaveRestoreHook
\column{B}{@{}>{\hspre}l<{\hspost}@{}}%
\column{5}{@{}>{\hspre}l<{\hspost}@{}}%
\column{7}{@{}>{\hspre}l<{\hspost}@{}}%
\column{11}{@{}>{\hspre}l<{\hspost}@{}}%
\column{18}{@{}>{\hspre}c<{\hspost}@{}}%
\column{18E}{@{}l@{}}%
\column{21}{@{}>{\hspre}l<{\hspost}@{}}%
\column{E}{@{}>{\hspre}l<{\hspost}@{}}%
\>[5]{}\Varid{makeDimLess}{}\<[18]%
\>[18]{}\ \mathop{:}\ {}\<[18E]%
\>[21]{}\{\mskip1.5mu \Varid{d}\ \mathop{:}\ \Conid{D}\mskip1.5mu\}\to \{\mskip1.5mu \Varid{k}\ \mathop{:}\ \mathbb{N}\mskip1.5mu\}\to \{\mskip1.5mu \Varid{ds}\ \mathop{:}\ \Conid{Vec}\;\Varid{k}\;\Conid{D}\mskip1.5mu\}\to {}\<[E]%
\\
\>[21]{}(\Varid{q}\ \mathop{:}\ \Conid{Q}\;\Varid{d})\to (\Varid{qs}\ \mathop{:}\ \Conid{Vec}_{\scalebox{.5}{Q}}\;\Varid{k}\;\Varid{ds})\to \Conid{IsDimDep}\;\Varid{q}\;\Varid{qs}\to \Conid{Q}\;\Conid{DimLess}{}\<[E]%
\\
\>[5]{}\Varid{makeDimLess}\;\{\mskip1.5mu \Varid{d}\mskip1.5mu\}\;\{\mskip1.5mu \Varid{ds}\mskip1.5mu\}\;\Varid{q}\;\Varid{qs}\;(\Conid{Evidence}\;(\Varid{p},\Varid{ps})\;\Varid{prf})\mathrel{=}{}\<[E]%
\\
\>[5]{}\hsindent{2}{}\<[7]%
\>[7]{}\mathbf{let}\;\Varid{comp}\mathrel{=}\Varid{pow}\;\Varid{q}\;\Varid{p}\mathbin{/}\Varid{prodPows}\;\Varid{qs}\;\Varid{ps}{}\<[E]%
\\
\>[5]{}\hsindent{2}{}\<[7]%
\>[7]{}\mathbf{in}\;{}\<[11]%
\>[11]{}\Varid{replace}\;(\Varid{dimMakeDimLessCompIsDimLess}\;\Varid{d}\;\Varid{ds}\;(\Conid{Evidence}\;(\Varid{p},\Varid{ps})\;\Varid{prf}))\;\Varid{comp}{}\<[E]%
\ColumnHook
\end{hscode}\resethooks
\noindent
In \ensuremath{\Varid{makeDimLess}}, we have applied the function
\ensuremath{\Varid{dimMakeDimLessCompIsDimLess}}. This takes a dimension \ensuremath{\Varid{d}}, a vector of
dimensions \ensuremath{\Varid{ds}} and evidence \ensuremath{\Varid{e}\ \mathop{:}\ \Conid{IsDep}\;\Varid{d}\;\Varid{ds}}. It returns a proof that
\ensuremath{\Varid{comp}\mathrel{=}\Varid{pow}\;\Varid{q}\;\Varid{p}\mathbin{/}\Varid{prodPows}\;\Varid{qs}\;\Varid{ps}} is indeed dimensionless.
Implementing this proof requires proving that \ensuremath{\Varid{d}\mathbin{`\Conid{Over}`}\Varid{d}} is equal to
\ensuremath{\Conid{DimLess}} for arbitrary \ensuremath{\Varid{d}}:
\begin{hscode}\SaveRestoreHook
\column{B}{@{}>{\hspre}l<{\hspost}@{}}%
\column{5}{@{}>{\hspre}l<{\hspost}@{}}%
\column{E}{@{}>{\hspre}l<{\hspost}@{}}%
\>[5]{}\Varid{dodIsDimLess}\ \mathop{:}\ \{\mskip1.5mu \Varid{d}\ \mathop{:}\ \Conid{D}\mskip1.5mu\}\to \Varid{d}\mathbin{`\Conid{Over}`}\Varid{d}\mathrel{=}\Conid{DimLess}{}\<[E]%
\ColumnHook
\end{hscode}\resethooks
This is straightforward for our definition of \ensuremath{\Conid{D}} and suggests that
\ensuremath{\Conid{D}}-values form group. See \citet{doi:10.1142/9789811242380_0020} and
\cref{section:generalization} for a discussion of the algebraic
structure of \ensuremath{\Conid{D}}.

\paragraph*{Independence.} \ Remember that a condition for the function
$f$ of the Pi theorem to be reducible to the form of \cref{eq:pi1} is
that the parameters $a_1,\dots,a_k$ ``have independent dimensions''.
Perhaps not surprisingly, the idea is that \ensuremath{\Varid{qs}\ \mathop{:}\ \Conid{Vec}_{\scalebox{.5}{Q}}\;\Varid{k}\;\Varid{ds}} are
dimensionally independent iff expressing \ensuremath{\Conid{DimLess}} as a product of
powers of their dimension functions requires all exponents to be zero.
This is equivalent to saying the vectors associated with their dimensions
are linearly independent, see \citep[Section 1.1.5]{barenblatt1996scaling}:
\begin{hscode}\SaveRestoreHook
\column{B}{@{}>{\hspre}l<{\hspost}@{}}%
\column{5}{@{}>{\hspre}l<{\hspost}@{}}%
\column{18}{@{}>{\hspre}c<{\hspost}@{}}%
\column{18E}{@{}l@{}}%
\column{21}{@{}>{\hspre}l<{\hspost}@{}}%
\column{E}{@{}>{\hspre}l<{\hspost}@{}}%
\>[5]{}\Conid{AreDimIndep}{}\<[18]%
\>[18]{}\ \mathop{:}\ {}\<[18E]%
\>[21]{}\{\mskip1.5mu \Varid{k}\ \mathop{:}\ \mathbb{N}\mskip1.5mu\}\to \{\mskip1.5mu \Varid{ds}\ \mathop{:}\ \Conid{Vec}\;\Varid{k}\;\Conid{D}\mskip1.5mu\}\to \Conid{Vec}_{\scalebox{.5}{Q}}\;\Varid{k}\;\Varid{ds}\to \Conid{Type}{}\<[E]%
\\
\>[5]{}\Conid{AreDimIndep}\;\{\mskip1.5mu \Varid{ds}\mskip1.5mu\}\;\anonymous \mathrel{=}\Conid{AreIndep}\;\Varid{ds}{}\<[E]%
\ColumnHook
\end{hscode}\resethooks
In the definition of \ensuremath{\Conid{AreDimIndep}} we have applied the predicate
\ensuremath{\Conid{AreIndep}}. In mechanics, \ensuremath{\Varid{n}\mathrel{=}\mathrm{3}} and \ensuremath{\Conid{D}\mathrel{=}\Conid{Vec}\;\mathrm{3}\;\mathbb{Z}}, so that \ensuremath{\Conid{AreIndep}} is decidable
\begin{hscode}\SaveRestoreHook
\column{B}{@{}>{\hspre}l<{\hspost}@{}}%
\column{5}{@{}>{\hspre}l<{\hspost}@{}}%
\column{E}{@{}>{\hspre}l<{\hspost}@{}}%
\>[5]{}\Conid{AreIndep}\ \mathop{:}\ \{\mskip1.5mu \Varid{k}\ \mathop{:}\ \mathbb{N}\mskip1.5mu\}\to \Conid{Vec}\;\Varid{k}\;(\Conid{Vec}\;\mathrm{3}\;\mathbb{Z})\to \Conid{Type}{}\<[E]%
\\
\>[5]{}\Conid{AreIndep}\;\Varid{vs}\mathrel{=}\Varid{areLinIndep}\;\Varid{vs}\mathrel{=}\Conid{True}{}\<[E]%
\ColumnHook
\end{hscode}\resethooks
and a decision procedure can be implemented by pattern matching on the the length of \ensuremath{\Varid{vs}}
\begin{hscode}\SaveRestoreHook
\column{B}{@{}>{\hspre}l<{\hspost}@{}}%
\column{5}{@{}>{\hspre}l<{\hspost}@{}}%
\column{24}{@{}>{\hspre}l<{\hspost}@{}}%
\column{47}{@{}>{\hspre}c<{\hspost}@{}}%
\column{47E}{@{}l@{}}%
\column{50}{@{}>{\hspre}l<{\hspost}@{}}%
\column{E}{@{}>{\hspre}l<{\hspost}@{}}%
\>[5]{}\Varid{areLinIndep}\ \mathop{:}\ \{\mskip1.5mu \Varid{k}\ \mathop{:}\ \mathbb{N}\mskip1.5mu\}\to \Conid{Vec}\;\Varid{k}\;(\Conid{Vec}\;\mathrm{3}\;\mathbb{Z})\to \Conid{Bool}{}\<[E]%
\\
\>[5]{}\Varid{areLinIndep}\;{}\<[24]%
\>[24]{}\Conid{Nil}{}\<[47]%
\>[47]{}\mathrel{=}{}\<[47E]%
\>[50]{}\Conid{True}{}\<[E]%
\\
\>[5]{}\Varid{areLinIndep}\;(v_1\mathbin{::}\Conid{Nil}){}\<[47]%
\>[47]{}\mathrel{=}{}\<[47E]%
\>[50]{}\neg \;(v_1\doubleequals[\mskip1.5mu \mathrm{0},\mathrm{0},\mathrm{0}\mskip1.5mu]){}\<[E]%
\\
\>[5]{}\Varid{areLinIndep}\;(v_1\mathbin{::}v_2\mathbin{::}\Conid{Nil}){}\<[47]%
\>[47]{}\mathrel{=}{}\<[47E]%
\>[50]{}\neg \;(\Varid{areCollinear}\;v_1\;v_2){}\<[E]%
\\
\>[5]{}\Varid{areLinIndep}\;(v_1\mathbin{::}v_2\mathbin{::}v_3\mathbin{::}\Conid{Nil}){}\<[47]%
\>[47]{}\mathrel{=}{}\<[47E]%
\>[50]{}\neg \;(\Varid{det}\;v_1\;v_2\;v_3\doubleequals\mathrm{0}){}\<[E]%
\\
\>[5]{}\Varid{areLinIndep}\;(v_1\mathbin{::}v_2\mathbin{::}v_3\mathbin{::}v_4\mathbin{::}\Varid{vs}){}\<[47]%
\>[47]{}\mathrel{=}{}\<[47E]%
\>[50]{}\Conid{False}{}\<[E]%
\ColumnHook
\end{hscode}\resethooks
where \ensuremath{\Varid{areCollinear}} and \ensuremath{\Varid{det}} are defined in terms of the standard dot
and cross products in \ensuremath{\Conid{Vec}\;\mathrm{3}\;\mathbb{Z}}. Thus, \ensuremath{\Conid{AreDimIndep}} can be
applied to assess the dimensional independence of physical quantities:
\begin{hscode}\SaveRestoreHook
\column{B}{@{}>{\hspre}l<{\hspost}@{}}%
\column{5}{@{}>{\hspre}l<{\hspost}@{}}%
\column{E}{@{}>{\hspre}l<{\hspost}@{}}%
\>[5]{}\Varid{check}_{11}\ \mathop{:}\ \Conid{AreDimIndep}\;[\mskip1.5mu \Varid{l},\Varid{g}\mskip1.5mu]{}\<[E]%
\\
\>[5]{}\Varid{check}_{11}\mathrel{=}\Conid{Refl}{}\<[E]%
\\[\blanklineskip]%
\>[5]{}\Varid{check}_{12}\ \mathop{:}\ \Conid{Not}\;(\Conid{AreDimIndep}\;[\mskip1.5mu \tau,\Varid{l},\Varid{g}\mskip1.5mu]){}\<[E]%
\\
\>[5]{}\Varid{check}_{12}\;\Conid{Refl}\;\Varid{impossible}{}\<[E]%
\ColumnHook
\end{hscode}\resethooks
We have encoded the notions of dimension function, physical quantity,
measurement, units of measurement and dimensional (in)dependence for
classical mechanics (and its subdomains) in Idris and
Agda\footnote{See \texttt{DSL1.agda} in the \texttt{agda} folder of \cite{Pi2023}.}.
With this minimal DSL, we can express dimensional
judgments and implement dimensionally consistent programs.

In the next section, we apply the DSL to formulate the covariance principle
and the Pi theorem in type theory, discuss its non-constructive nature
and present a method for applying the theorem to compute functions that
fulfil the covariance principle by construction.

\section{The covariance principle and verified applications of the Pi theorem}
\label{section:piexplained1}

We start by formalizing the covariance principle and the Pi theorem as
formulated in \cref{section:pi}. This strictly follows
\citet{barenblatt1996scaling} which, in turn, follows those of
\citet{Buckingham1914, Rayleigh1915, bridgman1922} and of standard DA
textbooks. Then, in \cref{subsection:piexplained1.2} we discuss the
non-constructive nature of classical proofs and, in
\cref{subsection:piexplained1.3}, we exploit the formalization of
\cref{subsection:piexplained1.1} to derive a method for ``applying'' the
Pi theorem (as this is routinely done in mathematical physics and
modelling) in a verified type theoretical setting.

\subsection{The covariance principle and the Pi theorem}
\label{subsection:piexplained1.1}
Following the principle that types can encode logical propositions
\citep{howard80}, our first step is to define a type, say \ensuremath{\Conid{Pi}}, such
that values of type \ensuremath{\Conid{Pi}} represent proofs of the Pi theorem.

In \cref{subsection:pi} we saw that the Pi theorem asserts that
a ``physical relationship'' \ensuremath{\Varid{f}} between a ``dimensional
quantity'' \ensuremath{\Varid{a}} and \ensuremath{\Varid{k}\mathbin{+}\Varid{m}} ``dimensional governing parameters''
$a_1,\dots,a_k$ and $b_1,\dots,b_m$ that fulfil \cref{eq:pih0,eq:pih1}
satisfies \cref{eq:pi0,eq:pi1}.
Thus the theorem entails two conclusions, both quantified over functions
between physical quantities. We account for this through two
higher-order function types \ensuremath{\Conid{Pi}_{\ref{eq:pi0}},\Conid{Pi}_{\ref{eq:pi1}}\ \mathop{:}\ \Conid{Type}}
\begin{hscode}\SaveRestoreHook
\column{B}{@{}>{\hspre}l<{\hspost}@{}}%
\column{5}{@{}>{\hspre}l<{\hspost}@{}}%
\column{10}{@{}>{\hspre}c<{\hspost}@{}}%
\column{10E}{@{}l@{}}%
\column{13}{@{}>{\hspre}l<{\hspost}@{}}%
\column{E}{@{}>{\hspre}l<{\hspost}@{}}%
\>[5]{}\Conid{Pi}_{\ref{eq:pi0}}{}\<[10]%
\>[10]{}\mathrel{=}{}\<[10E]%
\>[13]{}(\Varid{f}\ \mathop{:}\ \Conid{Vec}_{\scalebox{.5}{Q}}\;\Varid{k}\;\Varid{ds}\to \Conid{Vec}_{\scalebox{.5}{Q}}\;\Varid{m}\;\Varid{ds'}\to \Conid{Q}\;\Varid{d})\to \dots{}\<[E]%
\ColumnHook
\end{hscode}\resethooks
and similarly for \ensuremath{\Conid{Pi}_{\ref{eq:pi1}}} with implicit parameters \ensuremath{\Varid{k},\Varid{m}\ \mathop{:}\ \mathbb{N}}, \ensuremath{\Varid{ds}\ \mathop{:}\ \Conid{Vec}\;\Varid{k}\;\Conid{D}}, \ensuremath{\Varid{ds'}\ \mathop{:}\ \Conid{Vec}\;\Varid{m}\;\Conid{D}} and \ensuremath{\Varid{d}\ \mathop{:}\ \Conid{D}}. As discussed in
\cref{subsection:pi}, the term ``physical relationship'' is used by
\citet{barenblatt1996scaling} to denote a function that fulfils the
``covariance principle''.

We have seen in \cref{subsection:df} that the covariance principle (or
principle of relativity of measurements) posits that there is no
privileged system of units of measurement or, equivalently, that all
systems are equally good.
So far, we have formalized the notion of covariance for dimension
functions (through the specification \ref{eq:df0}) and we have argued
that, for elementary binary operations on physical quantities, the
covariance principle boils down to the requirement that \ensuremath{\mu_u} is a
homomorphism, see \cref{subsection:quantities}.

In general, a function \ensuremath{\Varid{f}\ \mathop{:}\ \Conid{Vec}_{\scalebox{.5}{Q}}\;\Varid{m}\;\Varid{ds}\to \Conid{Q}\;\Varid{d}} fulfils the covariance
principle iff there exists a representation \ensuremath{\rho_{\!f}\ \mathop{:}\ \Conid{Vec}\;\Varid{m}\;\Real\to \Real}
such that the diagram in \cref{figure:covariance} commutes.
\begin{figure}[h]
\begin{center}
\begin{tikzcd}[row sep=large, column sep = huge]
\ensuremath{\Conid{Vec}_{\scalebox{.5}{Q}}\;\Varid{m}\;\Varid{ds}} \arrow[r, "\ensuremath{\Varid{map}_{\scalebox{.5}{Q}}\;\mu_u}"] \arrow[d, "\ensuremath{\Varid{f}}"]
& \ensuremath{\Conid{Vec}\;\Varid{m}\;\Real} \arrow[d, "\ensuremath{\rho_{\!f}}"] \\
\ensuremath{\Conid{Q}\;\Varid{d}} \arrow[r, "\ensuremath{\mu_u}"]
& \ensuremath{\Real}
\end{tikzcd}
\end{center}
\caption{The covariance principle (or principle of relativity of
  measurements) for a function between physical quantities.}
\Description{A commutative square illustrating the covariance principle. It shows two equivalent paths from 'QVect m ds' to 'Real'. The first path goes right to 'Vect m Real' via 'mapQ mu', then down to 'Real' via 'rhof'. The second path goes down to 'Q d' via 'f', then right to 'Real' via 'mu'.}
\label{figure:covariance}
\end{figure}
Here \ensuremath{\Varid{map}_{\scalebox{.5}{Q}}\;\mu_u} is the function that applies \ensuremath{\mu_u} to the
physical quantities of a \ensuremath{\Conid{Vec}_{\scalebox{.5}{Q}}} and \ensuremath{\Varid{u}} is an arbitrary system of units
of measurement.
We can specialize this notion to binary operations between physical
quantities, for example multiplication
\DONE{Avoid page break in this code block}
\begin{hscode}\SaveRestoreHook
\column{B}{@{}>{\hspre}l<{\hspost}@{}}%
\column{5}{@{}>{\hspre}l<{\hspost}@{}}%
\column{22}{@{}>{\hspre}c<{\hspost}@{}}%
\column{22E}{@{}l@{}}%
\column{25}{@{}>{\hspre}l<{\hspost}@{}}%
\column{33}{@{}>{\hspre}l<{\hspost}@{}}%
\column{41}{@{}>{\hspre}c<{\hspost}@{}}%
\column{41E}{@{}l@{}}%
\column{45}{@{}>{\hspre}l<{\hspost}@{}}%
\column{E}{@{}>{\hspre}l<{\hspost}@{}}%
\>[5]{}\Varid{isCovariantMult}{}\<[22]%
\>[22]{}\ \mathop{:}\ {}\<[22E]%
\>[25]{}\Conid{Exists}\;{}\<[33]%
\>[33]{}(\Real\to \Real\to \Real)\;{}\<[E]%
\\
\>[33]{}(\lambda \rho{}\<[41]%
\>[41]{}\Rightarrow{}\<[41E]%
\>[45]{}(\Varid{u}\ \mathop{:}\ \Conid{Units})\to \{\mskip1.5mu d_1,d_2\ \mathop{:}\ \Conid{D}\mskip1.5mu\}\to {}\<[E]%
\\
\>[45]{}(q_1\ \mathop{:}\ \Conid{Q}\;d_1)\to (q_2\ \mathop{:}\ \Conid{Q}\;d_2)\to {}\<[E]%
\\
\>[45]{}\mu_{\Varid{u}}\;(q_1\ensuremath{\cdot}q_2)\mathrel{=}\rho\;(\mu_{\Varid{u}}\;q_1)\;(\mu_{\Varid{u}}\;q_2)){}\<[E]%
\ColumnHook
\end{hscode}\resethooks
and apply the results from \cref{subsection:quantities} to show that
multiplication between physical quantities is indeed covariant:
\begin{hscode}\SaveRestoreHook
\column{B}{@{}>{\hspre}l<{\hspost}@{}}%
\column{5}{@{}>{\hspre}l<{\hspost}@{}}%
\column{E}{@{}>{\hspre}l<{\hspost}@{}}%
\>[5]{}\Varid{isCovariantMult}\mathrel{=}\Conid{Evidence}\;(\ensuremath{\cdot})\;\mu{}\Conid{HomMult}{}\<[E]%
\ColumnHook
\end{hscode}\resethooks
Similarly, we can encode the requirement that the physical relationship
\ensuremath{\Varid{f}} of the Pi theorem fulfils the covariance principle in terms of a
predicate
\begin{hscode}\SaveRestoreHook
\column{B}{@{}>{\hspre}l<{\hspost}@{}}%
\column{5}{@{}>{\hspre}l<{\hspost}@{}}%
\column{7}{@{}>{\hspre}l<{\hspost}@{}}%
\column{15}{@{}>{\hspre}l<{\hspost}@{}}%
\column{18}{@{}>{\hspre}c<{\hspost}@{}}%
\column{18E}{@{}l@{}}%
\column{21}{@{}>{\hspre}l<{\hspost}@{}}%
\column{24}{@{}>{\hspre}c<{\hspost}@{}}%
\column{24E}{@{}l@{}}%
\column{28}{@{}>{\hspre}l<{\hspost}@{}}%
\column{42}{@{}>{\hspre}l<{\hspost}@{}}%
\column{E}{@{}>{\hspre}l<{\hspost}@{}}%
\>[5]{}\Conid{IsCovariant}{}\<[18]%
\>[18]{}\ \mathop{:}\ {}\<[18E]%
\>[21]{}\{\mskip1.5mu \Varid{k},\Varid{m}\ \mathop{:}\ \mathbb{N}\mskip1.5mu\}\to \{\mskip1.5mu \Varid{ds}{}\<[42]%
\>[42]{}\ \mathop{:}\ \Conid{Vec}\;\Varid{k}\;\Conid{D}\mskip1.5mu\}\to \{\mskip1.5mu \Varid{ds'}\ \mathop{:}\ \Conid{Vec}\;\Varid{m}\;\Conid{D}\mskip1.5mu\}\to \{\mskip1.5mu \Varid{d}\ \mathop{:}\ \Conid{D}\mskip1.5mu\}\to {}\<[E]%
\\
\>[21]{}(\Varid{f}\ \mathop{:}\ \Conid{Vec}_{\scalebox{.5}{Q}}\;\Varid{k}\;\Varid{ds}\to \Conid{Vec}_{\scalebox{.5}{Q}}\;\Varid{m}\;\Varid{ds'}\to \Conid{Q}\;\Varid{d})\to \Conid{Type}{}\<[E]%
\\[\blanklineskip]%
\>[5]{}\Conid{IsCovariant}\;\{\mskip1.5mu \Varid{k}\mskip1.5mu\}\;\{\mskip1.5mu \Varid{m}\mskip1.5mu\}\;\{\mskip1.5mu \Varid{ds}\mskip1.5mu\}\;\{\mskip1.5mu \Varid{ds'}\mskip1.5mu\}\;\{\mskip1.5mu \Varid{d}\mskip1.5mu\}\;\Varid{f}\mathrel{=}{}\<[E]%
\\
\>[5]{}\hsindent{2}{}\<[7]%
\>[7]{}\Conid{Exists}\;{}\<[15]%
\>[15]{}(\Conid{Vec}\;\Varid{k}\;\Real\to \Conid{Vec}\;\Varid{m}\;\Real\to \Real)\;{}\<[E]%
\\
\>[15]{}(\lambda \rho_{\!f}{}\<[24]%
\>[24]{}\Rightarrow{}\<[24E]%
\>[28]{}(\Varid{u}\ \mathop{:}\ \Conid{Units})\to (\Varid{as}\ \mathop{:}\ \Conid{Vec}_{\scalebox{.5}{Q}}\;\Varid{k}\;\Varid{ds})\to (\Varid{bs}\ \mathop{:}\ \Conid{Vec}_{\scalebox{.5}{Q}}\;\Varid{m}\;\Varid{ds'})\to {}\<[E]%
\\
\>[28]{}\mu_{\Varid{u}}\;(\Varid{f}\;\Varid{as}\;\Varid{bs})\mathrel{=}\rho_{\!f}\;(\Varid{map}_{\scalebox{.5}{Q}}\;\mu_{\Varid{u}}\;\Varid{as})\;(\Varid{map}_{\scalebox{.5}{Q}}\;\mu_{\Varid{u}}\;\Varid{bs})){}\<[E]%
\ColumnHook
\end{hscode}\resethooks
and apply \ensuremath{\Conid{IsCovariant}} to encode the first assumption of the theorem:
\begin{hscode}\SaveRestoreHook
\column{B}{@{}>{\hspre}l<{\hspost}@{}}%
\column{5}{@{}>{\hspre}l<{\hspost}@{}}%
\column{10}{@{}>{\hspre}c<{\hspost}@{}}%
\column{10E}{@{}l@{}}%
\column{13}{@{}>{\hspre}l<{\hspost}@{}}%
\column{E}{@{}>{\hspre}l<{\hspost}@{}}%
\>[5]{}\Conid{Pi}_{\ref{eq:pi0}}{}\<[10]%
\>[10]{}\mathrel{=}{}\<[10E]%
\>[13]{}(\Varid{f}\ \mathop{:}\ \Conid{Vec}_{\scalebox{.5}{Q}}\;\Varid{k}\;\Varid{ds}\to \Conid{Vec}_{\scalebox{.5}{Q}}\;\Varid{m}\;\Varid{ds'}\to \Conid{Q}\;\Varid{d})\to (h_1\ \mathop{:}\ \Conid{IsCovariant}\;\Varid{f})\to \dots{}\<[E]%
\ColumnHook
\end{hscode}\resethooks
Next, we need to formalize the two assumptions \cref{eq:pih0,eq:pih1}
about the arguments of \ensuremath{\Varid{f}}. The first one states that $a_1,\dots,a_k$
``have independent dimensions''. We have seen how to formalize this
assumption in \cref{subsection:dind}:
\begin{hscode}\SaveRestoreHook
\column{B}{@{}>{\hspre}l<{\hspost}@{}}%
\column{3}{@{}>{\hspre}l<{\hspost}@{}}%
\column{8}{@{}>{\hspre}c<{\hspost}@{}}%
\column{8E}{@{}l@{}}%
\column{11}{@{}>{\hspre}l<{\hspost}@{}}%
\column{E}{@{}>{\hspre}l<{\hspost}@{}}%
\>[3]{}\Conid{Pi}_{\ref{eq:pi0}}{}\<[8]%
\>[8]{}\mathrel{=}{}\<[8E]%
\>[11]{}(\Varid{f}\ \mathop{:}\ \Conid{Vec}_{\scalebox{.5}{Q}}\;\Varid{k}\;\Varid{ds}\to \Conid{Vec}_{\scalebox{.5}{Q}}\;\Varid{m}\;\Varid{ds'}\to \Conid{Q}\;\Varid{d})\to (h_1\ \mathop{:}\ \Conid{IsCovariant}\;\Varid{f})\to {}\<[E]%
\\
\>[11]{}(h_2\ \mathop{:}\ \Conid{AreIndep}\;\Varid{ds})\to \dots{}\<[E]%
\ColumnHook
\end{hscode}\resethooks
The second assumption of the Pi theorem, \cref{eq:pih1}, specifies \ensuremath{\Varid{m}}
equalities between dimension functions. We have seen that equality
between dimension functions boils down to equality in \ensuremath{\mathbb{Z}^n} (in
mechanics \ensuremath{\Varid{n}\mathrel{=}\mathrm{3}}) and is thus decidable.

In \cref{subsection:dind}, we have also seen that the exponents
in \cref{eq:pih1} are rational numbers and that we can rewrite these
equalities as
\begin{equation}
[b_i]^{p_{i}} = [a_1]^{p_{i1}} \dots [a_k]^{p_{ik}} \quad i = 1,\dots,m
\label{eq:pih1rev}
\end{equation}
with integers $p_{i},p_{i,1},\dots,p_{i,k}$ as we have done for the
simple pendulum example. This states that the dimension functions of the
physical quantities of the second argument \ensuremath{\Varid{bs}} of \ensuremath{\Varid{f}} can be expressed
as products of powers of the dimension functions of the physical
quantities of the first argument
\begin{hscode}\SaveRestoreHook
\column{B}{@{}>{\hspre}l<{\hspost}@{}}%
\column{5}{@{}>{\hspre}l<{\hspost}@{}}%
\column{10}{@{}>{\hspre}c<{\hspost}@{}}%
\column{10E}{@{}l@{}}%
\column{13}{@{}>{\hspre}l<{\hspost}@{}}%
\column{E}{@{}>{\hspre}l<{\hspost}@{}}%
\>[5]{}\Conid{Pi}_{\ref{eq:pi0}}{}\<[10]%
\>[10]{}\mathrel{=}{}\<[10E]%
\>[13]{}(\Varid{f}\ \mathop{:}\ \Conid{Vec}_{\scalebox{.5}{Q}}\;\Varid{k}\;\Varid{ds}\to \Conid{Vec}_{\scalebox{.5}{Q}}\;\Varid{m}\;\Varid{ds'}\to \Conid{Q}\;\Varid{d})\to (h_1\ \mathop{:}\ \Conid{IsCovariant}\;\Varid{f})\to {}\<[E]%
\\
\>[13]{}(h_2\ \mathop{:}\ \Conid{AreIndep}\;\Varid{ds})\to (h_3\ \mathop{:}\ \Conid{AreDep}\;\Varid{ds'}\;\Varid{ds})\to \dots{}\<[E]%
\ColumnHook
\end{hscode}\resethooks
where \ensuremath{h_3\ \mathop{:}\ \Conid{AreDep}\;\Varid{ds'}\;\Varid{ds}} is a vector of \ensuremath{\Conid{IsDep}} proofs, one for each
element of \ensuremath{\Varid{ds'}}:
\begin{hscode}\SaveRestoreHook
\column{B}{@{}>{\hspre}l<{\hspost}@{}}%
\column{7}{@{}>{\hspre}l<{\hspost}@{}}%
\column{E}{@{}>{\hspre}l<{\hspost}@{}}%
\>[7]{}\Conid{AreDep}\ \mathop{:}\ \{\mskip1.5mu \Varid{m},\Varid{k}\ \mathop{:}\ \mathbb{N}\mskip1.5mu\}\to (\Varid{ds'}\ \mathop{:}\ \Conid{Vec}\;\Varid{m}\;\Conid{D})\to (\Varid{ds}\ \mathop{:}\ \Conid{Vec}\;\Varid{k}\;\Conid{D})\to \Conid{Type}{}\<[E]%
\\
\>[7]{}\Conid{AreDep}\;\Varid{ds'}\;\Varid{ds}\mathrel{=}\Conid{All}\;(\lambda \Varid{d'}\Rightarrow\Conid{IsDep}\;\Varid{d'}\;\Varid{ds})\;\Varid{ds'}{}\<[E]%
\ColumnHook
\end{hscode}\resethooks
In the \ensuremath{\Conid{Pi}} theorem, the parameters that are turned into dimensionless
quantities are $b_1,\dots,b_m$ (with dimensions \ensuremath{\Varid{ds'}}).
They are dimensionally dependent on the dimensionally independent ones
$a_1,\dots,a_k$ (with dimensions \ensuremath{\Varid{ds}}).
From this angle, the two hypotheses \ensuremath{h_2\ \mathop{:}\ \Conid{AreIndep}\;\Varid{ds}} and \ensuremath{h_3\ \mathop{:}\ \Conid{AreDep}\;\Varid{ds'}\;\Varid{ds}} completely determine how the \ensuremath{\Varid{k}\mathbin{+}\Varid{m}} parameters are
split.
In general there can be several choices, and the
user of the \ensuremath{\Conid{Pi}} theorem decides.

With these premises, the Pi theorem warrants the existence of
exponents $p_1,\dots,p_k$ and of a function $\Phi$ such that the
equalities (\ref{eq:pi0}) and (\ref{eq:pi1}) hold. As for
\cref{eq:pih1rev}, these are rational numbers but we can reformulate
\cref{eq:pi0} as:
\begin{equation}
[a]^{p} = [a_1]^{p_1} \dots [a_k]^{p_k}
\label{eq:pi0rev}
\end{equation}
with integers exponents $p,p_1,\dots,p_k$ and the first conclusion of
the Pi theorem (with all its implicit arguments) as
\begin{hscode}\SaveRestoreHook
\column{B}{@{}>{\hspre}l<{\hspost}@{}}%
\column{5}{@{}>{\hspre}l<{\hspost}@{}}%
\column{10}{@{}>{\hspre}c<{\hspost}@{}}%
\column{10E}{@{}l@{}}%
\column{13}{@{}>{\hspre}l<{\hspost}@{}}%
\column{34}{@{}>{\hspre}l<{\hspost}@{}}%
\column{E}{@{}>{\hspre}l<{\hspost}@{}}%
\>[5]{}\Conid{Pi}_{\ref{eq:pi0}}{}\<[10]%
\>[10]{}\ \mathop{:}\ {}\<[10E]%
\>[13]{}\Conid{Type}{}\<[E]%
\\
\>[5]{}\Conid{Pi}_{\ref{eq:pi0}}{}\<[10]%
\>[10]{}\mathrel{=}{}\<[10E]%
\>[13]{}\{\mskip1.5mu \Varid{k},\Varid{m}\ \mathop{:}\ \mathbb{N}\mskip1.5mu\}\to \{\mskip1.5mu \Varid{ds}{}\<[34]%
\>[34]{}\ \mathop{:}\ \Conid{Vec}\;\Varid{k}\;\Conid{D}\mskip1.5mu\}\to \{\mskip1.5mu \Varid{ds'}\ \mathop{:}\ \Conid{Vec}\;\Varid{m}\;\Conid{D}\mskip1.5mu\}\to \{\mskip1.5mu \Varid{d}\ \mathop{:}\ \Conid{D}\mskip1.5mu\}\to {}\<[E]%
\\
\>[13]{}(\Varid{f}\ \mathop{:}\ \Conid{Vec}_{\scalebox{.5}{Q}}\;\Varid{k}\;\Varid{ds}\to \Conid{Vec}_{\scalebox{.5}{Q}}\;\Varid{m}\;\Varid{ds'}\to \Conid{Q}\;\Varid{d})\to (h_1\ \mathop{:}\ \Conid{IsCovariant}\;\Varid{f})\to {}\<[E]%
\\
\>[13]{}(h_2\ \mathop{:}\ \Conid{AreIndep}\;\Varid{ds})\to (h_3\ \mathop{:}\ \Conid{AreDep}\;\Varid{ds'}\;\Varid{ds})\to \Conid{IsDep}\;\Varid{d}\;\Varid{ds}{}\<[E]%
\ColumnHook
\end{hscode}\resethooks
The second conclusion of the Pi theorem is \cref{eq:pi1}. This states
the existence of a function \ensuremath{\Phi} that allows one to express \ensuremath{\Varid{f}\;\Varid{as}\;\Varid{bs}}
to the power of \ensuremath{\Varid{p}} as a product of powers of the \ensuremath{\Varid{as}} times \ensuremath{\Phi}
applied to the dimensionless ``$\Pi$'' fractions of \cref{eq:pi1}:
\begin{hscode}\SaveRestoreHook
\column{B}{@{}>{\hspre}l<{\hspost}@{}}%
\column{5}{@{}>{\hspre}l<{\hspost}@{}}%
\column{10}{@{}>{\hspre}c<{\hspost}@{}}%
\column{10E}{@{}l@{}}%
\column{13}{@{}>{\hspre}l<{\hspost}@{}}%
\column{21}{@{}>{\hspre}c<{\hspost}@{}}%
\column{21E}{@{}l@{}}%
\column{24}{@{}>{\hspre}l<{\hspost}@{}}%
\column{31}{@{}>{\hspre}c<{\hspost}@{}}%
\column{31E}{@{}l@{}}%
\column{34}{@{}>{\hspre}l<{\hspost}@{}}%
\column{35}{@{}>{\hspre}l<{\hspost}@{}}%
\column{40}{@{}>{\hspre}l<{\hspost}@{}}%
\column{49}{@{}>{\hspre}c<{\hspost}@{}}%
\column{49E}{@{}l@{}}%
\column{52}{@{}>{\hspre}l<{\hspost}@{}}%
\column{84}{@{}>{\hspre}c<{\hspost}@{}}%
\column{84E}{@{}l@{}}%
\column{E}{@{}>{\hspre}l<{\hspost}@{}}%
\>[5]{}\Conid{Pi}_{\ref{eq:pi1}}{}\<[10]%
\>[10]{}\ \mathop{:}\ {}\<[10E]%
\>[13]{}\Conid{Type}{}\<[E]%
\\
\>[5]{}\Conid{Pi}_{\ref{eq:pi1}}{}\<[10]%
\>[10]{}\mathrel{=}{}\<[10E]%
\>[13]{}\{\mskip1.5mu \Varid{k},\Varid{m}\ \mathop{:}\ \mathbb{N}\mskip1.5mu\}\to \{\mskip1.5mu \Varid{ds}{}\<[34]%
\>[34]{}\ \mathop{:}\ \Conid{Vec}\;\Varid{k}\;\Conid{D}\mskip1.5mu\}\to \{\mskip1.5mu \Varid{ds'}\ \mathop{:}\ \Conid{Vec}\;\Varid{m}\;\Conid{D}\mskip1.5mu\}\to \{\mskip1.5mu \Varid{d}\ \mathop{:}\ \Conid{D}\mskip1.5mu\}\to {}\<[E]%
\\
\>[13]{}(\Varid{f}\ \mathop{:}\ \Conid{Vec}_{\scalebox{.5}{Q}}\;\Varid{k}\;\Varid{ds}\to \Conid{Vec}_{\scalebox{.5}{Q}}\;\Varid{m}\;\Varid{ds'}\to \Conid{Q}\;\Varid{d})\to (h_1\ \mathop{:}\ \Conid{IsCovariant}\;\Varid{f})\to {}\<[E]%
\\
\>[13]{}(h_2\ \mathop{:}\ \Conid{AreIndep}\;\Varid{ds})\to (h_3\ \mathop{:}\ \Conid{AreDep}\;\Varid{ds'}\;\Varid{ds})\to {}\<[E]%
\\
\>[13]{}\Conid{Exists}\;{}\<[21]%
\>[21]{}({}\<[21E]%
\>[24]{}\Conid{Vec}_{\scalebox{.5}{Q}}\;\Varid{m}\;(\Varid{replicate}\;\Varid{m}\;\Conid{DimLess})\to \Conid{Q}\;\Conid{DimLess})\;{}\<[E]%
\\
\>[21]{}({}\<[21E]%
\>[24]{}\lambda \Phi{}\<[31]%
\>[31]{}\Rightarrow{}\<[31E]%
\>[35]{}(\Varid{as}\ \mathop{:}\ \Conid{Vec}_{\scalebox{.5}{Q}}\;\Varid{k}\;\Varid{ds})\to (\Varid{bs}\ \mathop{:}\ \Conid{Vec}_{\scalebox{.5}{Q}}\;\Varid{m}\;\Varid{ds'})\to {}\<[84]%
\>[84]{}~{}\<[84E]%
\\
\>[35]{}\mathbf{let}\;{}\<[40]%
\>[40]{}(\Varid{p},\Varid{ps}){}\<[49]%
\>[49]{}\mathrel{=}{}\<[49E]%
\>[52]{}\Varid{fst}\;(\pi_{\ref{eq:pi0}}\;\Varid{f}\;h_1\;h_2\;h_3){}\<[E]%
\\
\>[40]{}\Pi\Varid{s}{}\<[49]%
\>[49]{}\mathrel{=}{}\<[49E]%
\>[52]{}\Varid{makeAllDimLess}\;\Varid{bs}\;\Varid{as}\;h_3{}\<[E]%
\\
\>[35]{}\mathbf{in}\;{}\<[40]%
\>[40]{}\Varid{pow}\;(\Varid{f}\;\Varid{as}\;\Varid{bs})\;\Varid{p}\mathrel{=}\Varid{prodPows}\;\Varid{as}\;\Varid{ps}\ensuremath{\cdot}\Phi\;\Pi\Varid{s}{}\<[84]%
\>[84]{}){}\<[84E]%
\ColumnHook
\end{hscode}\resethooks
Thus, both conclusions of the Pi theorem are existential types (remember
the definition of \ensuremath{\Conid{IsDep}\;\Varid{d}\;\Varid{ds}} from \cref{subsection:dind}) and the
second depends on the first through the integer exponents
$p,p_1,\dots,p_k$ denoted, in the code as \ensuremath{(\Varid{p},\Varid{ps})}.
The ``$\Pi$'' fractions \ensuremath{\Pi\Varid{s}} are computed by mapping \ensuremath{\Varid{makeDimLess}} from
\cref{subsection:dind} on $b_1,\dots,b_m$ alias \ensuremath{\Varid{bs}}, see the Idris and
Agda code in \citep{Pi2023}.
%
%
%
Alternatively, one could combine the two conclusions in a single, nested
existential type.

\subsection{The non-constructive nature of the Pi theorem}
\label{subsection:piexplained1.2}

Before discussing the non-constructive nature of the Pi theorem, it is
worth pointing out two difficulties of the textbook formulation
discussed above.
The first one is technical. In the equality at the last line of the
definition of \ensuremath{\Conid{Pi}_{\ref{eq:pi1}}}
\begin{hscode}\SaveRestoreHook
\column{B}{@{}>{\hspre}l<{\hspost}@{}}%
\column{E}{@{}>{\hspre}l<{\hspost}@{}}%
\>[B]{}\Varid{pow}\;(\Varid{f}\;\Varid{as}\;\Varid{bs})\;\Varid{p}\mathrel{=}\Varid{prodPows}\;\Varid{as}\;\Varid{ps}\ensuremath{\cdot}\Phi\;\Pi\Varid{s}{}\<[E]%
\ColumnHook
\end{hscode}\resethooks
the left and the right hand side have different types. The left-hand
side has type \ensuremath{\Conid{Q}\;(\Varid{d}\mathbin{`\Conid{Pow}`}\Varid{p})} where \ensuremath{\Varid{d}} is the return type of \ensuremath{\Varid{f}}. The
right hand side has type \ensuremath{\Conid{Q}\;(\Conid{ProdPows}\;\Varid{ds}\;\Varid{ps}\mathbin{`\Conid{Times}`}\Conid{DimLess})}.
The second projection of the second projection of \ensuremath{\pi_{\ref{eq:pi0}}\;\Varid{f}\;h_1\;h_2\;h_3} is a
witness of the equality between these two types which needs to be
applied in order to formulate the Pi theorem in homogeneous
implementations of identity types \citep{martinlof1984}. For
example, in Agda with homogeneous equality the above formalization of
the Pi theorem reads
%
\begin{hscode}\SaveRestoreHook
\column{B}{@{}>{\hspre}l<{\hspost}@{}}%
\column{6}{@{}>{\hspre}c<{\hspost}@{}}%
\column{6E}{@{}l@{}}%
\column{9}{@{}>{\hspre}l<{\hspost}@{}}%
\column{11}{@{}>{\hspre}l<{\hspost}@{}}%
\column{14}{@{}>{\hspre}l<{\hspost}@{}}%
\column{20}{@{}>{\hspre}c<{\hspost}@{}}%
\column{20E}{@{}l@{}}%
\column{24}{@{}>{\hspre}l<{\hspost}@{}}%
\column{27}{@{}>{\hspre}l<{\hspost}@{}}%
\column{29}{@{}>{\hspre}l<{\hspost}@{}}%
\column{34}{@{}>{\hspre}l<{\hspost}@{}}%
\column{35}{@{}>{\hspre}c<{\hspost}@{}}%
\column{35E}{@{}l@{}}%
\column{38}{@{}>{\hspre}l<{\hspost}@{}}%
\column{39}{@{}>{\hspre}c<{\hspost}@{}}%
\column{39E}{@{}l@{}}%
\column{42}{@{}>{\hspre}l<{\hspost}@{}}%
\column{E}{@{}>{\hspre}l<{\hspost}@{}}%
\>[B]{}\Conid{Pi}_{\ref{eq:pi1}}{}\<[6]%
\>[6]{}\ \mathop{:}\ {}\<[6E]%
\>[9]{}\Conid{Set}{}\<[14]%
\>[14]{}\mbox{\onelinecomment  Agda version}{}\<[E]%
\\
\>[B]{}\Conid{Pi}_{\ref{eq:pi1}}{}\<[6]%
\>[6]{}\mathrel{=}{}\<[6E]%
\>[9]{}\{\mskip1.5mu \Varid{k}\;\Varid{m}\ \mathop{:}\ \mathbb{N} \mskip1.5mu\}\to \{\mskip1.5mu \Varid{ds}{}\<[27]%
\>[27]{}\ \mathop{:}\ \Conid{Vec}\;\Conid{D}\;\Varid{k}\mskip1.5mu\}\to \{\mskip1.5mu \Varid{ds'}\ \mathop{:}\ \Conid{Vec}\;\Conid{D}\;\Varid{m}\mskip1.5mu\}\to \{\mskip1.5mu \Varid{d}\ \mathop{:}\ \Conid{D}\mskip1.5mu\}\to {}\<[E]%
\\
\>[9]{}(\Varid{f}\ \mathop{:}\ \Conid{Vec}_{\scalebox{.5}{Q}}\;\Varid{k}\;\Varid{ds}\to \Conid{Vec}_{\scalebox{.5}{Q}}\;\Varid{m}\;\Varid{ds'}\to \Conid{Q}\;\Varid{d})\to (h_1\ \mathop{:}\ \Conid{IsCovariant}\;\Varid{f})\to {}\<[E]%
\\
\>[9]{}(h_2\ \mathop{:}\ \Conid{AreIndep}\;\Varid{ds})\to (h_3\ \mathop{:}\ \Conid{AreDep}\;\Varid{ds'}\;\Varid{ds})\to {}\<[E]%
\\
\>[9]{}\Sigma \;(\Conid{Vec}_{\scalebox{.5}{Q}}\;\Varid{m}\;(\Varid{replicate}\;\Conid{DimLess})\to \Conid{Q}\;\Conid{DimLess})\;{}\<[E]%
\\
\>[9]{}\hsindent{2}{}\<[11]%
\>[11]{}(\lambda \Phi{}\<[20]%
\>[20]{}\to {}\<[20E]%
\>[24]{}(\Varid{as}\ \mathop{:}\ \Conid{Vec}_{\scalebox{.5}{Q}}\;\Varid{k}\;\Varid{ds})\to (\Varid{bs}\ \mathop{:}\ \Conid{Vec}_{\scalebox{.5}{Q}}\;\Varid{m}\;\Varid{ds'})\to {}\<[E]%
\\
\>[24]{}\mathbf{let}\;{}\<[29]%
\>[29]{}\Varid{ddds}{}\<[35]%
\>[35]{}\ \mathop{:}\ {}\<[35E]%
\>[38]{}\Conid{IsDep}\;\Varid{d}\;\Varid{ds}{}\<[E]%
\\
\>[29]{}\Varid{ddds}{}\<[35]%
\>[35]{}\mathrel{=}{}\<[35E]%
\>[38]{}\pi_{\ref{eq:pi0}}\;\Varid{f}\;h_1\;h_2\;h_3{}\<[E]%
\\
\>[29]{}\Varid{p}{}\<[35]%
\>[35]{}\ \mathop{:}\ {}\<[35E]%
\>[38]{}\mathbb{Z}{}\<[E]%
\\
\>[29]{}\Varid{p}{}\<[35]%
\>[35]{}\mathrel{=}{}\<[35E]%
\>[38]{}\pi_1\;(\pi_1\;\Varid{ddds}){}\<[E]%
\\
\>[29]{}\Varid{ps}{}\<[35]%
\>[35]{}\ \mathop{:}\ {}\<[35E]%
\>[38]{}\Conid{Vec}\;\mathbb{Z}\;\Varid{k}{}\<[E]%
\\
\>[29]{}\Varid{ps}{}\<[35]%
\>[35]{}\mathrel{=}{}\<[35E]%
\>[38]{}\pi_2\;(\pi_1\;\Varid{ddds}){}\<[E]%
\\
\>[29]{}\Varid{prf}_{\!1}{}\<[35]%
\>[35]{}\ \mathop{:}\ {}\<[35E]%
\>[38]{}\Conid{Pow}\;\Varid{d}\;\Varid{p}\equiv\Conid{ProdPows}\;\Varid{ds}\;\Varid{ps}{}\<[E]%
\\
\>[29]{}\Varid{prf}_{\!1}{}\<[35]%
\>[35]{}\mathrel{=}{}\<[35E]%
\>[38]{}\pi_2\;(\pi_2\;\Varid{ddds}){}\<[E]%
\\
\>[29]{}\Varid{prf}_{\!2}{}\<[35]%
\>[35]{}\ \mathop{:}\ {}\<[35E]%
\>[38]{}\Conid{Pow}\;\Varid{d}\;\Varid{p}\equiv\Conid{ProdPows}\;\Varid{ds}\;\Varid{ps}\;\Conid{Times}\;\Conid{DimLess}{}\<[E]%
\\
\>[29]{}\Varid{prf}_{\!2}{}\<[35]%
\>[35]{}\mathrel{=}{}\<[35E]%
\>[38]{}\Varid{trans}\;\Varid{prf}_{\!1}\;(\Varid{sym}\;(\Varid{dTimesDimLessIsd}\;(\Conid{ProdPows}\;\Varid{ds}\;\Varid{ps}))){}\<[E]%
\\
\>[29]{}\Pi\Varid{s}{}\<[39]%
\>[39]{}\mathrel{=}{}\<[39E]%
\>[42]{}\Varid{makeAllDimLess}\;\Varid{bs}\;\Varid{as}\;h_3{}\<[E]%
\\
\>[29]{}\mathbf{in}\;{}\<[34]%
\>[34]{}\Varid{pow}\;(\Varid{f}\;\Varid{as}\;\Varid{bs})\;\Varid{p}{}\<[E]%
\\
\>[34]{}\equiv{}\<[E]%
\\
\>[34]{}\Varid{transport}\;\{\mskip1.5mu \Conid{P}\mathrel{=}\Conid{Q}\mskip1.5mu\}\;(\Varid{sym}\;\Varid{prf}_{\!2})\;(\Varid{prodPows}\;\Varid{as}\;\Varid{ps}\;\Conid{Q}\ensuremath{\cdot}\Conid{Q}\;\Phi\;\Pi\Varid{s})){}\<[E]%
\ColumnHook
\end{hscode}\resethooks
This is in principle not a problem. But it suggests that a one-to-one
translation (modulo integer arithmetic) of the textbook formulation
\cref{eq:pi1} of the Pi theorem is perhaps inadequate.

The second difficulty is conceptual. The Pi theorem starts from the
assumption that a ``physical relationship'' \ensuremath{\Varid{f}} that fulfills the
covariance principle exists between certain physical variables.
In the context of applications of the Pi theorem (Galileo, Newton,
Fourier, Maxwell, Reynolds and Rayleigh) until its formulation in 1914
\citep{Buckingham1914, Buckingham1915} and until today, the existence of
\ensuremath{\Varid{f}} has never been understood to be evidential: \ensuremath{\Varid{f}} is a sought physical
law (like Newton's second law or like the ideal gas law from
\cref{subsection:laws}) that needs to be identified by empirical
experiments or from first principles.

In this context the Pi theorem merely states that \emph{if} \ensuremath{\Varid{f}} would
exist, it would have to satisfy certain dimensional requirements and its
form would be constrained as expressed by \cref{eq:pi1}. In other words:
even if we were given the \ensuremath{\Varid{p}} and \ensuremath{\Varid{ps}} and evidence that \ensuremath{\Varid{d}\mathbin{`\Conid{Pow}`}\Varid{p}}
is equal to \ensuremath{\Conid{ProdPows}\;\Varid{ds}\;\Varid{ps}\mathbin{`\Conid{Times}`}\Conid{DimLess}}, that is, evidence \ensuremath{\pi_{\ref{eq:pi0}}\ \mathop{:}\ \Conid{Pi}_{\ref{eq:pi0}}}, it would make little sense to ``apply'' the Pi theorem to
``compute'' the shape function \ensuremath{\Phi}
because the input \ensuremath{\Varid{f}}, in fact, is not known.

This observation, too, suggests that applying the Pi theorem in type
theory should not be understood as the act of implementing \ensuremath{\Conid{Pi}_{\ref{eq:pi0}}} and
\ensuremath{\Conid{Pi}_{\ref{eq:pi1}}} and then apply these functions to suitable arguments: at least one
of these arguments, namely the function \ensuremath{\Varid{f}} is what we are trying to
identify.

Indeed, proofs of the Pi theorem are typically non-constructive and thus
\ensuremath{\Conid{Pi}_{\ref{eq:pi0}}} is not implementable. For example, the one sketched in Section 1
of \citep{barenblatt1996scaling} starts from the linear algebra lemma
\begin{lemma}
  \label{lemma1}
Let $as = a_1,\dots,a_k$ have independent dimensions. Then for any system of
units $u$, any positive real number $r$ and any $1 \leq j \leq k$ there
exists a system of units $u_j$ such that $\mu \ u_j \ a_j = r \cdot \mu \ u
\ a_j$ and $\mu \ u_j \ a_i = \mu \ u \ a_i$ for all $1 \leq i \leq k$,
$i \neq j$.
\end{lemma}
\noindent
and goes on as follows. Let \ensuremath{\Varid{bs}} = $b_1,\dots,b_m$. Assume \ensuremath{\Varid{f}\;(\Varid{as},\Varid{bs})} and \ensuremath{\Varid{as}}
are dimensionally independent and take $r \neq 1$. Then one has
\begin{equation*}
\ensuremath{\mu_{\Varid{u}_{0}}\;(\Varid{f}\;(\Varid{as},\Varid{bs}))\mathrel{=}\Varid{r}\ensuremath{\cdot}\mu_{\Varid{u}}\;(\Varid{f}\;(\Varid{as},\Varid{bs}))}
\end{equation*}
by \cref{lemma1} and also
\begin{equation*}
\ensuremath{\mu_{\Varid{u}_{0}}\;(\Varid{f}\;(\Varid{as},\Varid{bs}))\mathrel{=}\rho_{\!f}\;(\Varid{map}_{\scalebox{.5}{Q}}\;\mu_{\Varid{u}_{0}}\;\Varid{as},\Varid{map}_{\scalebox{.5}{Q}}\;\mu_{\Varid{u}_{0}}\;\Varid{bs})}
\end{equation*}
by the covariance principle. But by \cref{lemma1} one also has \ensuremath{\Varid{map}_{\scalebox{.5}{Q}}\;\mu_{\Varid{u}_{0}}\;\Varid{as}\mathrel{=}\Varid{map}_{\scalebox{.5}{Q}}\;\mu_{\Varid{u}}\;\Varid{as}}, and also \ensuremath{\Varid{map}_{\scalebox{.5}{Q}}\;\mu_{\Varid{u}_{0}}\;\Varid{bs}\mathrel{=}\Varid{map}_{\scalebox{.5}{Q}}\;\mu_{\Varid{u}}\;\Varid{bs}} (because \ensuremath{\Varid{bs}} are dimensionally dependent on \ensuremath{\Varid{as}}) and thus
\begin{equation*}
\ensuremath{\mu_{\Varid{u}_{0}}\;(\Varid{f}\;(\Varid{as},\Varid{bs}))\mathrel{=}\rho_{\!f}\;(\Varid{map}_{\scalebox{.5}{Q}}\;\mu_{\Varid{u}}\;\Varid{as},\Varid{map}_{\scalebox{.5}{Q}}\;\mu_{\Varid{u}}\;\Varid{bs})}
\end{equation*}
by congruence and, again by the covariance principle \ensuremath{\mu_{\Varid{u}_{0}}\;(\Varid{f}\;(\Varid{as},\Varid{bs}))\mathrel{=}\mu_{\Varid{u}}\;(\Varid{f}\;(\Varid{as},\Varid{bs}))} and thus $r = 1$. Contradiction.  Thus, it is not
the case that \ensuremath{\Varid{f}\;(\Varid{as},\Varid{bs})} and \ensuremath{\Varid{as}} have independent dimensions.

Even if \cref{lemma1} is implementable, the proof sketched by
\citet{barenblatt1996scaling} is non-constructive. One can derive a
contradiction from the assumption that \ensuremath{\Varid{f}\;(\Varid{as},\Varid{bs})} and \ensuremath{\Varid{as}} have
independent dimensions and thus implement a function of type \ensuremath{\neg \;(\Conid{IsDep}\;\Varid{d}\;\Varid{ds})\to \Conid{Void}}.

But such a function does not allow one to produce a value of type \ensuremath{\Conid{IsDep}\;\Varid{d}\;\Varid{ds}}. What is missing is double negation elimination or, equivalently,
the excluded middle.

\subsection{Verified applications of the Pi theorem}
\label{subsection:piexplained1.3}

The lack of implementable formulations of the Pi theorem should neither
be surprising nor concerning. In order to support applying the theorem
as this is routinely done in mathematical physics and modelling (remember
the discussion about Stommel's paper in \cref{subsection:laws}) in a
safe type theoretical setting, one does not need to be able to implement
\ensuremath{\Conid{Pi}_{\ref{eq:pi0}}}, not to mention \ensuremath{\Conid{Pi}_{\ref{eq:pi1}}}.

We only need to recognize that the Pi theorem, while not implementable
in type theory, encodes important restrictions on the return type and on
the form of functions of physical quantities and provide users with a
verified method for enforcing these restrictions.

A way of doing so that does not require modifications of the host
language can be defined in two steps. First, encode the type of
functions of physical quantities as constrained by the Pi theorem
through a data type:
\begin{hscode}\SaveRestoreHook
\column{B}{@{}>{\hspre}l<{\hspost}@{}}%
\column{5}{@{}>{\hspre}l<{\hspost}@{}}%
\column{7}{@{}>{\hspre}l<{\hspost}@{}}%
\column{12}{@{}>{\hspre}c<{\hspost}@{}}%
\column{12E}{@{}l@{}}%
\column{15}{@{}>{\hspre}l<{\hspost}@{}}%
\column{33}{@{}>{\hspre}l<{\hspost}@{}}%
\column{E}{@{}>{\hspre}l<{\hspost}@{}}%
\>[5]{}\mathbf{data}\;\Conid{PhysFunType}\ \mathop{:}\ (\Varid{k},\Varid{m}\ \mathop{:}\ \mathbb{N})\to \Conid{Vec}\;\Varid{k}\;\Conid{D}\to \Conid{Vec}\;\Varid{m}\;\Conid{D}\to \Conid{D}\to \Conid{Type}\;\mathbf{where}{}\<[E]%
\\
\>[5]{}\hsindent{2}{}\<[7]%
\>[7]{}\Conid{PFT}{}\<[12]%
\>[12]{}\ \mathop{:}\ {}\<[12E]%
\>[15]{}\{\mskip1.5mu \Varid{k}\ \mathop{:}\ \mathbb{N}\mskip1.5mu\}\to \{\mskip1.5mu \Varid{ds}{}\<[33]%
\>[33]{}\ \mathop{:}\ \Conid{Vec}\;\Varid{k}\;\Conid{D}\mskip1.5mu\}\to \{\mskip1.5mu \Varid{m}\ \mathop{:}\ \mathbb{N}\mskip1.5mu\}\to \{\mskip1.5mu \Varid{ds'}\ \mathop{:}\ \Conid{Vec}\;\Varid{m}\;\Conid{D}\mskip1.5mu\}\to \{\mskip1.5mu \Varid{d}\ \mathop{:}\ \Conid{D}\mskip1.5mu\}\to {}\<[E]%
\\
\>[15]{}\Conid{AreIndep}\;\Varid{ds}\to \Conid{AreDep}\;\Varid{ds'}\;\Varid{ds}\to \Conid{IsDep}\;\Varid{d}\;\Varid{ds}\to \Conid{PhysFunType}\;\Varid{k}\;\Varid{m}\;\Varid{ds}\;\Varid{ds'}\;\Varid{d}{}\<[E]%
\ColumnHook
\end{hscode}\resethooks
For the second step, it is useful to define the projections
%
\begin{hscode}\SaveRestoreHook
\column{B}{@{}>{\hspre}l<{\hspost}@{}}%
\column{5}{@{}>{\hspre}l<{\hspost}@{}}%
\column{13}{@{}>{\hspre}c<{\hspost}@{}}%
\column{13E}{@{}l@{}}%
\column{16}{@{}>{\hspre}l<{\hspost}@{}}%
\column{34}{@{}>{\hspre}l<{\hspost}@{}}%
\column{E}{@{}>{\hspre}l<{\hspost}@{}}%
\>[5]{}\Varid{areDep}{}\<[13]%
\>[13]{}\ \mathop{:}\ {}\<[13E]%
\>[16]{}\{\mskip1.5mu \Varid{k}\ \mathop{:}\ \mathbb{N}\mskip1.5mu\}\to \{\mskip1.5mu \Varid{ds}{}\<[34]%
\>[34]{}\ \mathop{:}\ \Conid{Vec}\;\Varid{k}\;\Conid{D}\mskip1.5mu\}\to \{\mskip1.5mu \Varid{m}\ \mathop{:}\ \mathbb{N}\mskip1.5mu\}\to \{\mskip1.5mu \Varid{ds'}\ \mathop{:}\ \Conid{Vec}\;\Varid{m}\;\Conid{D}\mskip1.5mu\}\to \{\mskip1.5mu \Varid{d}\ \mathop{:}\ \Conid{D}\mskip1.5mu\}\to {}\<[E]%
\\
\>[16]{}\Conid{PhysFunType}\;\Varid{k}\;\Varid{m}\;\Varid{ds}\;\Varid{ds'}\;\Varid{d}\to \Conid{AreDep}\;\Varid{ds'}\;\Varid{ds}{}\<[E]%
\\
\>[5]{}\Varid{areDep}\;(\Conid{PFT}\;h_2\;h_3\;\pi_{\ref{eq:pi0}})\mathrel{=}h_3{}\<[E]%
\ColumnHook
\end{hscode}\resethooks
\begin{hscode}\SaveRestoreHook
\column{B}{@{}>{\hspre}l<{\hspost}@{}}%
\column{5}{@{}>{\hspre}l<{\hspost}@{}}%
\column{12}{@{}>{\hspre}c<{\hspost}@{}}%
\column{12E}{@{}l@{}}%
\column{15}{@{}>{\hspre}l<{\hspost}@{}}%
\column{33}{@{}>{\hspre}l<{\hspost}@{}}%
\column{E}{@{}>{\hspre}l<{\hspost}@{}}%
\>[5]{}\Varid{isDep}{}\<[12]%
\>[12]{}\ \mathop{:}\ {}\<[12E]%
\>[15]{}\{\mskip1.5mu \Varid{k}\ \mathop{:}\ \mathbb{N}\mskip1.5mu\}\to \{\mskip1.5mu \Varid{ds}{}\<[33]%
\>[33]{}\ \mathop{:}\ \Conid{Vec}\;\Varid{k}\;\Conid{D}\mskip1.5mu\}\to \{\mskip1.5mu \Varid{m}\ \mathop{:}\ \mathbb{N}\mskip1.5mu\}\to \{\mskip1.5mu \Varid{ds'}\ \mathop{:}\ \Conid{Vec}\;\Varid{m}\;\Conid{D}\mskip1.5mu\}\to \{\mskip1.5mu \Varid{d}\ \mathop{:}\ \Conid{D}\mskip1.5mu\}\to {}\<[E]%
\\
\>[15]{}\Conid{PhysFunType}\;\Varid{k}\;\Varid{m}\;\Varid{ds}\;\Varid{ds'}\;\Varid{d}\to \Conid{IsDep}\;\Varid{d}\;\Varid{ds}{}\<[E]%
\\
\>[5]{}\Varid{isDep}\;(\Conid{PFT}\;h_2\;h_3\;\pi_{\ref{eq:pi0}})\mathrel{=}\pi_{\ref{eq:pi0}}{}\<[E]%
\ColumnHook
\end{hscode}\resethooks
and a function that computes the return type of functions of physical
quantities as encoded by \ensuremath{\Conid{PhysFunType}} values:
\begin{hscode}\SaveRestoreHook
\column{B}{@{}>{\hspre}l<{\hspost}@{}}%
\column{5}{@{}>{\hspre}l<{\hspost}@{}}%
\column{14}{@{}>{\hspre}c<{\hspost}@{}}%
\column{14E}{@{}l@{}}%
\column{17}{@{}>{\hspre}l<{\hspost}@{}}%
\column{35}{@{}>{\hspre}l<{\hspost}@{}}%
\column{E}{@{}>{\hspre}l<{\hspost}@{}}%
\>[5]{}\Conid{RetType}{}\<[14]%
\>[14]{}\ \mathop{:}\ {}\<[14E]%
\>[17]{}\{\mskip1.5mu \Varid{k}\ \mathop{:}\ \mathbb{N}\mskip1.5mu\}\to \{\mskip1.5mu \Varid{ds}{}\<[35]%
\>[35]{}\ \mathop{:}\ \Conid{Vec}\;\Varid{k}\;\Conid{D}\mskip1.5mu\}\to \{\mskip1.5mu \Varid{m}\ \mathop{:}\ \mathbb{N}\mskip1.5mu\}\to \{\mskip1.5mu \Varid{ds'}\ \mathop{:}\ \Conid{Vec}\;\Varid{m}\;\Conid{D}\mskip1.5mu\}\to \{\mskip1.5mu \Varid{d}\ \mathop{:}\ \Conid{D}\mskip1.5mu\}\to {}\<[E]%
\\
\>[17]{}\Conid{PhysFunType}\;\Varid{k}\;\Varid{m}\;\Varid{ds}\;\Varid{ds'}\;\Varid{d}\to \Conid{Type}{}\<[E]%
\\
\>[5]{}\Conid{RetType}\;\{\mskip1.5mu \Varid{ds}\mskip1.5mu\}\;\Varid{pft}\mathrel{=}\Conid{Q}\;(\Conid{ProdPows}\;\Varid{ds}\;(\Varid{snd}\;(\Varid{fst}\;(\Varid{isDep}\;\Varid{pft})))){}\<[E]%
\ColumnHook
\end{hscode}\resethooks
With these computations in place, we define an \ensuremath{\Varid{applyPi}} function that
constructs functions of physical quantities that are consistent with the
Pi theorem from arbitrary shape functions:
\begin{hscode}\SaveRestoreHook
\column{B}{@{}>{\hspre}l<{\hspost}@{}}%
\column{5}{@{}>{\hspre}l<{\hspost}@{}}%
\column{7}{@{}>{\hspre}l<{\hspost}@{}}%
\column{12}{@{}>{\hspre}l<{\hspost}@{}}%
\column{16}{@{}>{\hspre}l<{\hspost}@{}}%
\column{17}{@{}>{\hspre}c<{\hspost}@{}}%
\column{17E}{@{}l@{}}%
\column{20}{@{}>{\hspre}l<{\hspost}@{}}%
\column{34}{@{}>{\hspre}l<{\hspost}@{}}%
\column{E}{@{}>{\hspre}l<{\hspost}@{}}%
\>[5]{}\Varid{applyPi}\ \mathop{:}\ {}\<[16]%
\>[16]{}\{\mskip1.5mu \Varid{k}\ \mathop{:}\ \mathbb{N}\mskip1.5mu\}\to \{\mskip1.5mu \Varid{ds}{}\<[34]%
\>[34]{}\ \mathop{:}\ \Conid{Vec}\;\Varid{k}\;\Conid{D}\mskip1.5mu\}\to \{\mskip1.5mu \Varid{m}\ \mathop{:}\ \mathbb{N}\mskip1.5mu\}\to \{\mskip1.5mu \Varid{ds'}\ \mathop{:}\ \Conid{Vec}\;\Varid{m}\;\Conid{D}\mskip1.5mu\}\to \{\mskip1.5mu \Varid{d}\ \mathop{:}\ \Conid{D}\mskip1.5mu\}\to {}\<[E]%
\\
\>[16]{}(\Varid{pft}\ \mathop{:}\ \Conid{PhysFunType}\;\Varid{k}\;\Varid{m}\;\Varid{ds}\;\Varid{ds'}\;\Varid{d})\to (\Conid{Vec}\;\Varid{m}\;\Real\to \Real)\to {}\<[E]%
\\
\>[16]{}(\Conid{Vec}_{\scalebox{.5}{Q}}\;\Varid{k}\;\Varid{ds}\to \Conid{Vec}_{\scalebox{.5}{Q}}\;\Varid{m}\;\Varid{ds'}\to \Conid{RetType}\;\Varid{pft}){}\<[E]%
\\[\blanklineskip]%
\>[5]{}\Varid{applyPi}\;\{\mskip1.5mu \Varid{k}\mskip1.5mu\}\;\{\mskip1.5mu \Varid{ds}\mskip1.5mu\}\;\{\mskip1.5mu \Varid{m}\mskip1.5mu\}\;\{\mskip1.5mu \Varid{ds'}\mskip1.5mu\}\;\{\mskip1.5mu \Varid{d}\mskip1.5mu\}\;\Varid{pft}\;\varphi\;\Varid{as}\;\Varid{bs}\mathrel{=}{}\<[E]%
\\
\>[5]{}\hsindent{2}{}\<[7]%
\>[7]{}\mathbf{let}\;{}\<[12]%
\>[12]{}\Varid{ps}{}\<[17]%
\>[17]{}\mathrel{=}{}\<[17E]%
\>[20]{}\Varid{snd}\;(\Varid{fst}\;(\Varid{isDep}\;\Varid{pft})){}\<[E]%
\\
\>[12]{}\Varid{pis}{}\<[17]%
\>[17]{}\mathrel{=}{}\<[17E]%
\>[20]{}\Varid{map}_{\scalebox{.5}{Q}}\;\mu_{\Conid{SI}}\;(\Varid{makeAllDimLess}\;\Varid{bs}\;\Varid{as}\;(\Varid{areDep}\;\Varid{pft})){}\<[E]%
\\
\>[5]{}\hsindent{2}{}\<[7]%
\>[7]{}\mathbf{in}\;{}\<[12]%
\>[12]{}\Varid{prodPows}\;\Varid{as}\;\Varid{ps}\ensuremath{\triangleright}\varphi\;\Varid{pis}{}\<[E]%
\ColumnHook
\end{hscode}\resethooks
The construction follows directly from the formalization of the Pi
theorem of \cref{subsection:piexplained1.1}, see the definition of
\ensuremath{\Conid{Pi}_{\ref{eq:pi1}}}.

\DONE{Where is \ensuremath{\Varid{exponents}} defined?}
\DONE{Is this \ensuremath{\Varid{fst}\;(\Varid{exponents}\;h_2)} correct? In \ensuremath{\Conid{Pi}_{\ref{eq:pi1}}}, it looks like \ensuremath{\Varid{fst}\;(\pi_{\ref{eq:pi1}}\dots)}. Here, perhaps \ensuremath{\Varid{fst}\;(\Varid{isDep}\dots))}?}
Notice that, because the minimal DSL from \cref{section:dsl1} represents
dimension functions in terms of integer exponents, the physical quantity
\ensuremath{\Varid{applyPi}\;(\Conid{PFT}\;h_2\;h_3\;\pi_{\ref{eq:pi0}})\;\varphi\;\Varid{as}\;\Varid{bs}} is actually the \ensuremath{\Varid{fst}\;(\Varid{fst}\;h_2)}
power of \ensuremath{\Varid{f}\;\Varid{as}\;\Varid{bs}} as one can see from 
the definition
of \ensuremath{\Conid{Pi}_{\ref{eq:pi1}}}.

Notice also that the first argument taken by the data constructor \ensuremath{\Conid{PFT}},
a proof that the physical quantities \ensuremath{\Varid{as}} in \ensuremath{\Varid{applyPi}\;\Varid{pft}\;\varphi\;\Varid{as}\;\Varid{bs}} are
dimensionally independent, is not used in the definition of
\ensuremath{\Varid{applyPi}}. Its role is to prevent applications of the Pi theorem that
yield functions of physical quantities that are not consistent with the
Pi theorem.

The second argument of \ensuremath{\Varid{applyPi}} is a ``shape function'' \ensuremath{\varphi}. Its
type is \ensuremath{\Conid{Vec}\;\Varid{m}\;\Real\to \Real} rather than \ensuremath{\Conid{Vec}_{\scalebox{.5}{Q}}\;\Varid{m}\;(\Varid{replicate}\;\Varid{m}\;\Conid{DimLess})\to \Conid{Q}\;\Conid{DimLess}} as in the Pi theorem, see \ensuremath{\Conid{Pi}_{\ref{eq:pi1}}}. This is because of two
reasons: the first one is that functions of type \ensuremath{\Conid{Vec}_{\scalebox{.5}{Q}}\;\Varid{m}\;(\Varid{replicate}\;\Varid{m}\;\Conid{DimLess})\to \Conid{Q}\;\Conid{DimLess}} that respect the covariance principle are in fact
just functions of type \ensuremath{\Conid{Vec}\;\Varid{m}\;\Real\to \Real}.
The second reason is that, in applications of the Pi
theorem, the shape function is what needs to be identified from first
principles or approximated from empirical data, often via statistical
methods, approximation theory or machine learning. These methodologies
yield functions of real variables, not of physical quantities.

The idea of the approach encoded by \ensuremath{\Conid{PhysFunType}}-\ensuremath{\Varid{applyPi}} is that it
should yield \emph{verified} functions of physical quantities
\begin{hscode}\SaveRestoreHook
\column{B}{@{}>{\hspre}l<{\hspost}@{}}%
\column{5}{@{}>{\hspre}l<{\hspost}@{}}%
\column{21}{@{}>{\hspre}l<{\hspost}@{}}%
\column{39}{@{}>{\hspre}l<{\hspost}@{}}%
\column{E}{@{}>{\hspre}l<{\hspost}@{}}%
\>[5]{}\Varid{applyPiLemma}\ \mathop{:}\ {}\<[21]%
\>[21]{}\{\mskip1.5mu \Varid{k}\ \mathop{:}\ \mathbb{N}\mskip1.5mu\}\to \{\mskip1.5mu \Varid{ds}{}\<[39]%
\>[39]{}\ \mathop{:}\ \Conid{Vec}\;\Varid{k}\;\Conid{D}\mskip1.5mu\}\to \{\mskip1.5mu \Varid{m}\ \mathop{:}\ \mathbb{N}\mskip1.5mu\}\to \{\mskip1.5mu \Varid{ds'}\ \mathop{:}\ \Conid{Vec}\;\Varid{m}\;\Conid{D}\mskip1.5mu\}\to \{\mskip1.5mu \Varid{d}\ \mathop{:}\ \Conid{D}\mskip1.5mu\}\to {}\<[E]%
\\
\>[21]{}(\Varid{pft}\ \mathop{:}\ \Conid{PhysFunType}\;\Varid{k}\;\Varid{m}\;\Varid{ds}\;\Varid{ds'}\;\Varid{d})\to (\varphi\ \mathop{:}\ \Conid{Vec}\;\Varid{m}\;\Real\to \Real)\to {}\<[E]%
\\
\>[21]{}\Conid{IsCovariant}\;(\Varid{applyPi}\;\Varid{pft}\;\varphi){}\<[E]%
\ColumnHook
\end{hscode}\resethooks
Proving \ensuremath{\Varid{applyPiLemma}} is tedious but conceptually straightforward: we
have seen in \cref{subsection:piexplained1.1} that multiplication
between physical variables fulfills the covariance principle:
\ensuremath{\Varid{isCovariantMult}}. The same holds for division.
Integer powers and products of integer powers are just iterated
multiplication and division. Making a physical quantity dimensionless is
again division (of that quantity) by a suitable products of integer
powers (of other physical quantities).

Thus, the whole computation \ensuremath{\Varid{prodPows}\;\Varid{as}\;\Varid{ps}\ensuremath{\triangleright}\varphi\;\Varid{pis}} is a combination
of multiplication and division between physical quantities and scaling
and in \cref{subsection:quantities} we have seen that these operations
fulfil the covariance principle.
Indeed, it is not difficult to prove
\begin{joincode}%
\begin{hscode}\SaveRestoreHook
\column{B}{@{}>{\hspre}l<{\hspost}@{}}%
\column{5}{@{}>{\hspre}l<{\hspost}@{}}%
\column{7}{@{}>{\hspre}l<{\hspost}@{}}%
\column{15}{@{}>{\hspre}l<{\hspost}@{}}%
\column{21}{@{}>{\hspre}l<{\hspost}@{}}%
\column{38}{@{}>{\hspre}c<{\hspost}@{}}%
\column{38E}{@{}l@{}}%
\column{41}{@{}>{\hspre}l<{\hspost}@{}}%
\column{E}{@{}>{\hspre}l<{\hspost}@{}}%
\>[5]{}\Varid{powIsCovariant}{}\<[21]%
\>[21]{}\ \mathop{:}\ (\Varid{p}\ \mathop{:}\ \mathbb{Z}){}\<[38]%
\>[38]{}\to {}\<[38E]%
\\
\>[5]{}\hsindent{2}{}\<[7]%
\>[7]{}\Conid{Exists}\;{}\<[15]%
\>[15]{}(\Real\to \Real)\;(\lambda \rho\Rightarrow{}\<[41]%
\>[41]{}(\Varid{u}\ \mathop{:}\ \Conid{Units})\to \{\mskip1.5mu \Varid{d}\ \mathop{:}\ \Conid{D}\mskip1.5mu\}\to (\Varid{q}\ \mathop{:}\ \Conid{Q}\;\Varid{d})\to \mu_{\Varid{u}}\;(\Varid{pow}\;\Varid{q}\;\Varid{p})\mathrel{=}\rho\;(\mu_{\Varid{u}}\;\Varid{q})){}\<[E]%
\ColumnHook
\end{hscode}\resethooks
\begin{hscode}\SaveRestoreHook
\column{B}{@{}>{\hspre}l<{\hspost}@{}}%
\column{5}{@{}>{\hspre}l<{\hspost}@{}}%
\column{7}{@{}>{\hspre}l<{\hspost}@{}}%
\column{26}{@{}>{\hspre}c<{\hspost}@{}}%
\column{26E}{@{}l@{}}%
\column{29}{@{}>{\hspre}l<{\hspost}@{}}%
\column{44}{@{}>{\hspre}c<{\hspost}@{}}%
\column{44E}{@{}l@{}}%
\column{48}{@{}>{\hspre}l<{\hspost}@{}}%
\column{E}{@{}>{\hspre}l<{\hspost}@{}}%
\>[5]{}\Varid{prodPowsIsCovariant}{}\<[26]%
\>[26]{}\ \mathop{:}\ {}\<[26E]%
\>[29]{}\{\mskip1.5mu \Varid{k}\ \mathop{:}\ \mathbb{N}\mskip1.5mu\}\to (\Varid{ps}\ \mathop{:}\ \Conid{Vec}\;\Varid{k}\;\mathbb{Z})\to {}\<[E]%
\\
\>[5]{}\hsindent{2}{}\<[7]%
\>[7]{}\Conid{Exists}\;(\Conid{Vec}\;\Varid{k}\;\Real\to \Real)\;(\lambda \rho{}\<[44]%
\>[44]{}\Rightarrow{}\<[44E]%
\>[48]{}(\Varid{u}\ \mathop{:}\ \Conid{Units})\to \{\mskip1.5mu \Varid{ds}\ \mathop{:}\ \Conid{Vec}\;\Varid{k}\;\Conid{D}\mskip1.5mu\}\to (\Varid{as}\ \mathop{:}\ \Conid{Vec}_{\scalebox{.5}{Q}}\;\Varid{k}\;\Varid{ds})\to {}\<[E]%
\\
\>[48]{}\mu_{\Varid{u}}\;(\Varid{prodPows}\;\Varid{as}\;\Varid{ps})\mathrel{=}\rho\;(\Varid{map}_{\scalebox{.5}{Q}}\;\mu_{\Varid{u}}\;\Varid{as})){}\<[E]%
\ColumnHook
\end{hscode}\resethooks
\begin{hscode}\SaveRestoreHook
\column{B}{@{}>{\hspre}l<{\hspost}@{}}%
\column{5}{@{}>{\hspre}l<{\hspost}@{}}%
\column{7}{@{}>{\hspre}l<{\hspost}@{}}%
\column{15}{@{}>{\hspre}l<{\hspost}@{}}%
\column{23}{@{}>{\hspre}c<{\hspost}@{}}%
\column{23E}{@{}l@{}}%
\column{27}{@{}>{\hspre}l<{\hspost}@{}}%
\column{E}{@{}>{\hspre}l<{\hspost}@{}}%
\>[5]{}\Varid{makeAllDimLessIsCovariant}\ \mathop{:}\ \{\mskip1.5mu \Varid{k}\ \mathop{:}\ \mathbb{N}\mskip1.5mu\}\to \{\mskip1.5mu \Varid{ds}\ \mathop{:}\ \Conid{Vec}\;\Varid{k}\;\Conid{D}\mskip1.5mu\}\to {}\<[E]%
\\
\>[5]{}\hsindent{2}{}\<[7]%
\>[7]{}\{\mskip1.5mu \Varid{m}\ \mathop{:}\ \mathbb{N}\mskip1.5mu\}\to \{\mskip1.5mu \Varid{ds'}\ \mathop{:}\ \Conid{Vec}\;\Varid{m}\;\Conid{D}\mskip1.5mu\}\to (\Varid{areDep}\ \mathop{:}\ \Conid{AreDep}\;\Varid{ds'}\;\Varid{ds})\to {}\<[E]%
\\
\>[5]{}\hsindent{2}{}\<[7]%
\>[7]{}\Conid{Exists}\;{}\<[15]%
\>[15]{}(\Conid{Vec}\;\Varid{k}\;\Real\to \Conid{Vec}\;\Varid{m}\;\Real\to \Conid{Vec}\;\Varid{m}\;\Real)\;{}\<[E]%
\\
\>[15]{}(\lambda \rho{}\<[23]%
\>[23]{}\Rightarrow{}\<[23E]%
\>[27]{}(\Varid{u}\ \mathop{:}\ \Conid{Units})\to (\Varid{as}\ \mathop{:}\ \Conid{Vec}_{\scalebox{.5}{Q}}\;\Varid{k}\;\Varid{ds})\to (\Varid{bs}\ \mathop{:}\ \Conid{Vec}_{\scalebox{.5}{Q}}\;\Varid{m}\;\Varid{ds'})\to {}\<[E]%
\\
\>[27]{}\Varid{map}_{\scalebox{.5}{Q}}\;\mu_{\Varid{u}}\;(\Varid{makeAllDimLess}\;\Varid{bs}\;\Varid{as}\;\Varid{areDep})\mathrel{=}\rho\;(\Varid{map}_{\scalebox{.5}{Q}}\;\mu_{\Varid{u}}\;\Varid{as})\;(\Varid{map}_{\scalebox{.5}{Q}}\;\mu_{\Varid{u}}\;\Varid{bs})){}\<[E]%
\ColumnHook
\end{hscode}\resethooks
\end{joincode}
%
\DONE{We have a lot of \ensuremath{\mu_{\Varid{u}}} that we have sometimes abbreviated to
  \ensuremath{\mu_u}, sometimes not. Trying to do so via main.fmt as a test.}
The proofs depend on four results from \cref{section:dsl1}:
\ensuremath{\Varid{powdfLemma}}, \ensuremath{\Varid{ipowHomMult}}, \ensuremath{\mu{}\Conid{HomMult}}, \ensuremath{\Varid{dfDimLessLemma}}. The
first two are postulated in Idris and fully implemented in Agda.

In implementing \ensuremath{\Varid{makeAllDimLessIsCovariant}} in Idris we have run into an
issue with the type checker but the Agda proof is fully implemented.
It is perhaps worth mentioning that, in Agda with homogeneous  Martin-Löf
equality, implementing \ensuremath{\Varid{makeAllDimLessIsCovariant}} involves proving
\begin{hscode}\SaveRestoreHook
\column{B}{@{}>{\hspre}l<{\hspost}@{}}%
\column{E}{@{}>{\hspre}l<{\hspost}@{}}%
\>[B]{}\mu\;\{\mskip1.5mu \Varid{d}\mathrel{=}\Conid{DimLess}\mskip1.5mu\}\;\Varid{u}\;(\Varid{transport}\;\{\mskip1.5mu \Conid{P}\mathrel{=}\Conid{Q}\mskip1.5mu\}\;\Varid{dmdd}\;\Varid{comp})\equiv\mu\;\{\mskip1.5mu \Varid{d}\mathrel{=}\Varid{dmd}\mskip1.5mu\}\;\Varid{u}\;\Varid{comp}{}\<[E]%
\ColumnHook
\end{hscode}\resethooks
where \ensuremath{\Varid{dmdd}\ \mathop{:}\ \Varid{dmd}\equiv\Conid{DimLess}}, and \ensuremath{\Varid{comp}\mathrel{=}\Varid{pow}\;\Varid{b}\;\Varid{p}\mathbin{/}\Varid{prodPows}\;\Varid{as}\;\Varid{ps}} is
of type \ensuremath{\Conid{Q}\;\Varid{dmd}} (with variables interpreted as usual). This step
requires congruence modulo transport

\begin{hscode}\SaveRestoreHook
\column{B}{@{}>{\hspre}l<{\hspost}@{}}%
\column{E}{@{}>{\hspre}l<{\hspost}@{}}%
\>[B]{}(\Varid{p}\ \mathop{:}\ \Varid{a}_{1}\equiv\Varid{a}_{2})\to \Varid{f}\;\Varid{a}_{1}\;\Conid{Pa}_1\equiv\Varid{f}\;\Varid{a}_{2}\;(\Varid{transport}\;\Varid{p}\;\Conid{Pa}_1){}\<[E]%
\ColumnHook
\end{hscode}\resethooks
which can be implemented without UIP. Finally, \ensuremath{\Varid{prodPowsIsCovariant}} (abbreviated to \ensuremath{\Varid{pPCov}} in the code) and
\ensuremath{\Varid{makeAllDimLessIsCovariant}} (abbreviated to \ensuremath{\Varid{mADLCov}}) are combined to prove that functions constructed
with \ensuremath{\Varid{applyPi}} are in fact covariant:
%
\begin{hscode}\SaveRestoreHook
\column{B}{@{}>{\hspre}l<{\hspost}@{}}%
\column{5}{@{}>{\hspre}l<{\hspost}@{}}%
\column{7}{@{}>{\hspre}l<{\hspost}@{}}%
\column{12}{@{}>{\hspre}l<{\hspost}@{}}%
\column{14}{@{}>{\hspre}l<{\hspost}@{}}%
\column{18}{@{}>{\hspre}c<{\hspost}@{}}%
\column{18E}{@{}l@{}}%
\column{19}{@{}>{\hspre}l<{\hspost}@{}}%
\column{21}{@{}>{\hspre}l<{\hspost}@{}}%
\column{24}{@{}>{\hspre}l<{\hspost}@{}}%
\column{25}{@{}>{\hspre}c<{\hspost}@{}}%
\column{25E}{@{}l@{}}%
\column{28}{@{}>{\hspre}l<{\hspost}@{}}%
\column{64}{@{}>{\hspre}l<{\hspost}@{}}%
\column{68}{@{}>{\hspre}l<{\hspost}@{}}%
\column{77}{@{}>{\hspre}l<{\hspost}@{}}%
\column{81}{@{}>{\hspre}l<{\hspost}@{}}%
\column{83}{@{}>{\hspre}c<{\hspost}@{}}%
\column{83E}{@{}l@{}}%
\column{86}{@{}>{\hspre}l<{\hspost}@{}}%
\column{88}{@{}>{\hspre}c<{\hspost}@{}}%
\column{88E}{@{}l@{}}%
\column{89}{@{}>{\hspre}c<{\hspost}@{}}%
\column{89E}{@{}l@{}}%
\column{91}{@{}>{\hspre}l<{\hspost}@{}}%
\column{E}{@{}>{\hspre}l<{\hspost}@{}}%
\>[5]{}\Varid{applyPiLemma}\;\{\mskip1.5mu \Varid{k}\mskip1.5mu\}\;\{\mskip1.5mu \Varid{ds}\mskip1.5mu\}\;\{\mskip1.5mu \Varid{m}\mskip1.5mu\}\;\{\mskip1.5mu \Varid{ds'}\mskip1.5mu\}\;\Varid{pft}\;\varphi\mathrel{=}{}\<[E]%
\\
\>[5]{}\hsindent{2}{}\<[7]%
\>[7]{}\mathbf{let}\;{}\<[12]%
\>[12]{}\Varid{f}{}\<[18]%
\>[18]{}\mathrel{=}{}\<[18E]%
\>[21]{}\Varid{applyPi}\;\Varid{pft}\;\varphi;{}\<[68]%
\>[68]{}\quad{}\<[81]%
\>[81]{}\Varid{ps}{}\<[88]%
\>[88]{}\mathrel{=}{}\<[88E]%
\>[91]{}\Varid{snd}\;(\Varid{fst}\;(\Varid{isDep}\;\Varid{pft})){}\<[E]%
\\
\>[12]{}\rho_1{}\<[18]%
\>[18]{}\mathrel{=}{}\<[18E]%
\>[21]{}\Varid{fst}\;(\Varid{pPCov}\;\Varid{ps});{}\<[68]%
\>[68]{}\quad{}\<[81]%
\>[81]{}\Varid{prf}_{\!1}{}\<[88]%
\>[88]{}\mathrel{=}{}\<[88E]%
\>[91]{}\Varid{snd}\;(\Varid{pPCov}\;\Varid{ps}){}\<[E]%
\\
\>[12]{}\rho_3{}\<[18]%
\>[18]{}\mathrel{=}{}\<[18E]%
\>[21]{}\Varid{fst}\;(\Varid{mADLCov}\;(\Varid{areDep}\;\Varid{pft}));{}\<[68]%
\>[68]{}\quad{}\<[81]%
\>[81]{}\Varid{prf}_{\!3}{}\<[88]%
\>[88]{}\mathrel{=}{}\<[88E]%
\>[91]{}\Varid{snd}\;(\Varid{mADLCov}\;(\Varid{areDep}\;\Varid{pft})){}\<[E]%
\\
\>[12]{}\rho{}\<[18]%
\>[18]{}\mathrel{=}{}\<[18E]%
\>[21]{}\lambda \Varid{xs}\Rightarrow\lambda \Varid{ys}\Rightarrow\rho_1\;\Varid{xs}\ensuremath{\cdot}\varphi\;(\rho_3\;\Varid{xs}\;\Varid{ys});{}\<[E]%
\\
\>[12]{}\Varid{prf}{}\<[18]%
\>[18]{}\ \mathop{:}\ {}\<[18E]%
\>[21]{}({}\<[24]%
\>[24]{}(\Varid{u}\ \mathop{:}\ \Conid{Units})\to (\Varid{as}\ \mathop{:}\ \Conid{Vec}_{\scalebox{.5}{Q}}\;\Varid{k}\;\Varid{ds})\to (\Varid{bs}\ \mathop{:}\ \Conid{Vec}_{\scalebox{.5}{Q}}\;\Varid{m}\;\Varid{ds'})\to {}\<[89]%
\>[89]{}{}\<[89E]%
\\
\>[24]{}\mu_{\Varid{u}}\;(\Varid{f}\;\Varid{as}\;\Varid{bs})\mathrel{=}\rho\;(\Varid{map}_{\scalebox{.5}{Q}}\;\mu_{\Varid{u}}\;\Varid{as})\;(\Varid{map}_{\scalebox{.5}{Q}}\;\mu_{\Varid{u}}\;\Varid{bs}){}\<[89]%
\>[89]{}){}\<[89E]%
\\
\>[18]{}\mathrel{=}{}\<[18E]%
\>[21]{}\lambda \Varid{u}\Rightarrow\lambda \Varid{as}\Rightarrow\lambda \Varid{bs}\Rightarrow{}\<[E]%
\\
\>[12]{}\hsindent{2}{}\<[14]%
\>[14]{}\mathbf{let}\;{}\<[19]%
\>[19]{}\Varid{comp}{}\<[25]%
\>[25]{}\mathrel{=}{}\<[25E]%
\>[28]{}\Varid{makeAllDimLess}\;\Varid{bs}\;\Varid{as}\;(\Varid{areDep}\;\Varid{pft});{}\<[64]%
\>[64]{}\quad{}\<[77]%
\>[77]{}\Varid{pis}{}\<[83]%
\>[83]{}\mathrel{=}{}\<[83E]%
\>[86]{}\Varid{map}_{\scalebox{.5}{Q}}\;\mu_{\Conid{SI}}\;\Varid{comp}{}\<[E]%
\\
\>[12]{}\hsindent{2}{}\<[14]%
\>[14]{}\mathbf{in}\;{}\<[19]%
\>[19]{}\mu_{\Varid{u}}\;(\Varid{f}\;\Varid{as}\;\Varid{bs}){}\<[E]%
\\
\>[19]{}\hsindent{2}{}\<[21]%
\>[21]{}=\hspace{-3pt}\{\; \mu{}\Conid{HomScalR}\;\Varid{u}\;(\Varid{prodPows}\;\Varid{as}\;\Varid{ps})\;(\varphi\;\Varid{pis})\;\}\hspace{-3pt}={}\<[E]%
\\
\>[19]{}(\mu_{\Varid{u}}\;(\Varid{prodPows}\;\Varid{as}\;\Varid{ps})\ensuremath{\cdot}\varphi\;\Varid{pis}){}\<[E]%
\\
\>[19]{}\hsindent{2}{}\<[21]%
\>[21]{}=\hspace{-3pt}\{\; \Varid{cong}\;\{\mskip1.5mu \Varid{f}\mathrel{=}\lambda \Varid{x}\Rightarrow\Varid{x}\ensuremath{\cdot}\varphi\;\Varid{pis}\mskip1.5mu\}\;(\Varid{prf}_{\!1}\;\Varid{u}\;\Varid{as})\;\}\hspace{-3pt}={}\<[E]%
\\
\>[19]{}(\rho_1\;(\Varid{map}_{\scalebox{.5}{Q}}\;\mu_{\Varid{u}}\;\Varid{as})\ensuremath{\cdot}\varphi\;\Varid{pis}){}\<[E]%
\\
\>[19]{}\hsindent{2}{}\<[21]%
\>[21]{}=\hspace{-3pt}\{\; \Varid{cong}\;\{\mskip1.5mu \Varid{f}\mathrel{=}\lambda \Varid{x}\Rightarrow\rho_1\;(\Varid{map}_{\scalebox{.5}{Q}}\;\mu_{\Varid{u}}\;\Varid{as})\ensuremath{\cdot}\varphi\;\Varid{x}\mskip1.5mu\}\;(\Varid{measVectDimLessLemma}\;\Varid{comp})\;\}\hspace{-3pt}={}\<[E]%
\\
\>[19]{}(\rho_1\;(\Varid{map}_{\scalebox{.5}{Q}}\;\mu_{\Varid{u}}\;\Varid{as})\ensuremath{\cdot}\varphi\;(\Varid{map}_{\scalebox{.5}{Q}}\;\mu_{\Varid{u}}\;\Varid{comp})){}\<[E]%
\\
\>[19]{}\hsindent{2}{}\<[21]%
\>[21]{}=\hspace{-3pt}\{\; \Varid{cong}\;\{\mskip1.5mu \Varid{f}\mathrel{=}\lambda \Varid{x}\Rightarrow\rho_1\;(\Varid{map}_{\scalebox{.5}{Q}}\;\mu_{\Varid{u}}\;\Varid{as})\ensuremath{\cdot}\varphi\;\Varid{x}\mskip1.5mu\}\;(\Varid{prf}_{\!3}\;\Varid{u}\;\Varid{as}\;\Varid{bs})\;\}\hspace{-3pt}={}\<[E]%
\\
\>[19]{}(\rho\;(\Varid{map}_{\scalebox{.5}{Q}}\;\mu_{\Varid{u}}\;\Varid{as})\;(\Varid{map}_{\scalebox{.5}{Q}}\;\mu_{\Varid{u}}\;\Varid{bs}))\;{}\<[E]%
\\
\>[19]{}\hsindent{2}{}\<[21]%
\>[21]{}\Conid{QED}{}\<[E]%
\\
\>[5]{}\hsindent{2}{}\<[7]%
\>[7]{}\mathbf{in}\;{}\<[12]%
\>[12]{}\Conid{Evidence}\;\rho\;\Varid{prf}{}\<[E]%
\ColumnHook
\end{hscode}\resethooks
Notice that, in order to apply \ensuremath{\Varid{applyPi}} and construct a physical law
that fulfills the covariance principle, one needs to implement a value
of type \ensuremath{\Conid{PhysFunType}}.

Doing so, requires computing evidence of \ensuremath{\Conid{AreIndep}\;\Varid{ds}}, \ensuremath{\Conid{AreDep}\;\Varid{ds'}\;\Varid{ds}}
and \ensuremath{\Conid{IsDep}\;\Varid{d}\;\Varid{ds}} for the dimensions of the first and of the second
argument \ensuremath{\Varid{ds}} and \ensuremath{\Varid{ds'}} of \ensuremath{\Varid{f}} and for its return type~\ensuremath{\Varid{d}}. This is
equivalent to proving the linear independence of \ensuremath{\Varid{ds}} and to solving \ensuremath{\Varid{m}\mathbin{+}\mathrm{1}} systems of linear equations for \ensuremath{\mathbb{Z}} values.
Formalizing the fragment of linear algebra necessary to accomplish this
task goes well beyond the scope of this paper but we point the
interested reader to \citep{daSilvaALA}.

A more powerful approach towards applying the Pi theorem and
constructing functions of physical quantities that fulfil the covariance
principle would require modifications of the host language. We discuss
this possibility in \cref{subsection:adv}.

We conclude this section by applying the \ensuremath{\Conid{PhysFunType}} and \ensuremath{\Varid{applyPi}}
approach outlined above to the pendulum example from
\cref{subsection:dind}.

First we need to encode the type of functions that computes how the
second power of the period of a pendulum depends on its mass, its length
and on the acceleration of gravity:
\begin{hscode}\SaveRestoreHook
\column{B}{@{}>{\hspre}l<{\hspost}@{}}%
\column{5}{@{}>{\hspre}l<{\hspost}@{}}%
\column{7}{@{}>{\hspre}l<{\hspost}@{}}%
\column{12}{@{}>{\hspre}c<{\hspost}@{}}%
\column{12E}{@{}l@{}}%
\column{15}{@{}>{\hspre}l<{\hspost}@{}}%
\column{E}{@{}>{\hspre}l<{\hspost}@{}}%
\>[5]{}\Varid{tau2t}{}\<[12]%
\>[12]{}\ \mathop{:}\ {}\<[12E]%
\>[15]{}\Conid{PhysFunType}\;\mathrm{3}\;\mathrm{0}\;[\mskip1.5mu \Conid{Mass},\Conid{Length},\Conid{Acceleration}\mskip1.5mu]\;[\mskip1.5mu \mskip1.5mu]\;\Conid{Time}{}\<[E]%
\\
\>[5]{}\Varid{tau2t}{}\<[12]%
\>[12]{}\mathrel{=}{}\<[12E]%
\>[15]{}\Conid{PFT}\;h_2\;h_3\;\pi_{\ref{eq:pi0}}\;\mathbf{where}{}\<[E]%
\\
\>[5]{}\hsindent{2}{}\<[7]%
\>[7]{}h_2{}\<[12]%
\>[12]{}\ \mathop{:}\ {}\<[12E]%
\>[15]{}\Conid{AreIndep}\;[\mskip1.5mu \Conid{Mass},\Conid{Length},\Conid{Acceleration}\mskip1.5mu]{}\<[E]%
\\
\>[5]{}\hsindent{2}{}\<[7]%
\>[7]{}h_2{}\<[12]%
\>[12]{}\mathrel{=}{}\<[12E]%
\>[15]{}\Conid{Refl}{}\<[E]%
\\
\>[5]{}\hsindent{2}{}\<[7]%
\>[7]{}h_3{}\<[12]%
\>[12]{}\ \mathop{:}\ {}\<[12E]%
\>[15]{}\Conid{AreDep}\;[\mskip1.5mu \mskip1.5mu]\;[\mskip1.5mu \Conid{Mass},\Conid{Length},\Conid{Acceleration}\mskip1.5mu]{}\<[E]%
\\
\>[5]{}\hsindent{2}{}\<[7]%
\>[7]{}h_3{}\<[12]%
\>[12]{}\mathrel{=}{}\<[12E]%
\>[15]{}[\mskip1.5mu \mskip1.5mu]{}\<[E]%
\\
\>[5]{}\hsindent{2}{}\<[7]%
\>[7]{}\pi_{\ref{eq:pi0}}{}\<[12]%
\>[12]{}\ \mathop{:}\ {}\<[12E]%
\>[15]{}\Conid{IsDep}\;\Conid{Time}\;[\mskip1.5mu \Conid{Mass},\Conid{Length},\Conid{Acceleration}\mskip1.5mu]{}\<[E]%
\\
\>[5]{}\hsindent{2}{}\<[7]%
\>[7]{}\pi_{\ref{eq:pi0}}{}\<[12]%
\>[12]{}\mathrel{=}{}\<[12E]%
\>[15]{}\Conid{Evidence}\;(\mathrm{2},[\mskip1.5mu \mathrm{0},\mathrm{1},\mathbin{-}\mathrm{1}\mskip1.5mu])\;(\Varid{not2eq0},\Conid{Refl}){}\<[E]%
\ColumnHook
\end{hscode}\resethooks
Then we apply the Pi theorem and construct a function that computes the
second power of the period of a pendulum:
\begin{hscode}\SaveRestoreHook
\column{B}{@{}>{\hspre}l<{\hspost}@{}}%
\column{5}{@{}>{\hspre}l<{\hspost}@{}}%
\column{7}{@{}>{\hspre}l<{\hspost}@{}}%
\column{E}{@{}>{\hspre}l<{\hspost}@{}}%
\>[5]{}\Varid{tau2}\ \mathop{:}\ \Conid{Vec}_{\scalebox{.5}{Q}}\;\mathrm{3}\;[\mskip1.5mu \Conid{Mass},\Conid{Length},\Conid{Acceleration}\mskip1.5mu]\to \Conid{Vec}_{\scalebox{.5}{Q}}\;\mathrm{0}\;[\mskip1.5mu \mskip1.5mu]\to \Conid{RetType}\;\Varid{tau2t}{}\<[E]%
\\
\>[5]{}\Varid{tau2}\mathrel{=}\Varid{applyPi}\;\Varid{tau2t}\;\varphi\;\mathbf{where}{}\<[E]%
\\
\>[5]{}\hsindent{2}{}\<[7]%
\>[7]{}\varphi\ \mathop{:}\ \Conid{Vec}\;\mathrm{0}\;\Real\to \Real{}\<[E]%
\\
\>[5]{}\hsindent{2}{}\<[7]%
\>[7]{}\varphi\;[\mskip1.5mu \mskip1.5mu]\mathrel{=}\Varid{pow}\;(\mathrm{2}\ensuremath{\cdot}\pi)\;\mathrm{2.0}{}\<[E]%
\ColumnHook
\end{hscode}\resethooks
One can check that measurements of \ensuremath{\Varid{tau2}\;\Varid{m}\;\Varid{l}\;\Varid{g}} do not depend on the
mass \ensuremath{\Varid{m}} of the pendulum, are proportional to its length \ensuremath{\Varid{l}}
and inversely proportional to the gravity \ensuremath{\Varid{g}}.

The minimal DSL presented in \cref{section:dsl1}, the formulations of
the covariance principle and of the Pi theorem and the two-steps
approach towards constructing functions of physical quantities that
respect this principle are the main contributions of this paper towards
making mathematical physics (functional programming) more accessible to
computer scientists (modelers and physicists).

In the next section we discuss possible generalizations and desiderata
that go beyond the scope of this manuscript.

\section{Possible generalizations and extension}
\label{section:generalization}

We discuss possible generalizations and extensions of the DSL for
dimensionally consistent programming presented in \cref{section:dsl1}.
Some of these extensions (\cref{subsection:fun}) can be implemented
straightforwardly, while some (\cref{subsection:dpq}) come with substantial
disadvantages and are perhaps not worth being pursued. Other extensions
(\cref{subsection:adv}) go beyond the scope of this paper.

\subsection{Dimensions and physical quantities}
\label{subsection:dpq}

In \cref{section:dsl1} we have introduced the concrete data types \ensuremath{\Conid{D}}
and \ensuremath{\Conid{Q}} and encoded the notions of dimensions and of physical quantities
in the domain of mechanics and for the \ensuremath{\Conid{LTE}} (lengths, times and masses)
class of units of measurement.
This has allowed us to formalize basic notions of DA in type theory and
to apply dependent types to ensure the dimensional consistency of
expressions involving physical quantities (\cref{section:dsl1}) and
assist the formulation of relationships between physical quantities that
provably satisfy the covariance principle (\cref{section:piexplained1}).
In doing so, we have exploited a number of properties that values of
type \ensuremath{\Conid{D}} fulfilled by definition. Most prominently, the fact that equality of
dimensions is decidable. In \cref{subsection:dind} we also suggested
that \ensuremath{\Conid{D}} together with the binary operation \ensuremath{\Conid{Times}} form a group.
\paragraph*{The algebraic structure of dimensions.} \ It seems natural
to generalize the approach of \cref{section:dsl1} by putting forward the
algebraic structure of dimensions. As we will see, this has both
advantages and disadvantages. Again, we first discuss the generalization
in the domain of mechanics and for the \ensuremath{\Conid{LTE}} class of units of
measurement.

As done in \citep{botta_brede_jansson_richter_2021} for the notions of
functor and monad, we discuss the operations required for a type to be a
dimension as well as their laws through an Idris type class.
Encoding the algebraic structure of dimensions through type classes
and attempting generic implementations of \ensuremath{\Varid{df}} and of dimensional
judgments requires introducing a few language-specific details but is
a well-established method for discovering the potential drawbacks of
more abstract approaches than the one proposed in \cref{section:dsl1}.
In Idris, type classes are introduced through the \ensuremath{\mathbf{interface}}
keyword. For example
\begin{hscode}\SaveRestoreHook
\column{B}{@{}>{\hspre}l<{\hspost}@{}}%
\column{3}{@{}>{\hspre}l<{\hspost}@{}}%
\column{5}{@{}>{\hspre}l<{\hspost}@{}}%
\column{E}{@{}>{\hspre}l<{\hspost}@{}}%
\>[3]{}\mathbf{interface}\;\Conid{DecEq}\;\Varid{t}\;\mathbf{where}{}\<[E]%
\\
\>[3]{}\hsindent{2}{}\<[5]%
\>[5]{}\Varid{decEq}\ \mathop{:}\ (\Varid{x}_{1}\ \mathop{:}\ \Varid{t})\to (\Varid{x}_{2}\ \mathop{:}\ \Varid{t})\to \Conid{Dec}\;(\Varid{x}_{1}\mathrel{=}\Varid{x}_{2}){}\<[E]%
\ColumnHook
\end{hscode}\resethooks
explains what it means for a type \ensuremath{\Varid{t}} to be in \ensuremath{\Conid{DecEq}}, the class of
types for which propositional equality is decidable. The data
constructor \ensuremath{\Conid{Dec}} in the definition of \ensuremath{\Conid{DecEq}} is defined as
\begin{hscode}\SaveRestoreHook
\column{B}{@{}>{\hspre}l<{\hspost}@{}}%
\column{3}{@{}>{\hspre}l<{\hspost}@{}}%
\column{5}{@{}>{\hspre}l<{\hspost}@{}}%
\column{10}{@{}>{\hspre}c<{\hspost}@{}}%
\column{10E}{@{}l@{}}%
\column{13}{@{}>{\hspre}l<{\hspost}@{}}%
\column{E}{@{}>{\hspre}l<{\hspost}@{}}%
\>[3]{}\mathbf{data}\;\Conid{Dec}\ \mathop{:}\ \Conid{Type}\to \Conid{Type}\;\mathbf{where}{}\<[E]%
\\
\>[3]{}\hsindent{2}{}\<[5]%
\>[5]{}\Conid{Yes}{}\<[10]%
\>[10]{}\ \mathop{:}\ {}\<[10E]%
\>[13]{}(\Varid{prf}\ \mathop{:}\ \Varid{prop})\to \Conid{Dec}\;\Varid{prop}{}\<[E]%
\\
\>[3]{}\hsindent{2}{}\<[5]%
\>[5]{}\Conid{No}{}\<[10]%
\>[10]{}\ \mathop{:}\ {}\<[10E]%
\>[13]{}(\Varid{contra}\ \mathop{:}\ \Varid{prop}\to \Conid{Void})\to \Conid{Dec}\;\Varid{prop}{}\<[E]%
\ColumnHook
\end{hscode}\resethooks
A value of type \ensuremath{\Conid{Dec}\;\Varid{prop}} can only be constructed in two ways: either
by providing a proof of \ensuremath{\Varid{prop}} (a value of type \ensuremath{\Varid{prop}}) or by providing
a proof of \ensuremath{\Conid{Not}\;\Varid{prop}} (a function that maps values of type \ensuremath{\Varid{prop}} to
values of the empty type, that is, a contradiction). Thus, a value of
type \ensuremath{\Conid{Dec}\;(\Varid{x}_{1}\mathrel{=}\Varid{x}_{2})} is either a proof of \ensuremath{\Varid{x}_{1}\mathrel{=}\Varid{x}_{2}} or a proof of \ensuremath{\Conid{Not}\;(\Varid{x}_{1}\mathrel{=}\Varid{x}_{2})} which is what it means for the equality to be decidable.

Similarly, we can explain what it means for a type \ensuremath{\Conid{D}} to encode the notion of
dimension through a \ensuremath{\Conid{Dimension}} interface. As discussed in
\cref{subsection:df}, we need dimensional judgments or, more precisely,
equality in \ensuremath{\Conid{D}}, to be decidable.
This can be expressed by introducing \ensuremath{\Conid{Dimension}} as a \emph{refinement} of
\ensuremath{\Conid{DecEq}}:
\savecolumns
\begin{hscode}\SaveRestoreHook
\column{B}{@{}>{\hspre}l<{\hspost}@{}}%
\column{5}{@{}>{\hspre}l<{\hspost}@{}}%
\column{E}{@{}>{\hspre}l<{\hspost}@{}}%
\>[5]{}\mathbf{interface}\;\Conid{DecEq}\;\Conid{D}\Rightarrow\Conid{Dimension}\;\Conid{D}\;\mathbf{where}{}\<[E]%
\ColumnHook
\end{hscode}\resethooks
Perhaps confusingly, this says that \ensuremath{\Conid{Dimension}\;\Conid{D}} implies \ensuremath{\Conid{DecEq}\;\Conid{D}} or, in
other words, that being in \ensuremath{\Conid{DecEq}} is a necessary condition for being in
\ensuremath{\Conid{Dimension}}.
This condition is certainly not sufficient. We have seen in
\cref{subsection:df} that, as a minimum, we need to be able to define
dimensionless physical quantities and the 3 fundamental dimensions of
the \ensuremath{\Conid{LTE}} class:
\restorecolumns
\begin{hscode}\SaveRestoreHook
\column{B}{@{}>{\hspre}l<{\hspost}@{}}%
\column{7}{@{}>{\hspre}l<{\hspost}@{}}%
\column{E}{@{}>{\hspre}l<{\hspost}@{}}%
\>[7]{}\Conid{DimLess}\ \mathop{:}\ \Conid{D};\quad\Conid{Length}\ \mathop{:}\ \Conid{D};\quad\Conid{Time}\ \mathop{:}\ \Conid{D};\quad\Conid{Mass}\ \mathop{:}\ \Conid{D}{}\<[E]%
\ColumnHook
\end{hscode}\resethooks
Further, we need the \ensuremath{\Conid{Times}} and \ensuremath{\Conid{Over}} combinators
\restorecolumns
\begin{hscode}\SaveRestoreHook
\column{B}{@{}>{\hspre}l<{\hspost}@{}}%
\column{7}{@{}>{\hspre}l<{\hspost}@{}}%
\column{16}{@{}>{\hspre}c<{\hspost}@{}}%
\column{16E}{@{}l@{}}%
\column{19}{@{}>{\hspre}l<{\hspost}@{}}%
\column{E}{@{}>{\hspre}l<{\hspost}@{}}%
\>[7]{}\Conid{Times}{}\<[16]%
\>[16]{}\ \mathop{:}\ {}\<[16E]%
\>[19]{}\Conid{D}\to \Conid{D}\to \Conid{D}{}\<[E]%
\\
\>[7]{}\Conid{Over}{}\<[16]%
\>[16]{}\ \mathop{:}\ {}\<[16E]%
\>[19]{}\Conid{D}\to \Conid{D}\to \Conid{D}{}\<[E]%
\ColumnHook
\end{hscode}\resethooks
It is time to put forward some axioms. In \cref{subsection:dind} we
mentioned that \ensuremath{\Varid{d}\mathbin{`\Conid{Over}`}\Varid{d}} is equal to \ensuremath{\Conid{DimLess}} (for any \ensuremath{\Varid{d}\ \mathop{:}\ \Conid{D}}) and
that \ensuremath{\Conid{D}} is a group. The idea is that \ensuremath{\Conid{D}} together with the \ensuremath{\Conid{Times}}
operation is the free Abelian group generated by the fundamental
dimensions (which are also required to be not equal).  Thus, writing
\ensuremath{(\ensuremath{\cdot})} for \ensuremath{\Conid{Times}}, \ensuremath{(\mathbin{/})} for \ensuremath{\Conid{Over}}, and \ensuremath{\mathrm{1}} for \ensuremath{\Conid{DimLess}} we have
\DONE{I think these are easier to read with 0 for DimLess, * for `Times`,
  etc. Here is one way of doing that.}

\begin{hscode}\SaveRestoreHook
\column{B}{@{}>{\hspre}l<{\hspost}@{}}%
\column{7}{@{}>{\hspre}l<{\hspost}@{}}%
\column{36}{@{}>{\hspre}c<{\hspost}@{}}%
\column{36E}{@{}l@{}}%
\column{39}{@{}>{\hspre}l<{\hspost}@{}}%
\column{E}{@{}>{\hspre}l<{\hspost}@{}}%
\>[7]{}\Varid{isCommutativeTimes}{}\<[36]%
\>[36]{}\ \mathop{:}\ {}\<[36E]%
\>[39]{}\{\mskip1.5mu d_1,d_2\ \mathop{:}\ \Conid{D}\mskip1.5mu\}\to d_1\ensuremath{\cdot}d_2\mathrel{=}d_2\ensuremath{\cdot}d_1{}\<[E]%
\\[\blanklineskip]%
\>[7]{}\Varid{isAssociativeTimes}{}\<[36]%
\>[36]{}\ \mathop{:}\ {}\<[36E]%
\>[39]{}\{\mskip1.5mu d_1,d_2,d_3\ \mathop{:}\ \Conid{D}\mskip1.5mu\}\to (d_1\ensuremath{\cdot}d_2)\ensuremath{\cdot}d_3\mathrel{=}d_1\ensuremath{\cdot}(d_2\ensuremath{\cdot}d_3){}\<[E]%
\\[\blanklineskip]%
\>[7]{}\Varid{isLeftIdentityDimLess}{}\<[36]%
\>[36]{}\ \mathop{:}\ {}\<[36E]%
\>[39]{}\{\mskip1.5mu \Varid{d}\ \mathop{:}\ \Conid{D}\mskip1.5mu\}\to 1\ensuremath{\cdot}\Varid{d}\mathrel{=}\Varid{d}{}\<[E]%
\\[\blanklineskip]%
\>[7]{}\Varid{isRightIdentityDimLess}{}\<[36]%
\>[36]{}\ \mathop{:}\ {}\<[36E]%
\>[39]{}\{\mskip1.5mu \Varid{d}\ \mathop{:}\ \Conid{D}\mskip1.5mu\}\to \Varid{d}\ensuremath{\cdot}1\mathrel{=}\Varid{d}{}\<[E]%
\\[\blanklineskip]%
\>[7]{}\Varid{isLeftInverseDimLessOver}{}\<[36]%
\>[36]{}\ \mathop{:}\ {}\<[36E]%
\>[39]{}\{\mskip1.5mu \Varid{d}\ \mathop{:}\ \Conid{D}\mskip1.5mu\}\to (1/\Varid{d})\ensuremath{\cdot}\Varid{d}\mathrel{=}1{}\<[E]%
\\[\blanklineskip]%
\>[7]{}\Varid{isRightInverseDimLessOver}{}\<[36]%
\>[36]{}\ \mathop{:}\ {}\<[36E]%
\>[39]{}\{\mskip1.5mu \Varid{d}\ \mathop{:}\ \Conid{D}\mskip1.5mu\}\to \Varid{d}\ensuremath{\cdot}(1/\Varid{d})\mathrel{=}1{}\<[E]%
\ColumnHook
\end{hscode}\resethooks

\noindent
In order to derive \ensuremath{\Varid{d}/\Varid{d}\mathrel{=}1} one also needs
\ensuremath{\Conid{Times}} to associate with \ensuremath{\Conid{Over}} \citep{gibbonsPhD1991}:

\begin{hscode}\SaveRestoreHook
\column{B}{@{}>{\hspre}l<{\hspost}@{}}%
\column{7}{@{}>{\hspre}l<{\hspost}@{}}%
\column{E}{@{}>{\hspre}l<{\hspost}@{}}%
\>[7]{}\Varid{noPrec}\ \mathop{:}\ \{\mskip1.5mu d_1,d_2,d_3\ \mathop{:}\ \Conid{D}\mskip1.5mu\}\to (d_1\ensuremath{\cdot}d_2)/d_3\mathrel{=}d_1\ensuremath{\cdot}(d_2/d_3){}\<[E]%
\ColumnHook
\end{hscode}\resethooks

\DONE{We may even use this shorter notation further down.}%

\noindent
With \ensuremath{\Conid{DimLess}}, \ensuremath{\Conid{Times}} and \ensuremath{\Conid{Over}} one can implement the functions \ensuremath{\Conid{Pow}}
and \ensuremath{\Conid{ProdPows}} from \cref{subsection:dind} generically

\begin{hscode}\SaveRestoreHook
\column{B}{@{}>{\hspre}l<{\hspost}@{}}%
\column{5}{@{}>{\hspre}l<{\hspost}@{}}%
\column{7}{@{}>{\hspre}l<{\hspost}@{}}%
\column{14}{@{}>{\hspre}c<{\hspost}@{}}%
\column{14E}{@{}l@{}}%
\column{15}{@{}>{\hspre}l<{\hspost}@{}}%
\column{17}{@{}>{\hspre}l<{\hspost}@{}}%
\column{19}{@{}>{\hspre}l<{\hspost}@{}}%
\column{26}{@{}>{\hspre}l<{\hspost}@{}}%
\column{30}{@{}>{\hspre}c<{\hspost}@{}}%
\column{30E}{@{}l@{}}%
\column{33}{@{}>{\hspre}l<{\hspost}@{}}%
\column{37}{@{}>{\hspre}c<{\hspost}@{}}%
\column{37E}{@{}l@{}}%
\column{40}{@{}>{\hspre}l<{\hspost}@{}}%
\column{42}{@{}>{\hspre}l<{\hspost}@{}}%
\column{E}{@{}>{\hspre}l<{\hspost}@{}}%
\>[5]{}\Conid{Pow}{}\<[14]%
\>[14]{}\ \mathop{:}\ {}\<[14E]%
\>[17]{}\{\mskip1.5mu \Conid{D}\ \mathop{:}\ \Conid{Type}\mskip1.5mu\}\to \Conid{Dimension}\;\Conid{D}\Rightarrow\Conid{D}\to \mathbb{Z}\to \Conid{D}{}\<[E]%
\\
\>[5]{}\Conid{Pow}\;\Varid{d}\;\Varid{n}{}\<[14]%
\>[14]{}\mathrel{=}{}\<[14E]%
\>[17]{}\Varid{pow}\;\Varid{d}\;(\Varid{integerRec}\;\Varid{n})\;\mathbf{where}{}\<[E]%
\\
\>[5]{}\hsindent{2}{}\<[7]%
\>[7]{}\Varid{pow}\ \mathop{:}\ \{\mskip1.5mu \Varid{n}\ \mathop{:}\ \mathbb{Z}\mskip1.5mu\}\to \Conid{D}\to \Conid{IntegerRec}\;\Varid{n}\to \Conid{D}{}\<[E]%
\\
\>[5]{}\hsindent{2}{}\<[7]%
\>[7]{}\Varid{pow}\;\Varid{d}\;{}\<[19]%
\>[19]{}\Conid{IntegerZ}{}\<[30]%
\>[30]{}\mathrel{=}{}\<[30E]%
\>[33]{}1{}\<[E]%
\\
\>[5]{}\hsindent{2}{}\<[7]%
\>[7]{}\Varid{pow}\;\Varid{d}\;(\Conid{IntegerSucc}\;\Varid{m}){}\<[30]%
\>[30]{}\mathrel{=}{}\<[30E]%
\>[33]{}\Varid{pow}\;\Varid{d}\;\Varid{m}\ensuremath{\cdot}\Varid{d}{}\<[E]%
\\
\>[5]{}\hsindent{2}{}\<[7]%
\>[7]{}\Varid{pow}\;\Varid{d}\;(\Conid{IntegerPred}\;\Varid{m}){}\<[30]%
\>[30]{}\mathrel{=}{}\<[30E]%
\>[33]{}\Varid{pow}\;\Varid{d}\;\Varid{m}{}\<[42]%
\>[42]{}/\Varid{d}{}\<[E]%
\\[\blanklineskip]%
\>[5]{}\Conid{ProdPows}\ \mathop{:}\ \{\mskip1.5mu \Varid{n}\ \mathop{:}\ \mathbb{N}\mskip1.5mu\}\to \{\mskip1.5mu \Conid{D}\ \mathop{:}\ \Conid{Type}\mskip1.5mu\}\to \Conid{Dimension}\;\Conid{D}\Rightarrow\Conid{Vec}\;\Varid{n}\;\Conid{D}\to \Conid{Vec}\;\Varid{n}\;\mathbb{Z}\to \Conid{D}{}\<[E]%
\\
\>[5]{}\Conid{ProdPows}\;{}\<[15]%
\>[15]{}\Conid{Nil}\;{}\<[26]%
\>[26]{}\Conid{Nil}{}\<[37]%
\>[37]{}\mathrel{=}{}\<[37E]%
\>[40]{}1{}\<[E]%
\\
\>[5]{}\Conid{ProdPows}\;{}\<[15]%
\>[15]{}(\Varid{d}\mathbin{::}\Varid{ds})\;{}\<[26]%
\>[26]{}(\Varid{p}\mathbin{::}\Varid{ps}){}\<[37]%
\>[37]{}\mathrel{=}{}\<[37E]%
\>[40]{}\Conid{Pow}\;\Varid{d}\;\Varid{p}\ensuremath{\cdot}\Conid{ProdPows}\;\Varid{ds}\;\Varid{ps}{}\<[E]%
\ColumnHook
\end{hscode}\resethooks
and define derived dimensions as we did in \cref{subsection:df}:
%
\begin{hscode}\SaveRestoreHook
\column{B}{@{}>{\hspre}l<{\hspost}@{}}%
\column{5}{@{}>{\hspre}l<{\hspost}@{}}%
\column{19}{@{}>{\hspre}l<{\hspost}@{}}%
\column{E}{@{}>{\hspre}l<{\hspost}@{}}%
\>[5]{}\Conid{Velocity}{}\<[19]%
\>[19]{}\ \mathop{:}\ \{\mskip1.5mu \Conid{D}\ \mathop{:}\ \Conid{Type}\mskip1.5mu\}\to \Conid{Dimension}\;\Conid{D}\Rightarrow\Conid{D}{}\<[E]%
\\
\>[5]{}\Conid{Velocity}{}\<[19]%
\>[19]{}\mathrel{=}\Conid{Length}/\Conid{Time}{}\<[E]%
\ColumnHook
\end{hscode}\resethooks
\begin{hscode}\SaveRestoreHook
\column{B}{@{}>{\hspre}l<{\hspost}@{}}%
\column{5}{@{}>{\hspre}l<{\hspost}@{}}%
\column{19}{@{}>{\hspre}l<{\hspost}@{}}%
\column{E}{@{}>{\hspre}l<{\hspost}@{}}%
\>[5]{}\Conid{Acceleration}{}\<[19]%
\>[19]{}\ \mathop{:}\ \{\mskip1.5mu \Conid{D}\ \mathop{:}\ \Conid{Type}\mskip1.5mu\}\to \Conid{Dimension}\;\Conid{D}\Rightarrow\Conid{D}{}\<[E]%
\\
\>[5]{}\Conid{Acceleration}{}\<[19]%
\>[19]{}\mathrel{=}\Conid{Velocity}/\Conid{Time}{}\<[E]%
\\[\blanklineskip]%
\>[5]{}\Conid{Force}{}\<[19]%
\>[19]{}\ \mathop{:}\ \{\mskip1.5mu \Conid{D}\ \mathop{:}\ \Conid{Type}\mskip1.5mu\}\to \Conid{Dimension}\;\Conid{D}\Rightarrow\Conid{D}{}\<[E]%
\\
\>[5]{}\Conid{Force}{}\<[19]%
\>[19]{}\mathrel{=}\Conid{Mass}\ensuremath{\cdot}\Conid{Acceleration}{}\<[E]%
\\[\blanklineskip]%
\>[5]{}\Conid{Energy}{}\<[19]%
\>[19]{}\ \mathop{:}\ \{\mskip1.5mu \Conid{D}\ \mathop{:}\ \Conid{Type}\mskip1.5mu\}\to \Conid{Dimension}\;\Conid{D}\Rightarrow\Conid{D}{}\<[E]%
\\
\>[5]{}\Conid{Energy}{}\<[19]%
\>[19]{}\mathrel{=}\Conid{Mass}\ensuremath{\cdot}(\Conid{Velocity}\ensuremath{\cdot}\Conid{Velocity}){}\<[E]%
\\[\blanklineskip]%
\>[5]{}\Conid{Work}{}\<[19]%
\>[19]{}\ \mathop{:}\ \{\mskip1.5mu \Conid{D}\ \mathop{:}\ \Conid{Type}\mskip1.5mu\}\to \Conid{Dimension}\;\Conid{D}\Rightarrow\Conid{D}{}\<[E]%
\\
\>[5]{}\Conid{Work}{}\<[19]%
\>[19]{}\mathrel{=}\Conid{Force}\ensuremath{\cdot}\Conid{Length}{}\<[E]%
\ColumnHook
\end{hscode}\resethooks

\noindent
Notice, however, that the type of \ensuremath{\Conid{Velocity}}, \ensuremath{\Conid{Acceleration}}, etc. is
generic rather than specific. As a consequence, proving elementary
dimensional equalities requires some more work, as one would expect. For
example, in \cref{subsection:df}, we could assess the equivalence
between energy and mechanical work simply by
\begin{hscode}\SaveRestoreHook
\column{B}{@{}>{\hspre}l<{\hspost}@{}}%
\column{3}{@{}>{\hspre}l<{\hspost}@{}}%
\column{E}{@{}>{\hspre}l<{\hspost}@{}}%
\>[3]{}\Varid{check}_{1 }\ \mathop{:}\ \Conid{Energy}\mathrel{=}\Conid{Work}{}\<[E]%
\\
\>[3]{}\Varid{check}_{1 }\mathrel{=}\Conid{Refl}{}\<[E]%
\ColumnHook
\end{hscode}\resethooks
because the type of \ensuremath{\Conid{Energy}} and \ensuremath{\Conid{Work}} was fully \emph{defined}. A
similar proof based on the \emph{specification} \ensuremath{\Conid{Dimension}\;\Conid{D}} would look like
\DONE{Hard to read with the implicit type arguments to (=).}

\begin{hscode}\SaveRestoreHook
\column{B}{@{}>{\hspre}l<{\hspost}@{}}%
\column{5}{@{}>{\hspre}l<{\hspost}@{}}%
\column{13}{@{}>{\hspre}l<{\hspost}@{}}%
\column{17}{@{}>{\hspre}l<{\hspost}@{}}%
\column{40}{@{}>{\hspre}l<{\hspost}@{}}%
\column{E}{@{}>{\hspre}l<{\hspost}@{}}%
\>[5]{}\Varid{check}_{1 }\ \mathop{:}\ \{\mskip1.5mu \Conid{D}\ \mathop{:}\ \Conid{Type}\mskip1.5mu\}\to \Conid{Dimension}\;\Conid{D}\Rightarrow(\mathrel{=})\;\{\mskip1.5mu \Conid{A}\mathrel{=}\Conid{D}\mskip1.5mu\}\;\{\mskip1.5mu \Conid{B}\mathrel{=}\Conid{D}\mskip1.5mu\}\;\Conid{Energy}\;\Conid{Work}{}\<[E]%
\\
\>[5]{}\Varid{check}_{1 }{}\<[13]%
\>[13]{}\mathrel{=}{}\<[40]%
\>[40]{}(\Conid{Energy}){}\<[E]%
\\
\>[13]{}=\hspace{-3pt}\{\; {}\<[17]%
\>[17]{}\Conid{Refl}\;\}\hspace{-3pt}=\quad{}\<[40]%
\>[40]{}(\Conid{Mass}\ensuremath{\cdot}(\Conid{Velocity}\ensuremath{\cdot}\Conid{Velocity})){}\<[E]%
\\
\>[13]{}=\hspace{-3pt}\{\; {}\<[17]%
\>[17]{}\Conid{Refl}\;\}\hspace{-3pt}={}\<[40]%
\>[40]{}(\Conid{Mass}\ensuremath{\cdot}((\Conid{Length}/\Conid{Time})\ensuremath{\cdot}(\Conid{Length}/\Conid{Time}))){}\<[E]%
\\
\>[13]{}=\hspace{-3pt}\{\; {}\<[17]%
\>[17]{}\mathbf{?h_0}\;\}\hspace{-3pt}={}\<[40]%
\>[40]{}((\Conid{Mass}\ensuremath{\cdot}((\Conid{Length}/\Conid{Time})/\Conid{Time}))\ensuremath{\cdot}\Conid{Length}){}\<[E]%
\\
\>[13]{}=\hspace{-3pt}\{\; {}\<[17]%
\>[17]{}\Conid{Refl}\;\}\hspace{-3pt}={}\<[40]%
\>[40]{}((\Conid{Mass}\ensuremath{\cdot}(\Conid{Velocity}/\Conid{Time}))\ensuremath{\cdot}\Conid{Length}){}\<[E]%
\\
\>[13]{}=\hspace{-3pt}\{\; {}\<[17]%
\>[17]{}\Conid{Refl}\;\}\hspace{-3pt}={}\<[40]%
\>[40]{}((\Conid{Mass}\ensuremath{\cdot}\Conid{Acceleration})\ensuremath{\cdot}\Conid{Length}){}\<[E]%
\\
\>[13]{}=\hspace{-3pt}\{\; {}\<[17]%
\>[17]{}\Conid{Refl}\;\}\hspace{-3pt}={}\<[40]%
\>[40]{}(\Conid{Force}\ensuremath{\cdot}\Conid{Length}){}\<[E]%
\\
\>[13]{}=\hspace{-3pt}\{\; {}\<[17]%
\>[17]{}\Conid{Refl}\;\}\hspace{-3pt}={}\<[40]%
\>[40]{}(\Conid{Work}){}\<[E]%
\\
\>[13]{}\Conid{QED}{}\<[E]%
\ColumnHook
\end{hscode}\resethooks
We can omit all \ensuremath{\Conid{Refl}} steps, which are only there to guide the reader
(not the type checker) and perhaps make the type of \ensuremath{\Varid{check}_{1 }} more
readable.  However, filling in the \ensuremath{\mathbf{?h_0}} hole and completing the
proof requires invoking the axioms of \ensuremath{\Conid{Dimension}}, see the literate Idris code
that generates this document \citep{Pi2023}. Alas, we know that
implementing generic proofs can be awkward! \DONE{Add link to public
repo and name 'DSL.lidr'.}

Thus, a DSL for dimensionally consistent programming that does not rely
on a concrete representation of \ensuremath{\Conid{D}} like the one put forward in
\cref{section:dsl1}, would have to provide a library of proofs of
elementary equalities like the one between energy and work. Perhaps more
importantly, it would also have to provide proofs of elementary
inequalities, for example that \ensuremath{\Conid{Not}\;(\Conid{Force}\mathrel{=}\Conid{Energy})}. In
\cref{subsection:df}, we could assess this inequality by
\begin{hscode}\SaveRestoreHook
\column{B}{@{}>{\hspre}l<{\hspost}@{}}%
\column{3}{@{}>{\hspre}l<{\hspost}@{}}%
\column{E}{@{}>{\hspre}l<{\hspost}@{}}%
\>[3]{}\Varid{check}_{2 }\ \mathop{:}\ \Conid{Not}\;(\Conid{Force}\mathrel{=}\Conid{Energy}){}\<[E]%
\\
\>[3]{}\Varid{check}_{2 }\;\Conid{Refl}\;\Varid{impossible}{}\<[E]%
\ColumnHook
\end{hscode}\resethooks
Implementing a generic proof on the only basis that the type of \ensuremath{\Conid{Force}}
is equal to the type of \ensuremath{\Conid{Energy}} and that such type is in \ensuremath{\Conid{Dimension}} would
not be as easy. As a minimum, it would require extending the
\ensuremath{\Conid{Dimension}} interface with axioms that guarantee that the generators are not
equal.

Besides providing the basic grammar of the \ensuremath{\Conid{D}}-language, a data type in
\ensuremath{\Conid{Dimension}} also needs to provide a dimension function.
There are (at least) two ways of encoding this requirement. One is to
require \ensuremath{\Conid{Dimension}} to be equipped with a dimension function method
\begin{hscode}\SaveRestoreHook
\column{B}{@{}>{\hspre}l<{\hspost}@{}}%
\column{9}{@{}>{\hspre}l<{\hspost}@{}}%
\column{18}{@{}>{\hspre}c<{\hspost}@{}}%
\column{18E}{@{}l@{}}%
\column{21}{@{}>{\hspre}l<{\hspost}@{}}%
\column{E}{@{}>{\hspre}l<{\hspost}@{}}%
\>[9]{}\Varid{df}{}\<[18]%
\>[18]{}\ \mathop{:}\ {}\<[18E]%
\>[21]{}\Conid{D}\to (\Real_{+}^3\to \Real_{+}){}\<[E]%
\ColumnHook
\end{hscode}\resethooks
that fulfils the specifications corresponding to \ensuremath{\Varid{dfLemma1}}, \ensuremath{\Varid{dfLemma2}}
and \ensuremath{\Varid{dfLemma3}} from \cref{subsection:df}:
\begin{hscode}\SaveRestoreHook
\column{B}{@{}>{\hspre}l<{\hspost}@{}}%
\column{9}{@{}>{\hspre}l<{\hspost}@{}}%
\column{E}{@{}>{\hspre}l<{\hspost}@{}}%
\>[9]{}\Varid{dfSpec}_{1}\ \mathop{:}\ (\Varid{d}\ \mathop{:}\ \Conid{D})\to (\Varid{ls},\Varid{ls'}\ \mathop{:}\ \Real_{+}^3)\to \Varid{df}\;\Varid{d}\;\Varid{ls}\ensuremath{\cdot}\Varid{df}\;\Varid{d}\;\Varid{ls'}\mathrel{=}\Varid{df}\;\Varid{d}\;(\Varid{ls}\ensuremath{\cdot}\Varid{ls'}){}\<[E]%
\\[\blanklineskip]%
\>[9]{}\Varid{dfSpec}_{2}\ \mathop{:}\ (\Varid{d}\ \mathop{:}\ \Conid{D})\to \Varid{df}\;\Varid{d}\;\Varid{one3}\mathrel{=}\mathrm{1.0}{}\<[E]%
\\[\blanklineskip]%
\>[9]{}\Varid{dfSpec}_{3}\ \mathop{:}\ \{\mskip1.5mu \Varid{d}\ \mathop{:}\ \Conid{D}\mskip1.5mu\}\to \{\mskip1.5mu \Varid{u},\Varid{v},\Varid{w}\ \mathop{:}\ \Conid{Units}\mskip1.5mu\}\to \Varid{df}\;\Varid{d}\;(\Varid{fs}\;\Varid{u}\;\Varid{v})\ensuremath{\cdot}\Varid{df}\;\Varid{d}\;(\Varid{fs}\;\Varid{v}\;\Varid{w})\mathrel{=}\Varid{df}\;\Varid{d}\;(\Varid{fs}\;\Varid{u}\;\Varid{w}){}\<[E]%
\ColumnHook
\end{hscode}\resethooks
But instantiating \ensuremath{\Conid{Dimension}} with \ensuremath{\Varid{dfSpec}_{1}}, \ensuremath{\Varid{dfSpec}_{2}} and \ensuremath{\Varid{dfSpec}_{3}}
would have to rely on non-implementable assumptions (if \ensuremath{\Real_{+}} is
just an alias for floating-point numbers) or on a formalization of real
numbers. One way to circumvent this difficulty would be to restrict the
type of \ensuremath{\Varid{df}\;\Varid{d}} to \ensuremath{\Rational_{+}^3\to \Rational_{+}}. This is awkward and
conceptually unsatisfactory.

Alternatively, one could require \ensuremath{\Conid{Dimension}} to expose the integer
exponents of the dimension function \cref{eq:fun2}:
\begin{hscode}\SaveRestoreHook
\column{B}{@{}>{\hspre}l<{\hspost}@{}}%
\column{9}{@{}>{\hspre}l<{\hspost}@{}}%
\column{18}{@{}>{\hspre}c<{\hspost}@{}}%
\column{18E}{@{}l@{}}%
\column{21}{@{}>{\hspre}l<{\hspost}@{}}%
\column{E}{@{}>{\hspre}l<{\hspost}@{}}%
\>[9]{}\Varid{ds}{}\<[18]%
\>[18]{}\ \mathop{:}\ {}\<[18E]%
\>[21]{}\Conid{D}\to \Conid{Vec}\;\mathrm{3}\;\mathbb{Z}{}\<[E]%
\ColumnHook
\end{hscode}\resethooks
One could then define the dimension function associated with a \ensuremath{\Conid{D}} type
in \ensuremath{\Conid{Dimension}} on the basis of such exponents, as done in
\cref{subsection:quantities}. For example
\begin{hscode}\SaveRestoreHook
\column{B}{@{}>{\hspre}l<{\hspost}@{}}%
\column{5}{@{}>{\hspre}l<{\hspost}@{}}%
\column{7}{@{}>{\hspre}l<{\hspost}@{}}%
\column{E}{@{}>{\hspre}l<{\hspost}@{}}%
\>[5]{}\Varid{df}\ \mathop{:}\ \{\mskip1.5mu \Conid{D}\ \mathop{:}\ \Conid{Type}\mskip1.5mu\}\to \Conid{Dimension}\;\Conid{D}\Rightarrow\Conid{D}\to \Conid{Vec}\;\mathrm{3}\;\Real_{+}\to \Real_{+}{}\<[E]%
\\
\>[5]{}\Varid{df}\;\Varid{d}\;\Varid{ls}\mathrel{=}\Varid{foldr}\;(\ensuremath{\cdot})\;\mathrm{1.0}\;(\Varid{zipWith}\;\Varid{pow}\;\Varid{ls}\;\Varid{rds})\;\mathbf{where}{}\<[E]%
\\
\>[5]{}\hsindent{2}{}\<[7]%
\>[7]{}\Varid{rds}\ \mathop{:}\ \Conid{Vec}\;\mathrm{3}\;\Real{}\<[E]%
\\
\>[5]{}\hsindent{2}{}\<[7]%
\>[7]{}\Varid{rds}\mathrel{=}\Varid{map}\;\Varid{fromInt}\;(\Varid{ds}\;\Varid{d}){}\<[E]%
\ColumnHook
\end{hscode}\resethooks
The discussion above suggests that a DSL for DA and dimensionally
consistent programming should be based on a concrete implementation of
\ensuremath{\Conid{D}} like the one discussed in \cref{subsection:df}.
We argue that this conclusion holds even if we define \ensuremath{\Conid{Dimension}} as a
refinement of a \ensuremath{\Conid{Group}} type class. By a similar token, we argue that a
DSL for DA and dimensionally consistent programming should also be based
on a concrete implementation of the data type \ensuremath{\Conid{Q}} for physical
quantities, as proposed in \cref{subsection:quantities}.

\paragraph*{Beyond mechanics and the \ensuremath{\mathbf{LTE}} class.} \ Other parameters in
which it is natural to generalize the DSL from \cref{section:dsl1} are
the number of fundamental dimensions and the class of units of
measurement: we have introduced data types for dimensions, physical
quantities, etc. in the specific domain of mechanics (\ensuremath{\Varid{n}\mathrel{=}\mathrm{3}}
fundamental dimensions) and for the \ensuremath{\Conid{LTE}} (lengths, times and masses)
class of units of measurement. But the Pi theorem holds for an arbitrary
number of fundamental dimensions and, perhaps more importantly, for
arbitrary classes of units. It would be nice to have a general theory
that is parameterized on \ensuremath{\Varid{n}} and on the class of units and that can be
easily instantiated to other domains. The major obstacle towards such a
generalization is, as already mentioned, the need to formalize a
significant fraction of linear algebra. In \cref{section:dsl1} we have
defined the predicates \ensuremath{\Conid{IsDep}}, \ensuremath{\Conid{AreDep}}, \ensuremath{\Conid{AreIndep}} on \ensuremath{\Conid{D}}-values (and
the corresponding ones for physical quantities) for the specific case \ensuremath{\Varid{n}\mathrel{=}\mathrm{3}}. This is straightforward but implementing these predicates for a
generic \ensuremath{\Varid{n}} can only be done on the top of a library that formalizes the
basic notions of linear algebra. Building such a library is certainly
not trivial, see \citep{daSilvaALA} for a prototype implementation.

At the same time, it is probably worth keeping in mind the potential
danger associated with generalizations of DA: while the theory can be a
powerful methodology to reduce complexity and, up to a certain extent,
obtain physical laws ``for free'', it is not a ``generic'' tool. As Bridgman
has made very clear in Chapter 5 of his 1931 ``Dimensional Analysis''
book, applying DA to a domain whose fundamental laws have not yet been
formulated in a form independent of the size of the fundamental units
can be potentially very dangerous. At this early stage, we feel that the
best that we can do is to build small prototypes of consistent
``grammars'' of dimensions for specific domains and test how they work.

\medskip

Thus, our preliminary conclusion is that generalizing the approach of
\cref{section:dsl1} is useful to understand the algebraic structure of
dimension functions and the role of the number of fundamental dimensions
in the notions of dimensional (in)dependence that are at the core of the
Pi theorem but also comes with a number of practical disadvantages. This
is not really surprising if one keeps in mind that we still do not have
an established methodology for encoding fragments of well-understood
theories (for example, game theory, optimal control, linear algebra) in
dependently typed languages.

\subsection{Functions and their dimensions}
\label{subsection:fun}

As we have seen in
\cref{section:equations,section:dimensions,section:pi}, most
computations in mathematical physics involve operations on functions
between physical quantities. For example, the following function that
describes the position of a body moving in a constant gravitational
field:
\begin{hscode}\SaveRestoreHook
\column{B}{@{}>{\hspre}l<{\hspost}@{}}%
\column{E}{@{}>{\hspre}l<{\hspost}@{}}%
\>[B]{}\Varid{pos}\ \mathop{:}\ \Conid{Q}\;\Conid{Time}\to \Conid{Q}\;\Conid{Length}{}\<[E]%
\\
\>[B]{}\Varid{pos}\;\Varid{t}\mathrel{=}\mathrm{1}\mathbin{/}\mathrm{2}\ensuremath{\triangleleft}(\Varid{g}\ensuremath{\cdot}\Varid{pow}\;\Varid{t}\;\mathrm{2}){}\<[E]%
\ColumnHook
\end{hscode}\resethooks
Standard arithmetic operations between such functions can
be defined straightforwardly by lifting the corresponding operations on
\ensuremath{\Conid{Q}}-values. For example:
\begin{hscode}\SaveRestoreHook
\column{B}{@{}>{\hspre}l<{\hspost}@{}}%
\column{6}{@{}>{\hspre}c<{\hspost}@{}}%
\column{6E}{@{}l@{}}%
\column{9}{@{}>{\hspre}l<{\hspost}@{}}%
\column{E}{@{}>{\hspre}l<{\hspost}@{}}%
\>[B]{}(\mathbin{+}){}\<[6]%
\>[6]{}\ \mathop{:}\ {}\<[6E]%
\>[9]{}\{\mskip1.5mu d_1,d_2\ \mathop{:}\ \Conid{D}\mskip1.5mu\}\to (\Conid{Q}\;d_1\to \Conid{Q}\;d_2)\to (\Conid{Q}\;d_1\to \Conid{Q}\;d_2)\to (\Conid{Q}\;d_1\to \Conid{Q}\;d_2){}\<[E]%
\\
\>[B]{}(\mathbin{+})\;\Varid{f}_{1}\;\Varid{f}_{2}\mathrel{=}\lambda \Varid{q}\to\Varid{f}_{1}\;\Varid{q}\mathbin{+}\Varid{f}_{2}\;\Varid{q}{}\<[E]%
\ColumnHook
\end{hscode}\resethooks
Other operations, however, require some more care. Let \ensuremath{\Varid{pos'}} represent
the first derivative of \ensuremath{\Varid{pos}}. What should be the type of \ensuremath{\Varid{pos'}}? The
discussion at the end of \cref{section:dimensions} suggests that this
must be \ensuremath{\Conid{Q}\;\Conid{Time}\to \Conid{Q}\;\Conid{Velocity}}. We can build on the DSL of
\cref{section:dsl1} and specify types for dimensionally consistent
differentiation, for example
\begin{hscode}\SaveRestoreHook
\column{B}{@{}>{\hspre}l<{\hspost}@{}}%
\column{13}{@{}>{\hspre}c<{\hspost}@{}}%
\column{13E}{@{}l@{}}%
\column{16}{@{}>{\hspre}l<{\hspost}@{}}%
\column{E}{@{}>{\hspre}l<{\hspost}@{}}%
\>[B]{}\Varid{derivative}{}\<[13]%
\>[13]{}\ \mathop{:}\ {}\<[13E]%
\>[16]{}\{\mskip1.5mu d_0,d_1\ \mathop{:}\ \Conid{D}\mskip1.5mu\}\to (\Conid{Q}\;d_0\to \Conid{Q}\;d_1)\to \Conid{Q}\;d_0\to \Conid{Q}\;(d_1\mathbin{`\Conid{Over}`}d_0){}\<[E]%
\ColumnHook
\end{hscode}\resethooks
This is enough to support elementary dimensional judgments
\DONE{A bit confusing that the 2nd derivative of a body moving at constant speed is not zero!}
\begin{hscode}\SaveRestoreHook
\column{B}{@{}>{\hspre}l<{\hspost}@{}}%
\column{E}{@{}>{\hspre}l<{\hspost}@{}}%
\>[B]{}\Varid{pos''}\ \mathop{:}\ \Varid{typeOf}\;(\Varid{derivative}\;(\Varid{derivative}\;\Varid{pos})){}\<[E]%
\\
\>[B]{}\Varid{pos''}\;\Varid{t}\mathrel{=}\Varid{g}{}\<[E]%
\\[\blanklineskip]%
\>[B]{}\Varid{check}_{13}\ \mathop{:}\ \Varid{dimCodomain}\;\Varid{pos''}\mathrel{=}\Conid{Acceleration}{}\<[E]%
\\
\>[B]{}\Varid{check}_{13}\mathrel{=}\Conid{Refl}{}\<[E]%
\ColumnHook
\end{hscode}\resethooks
and reject definitions that are dimensionally inconsistent like  \ensuremath{\Varid{pos''}\;\Varid{t}\mathrel{=}\Varid{x}\mathbin{/}\Varid{t}}.
As one would expect, actually implementing \ensuremath{\Varid{derivative}} (and
dimensionally consistent operations for partial differentiation,
integration, ``nabla'' operators etc.) requires developing a small DSL
of elementary calculus for functions of \ensuremath{\Conid{Q}}-variables (for example, as
discussed in
\citep[Chapter 3]{JanssonIonescuBernardyDSLsofMathBook2022} for functions of real
variables) and thus involves making a number of non-trivial decisions.

\subsection{More advanced features: DA driven program derivation and
data analysis}
\label{subsection:adv}

Beside supporting program specification and verified programming,
dependently typed languages are also powerful tools for type-driven
program development \citep{idrisbook}. For example, the Idris system can
be queried interactively and asked to assist filling in holes like
\ensuremath{\mathbf{?h_0}}.
This suggests that, in principle, one should be able to exploit the
two-step construction discussed in \cref{subsection:piexplained1.3} to
make the type checker fit for assisting the implementation of physical
relationships that fulfil the Pi theorem.

For example, coming back to the simple pendulum example from
\cref{subsection:quantities,subsection:piexplained1.3}, we may want to
implement a function that computes the length \ensuremath{\Varid{penLen}\;\alpha\;\Varid{g}\;\Varid{m}\;\tau} of
a pendulum given the amplitude \ensuremath{\alpha} of the oscillations, the
acceleration of gravity \ensuremath{\Varid{g}}, its mass \ensuremath{\Varid{m}}, and its period of
oscillations \ensuremath{\tau}:
\begin{hscode}\SaveRestoreHook
\column{B}{@{}>{\hspre}l<{\hspost}@{}}%
\column{9}{@{}>{\hspre}c<{\hspost}@{}}%
\column{9E}{@{}l@{}}%
\column{12}{@{}>{\hspre}l<{\hspost}@{}}%
\column{E}{@{}>{\hspre}l<{\hspost}@{}}%
\>[B]{}\Varid{penLen}{}\<[9]%
\>[9]{}\ \mathop{:}\ {}\<[9E]%
\>[12]{}\Conid{Q}\;\Conid{DimLess}\to \Conid{Q}\;\Conid{Acceleration}\to \Conid{Q}\;\Conid{Mass}\to \Conid{Q}\;\Conid{Time}\to \Conid{Q}\;\Conid{Length}{}\<[E]%
\ColumnHook
\end{hscode}\resethooks
As a first step, we assess that \ensuremath{\Conid{Acceleration}}, \ensuremath{\Conid{Mass}} and \ensuremath{\Conid{Time}} are
independent and that \ensuremath{\Conid{DimLess}} depends on these three dimensions. This can
be done straightforwardly:
\begin{hscode}\SaveRestoreHook
\column{B}{@{}>{\hspre}l<{\hspost}@{}}%
\column{10}{@{}>{\hspre}c<{\hspost}@{}}%
\column{10E}{@{}l@{}}%
\column{13}{@{}>{\hspre}l<{\hspost}@{}}%
\column{E}{@{}>{\hspre}l<{\hspost}@{}}%
\>[B]{}\Varid{check}_{14}{}\<[10]%
\>[10]{}\ \mathop{:}\ {}\<[10E]%
\>[13]{}\Conid{AreIndep}\;[\mskip1.5mu \Conid{Acceleration},\Conid{Mass},\Conid{Time}\mskip1.5mu]{}\<[E]%
\\
\>[B]{}\Varid{check}_{14}{}\<[10]%
\>[10]{}\mathrel{=}{}\<[10E]%
\>[13]{}\Conid{Refl}{}\<[E]%
\\[\blanklineskip]%
\>[B]{}\Varid{check}_{15}{}\<[10]%
\>[10]{}\ \mathop{:}\ {}\<[10E]%
\>[13]{}\Conid{IsDep}\;\Conid{DimLess}\;[\mskip1.5mu \Conid{Acceleration},\Conid{Mass},\Conid{Time}\mskip1.5mu]{}\<[E]%
\\
\>[B]{}\Varid{check}_{15}{}\<[10]%
\>[10]{}\mathrel{=}{}\<[10E]%
\>[13]{}\Conid{Evidence}\;(\mathrm{1},[\mskip1.5mu \mathrm{0},\mathrm{0},\mathrm{0}\mskip1.5mu])\;(\Varid{not1eq0},\Conid{Refl}){}\<[E]%
\ColumnHook
\end{hscode}\resethooks
Then we define \ensuremath{\Varid{penLen}\;\alpha\;\Varid{g}\;\Varid{m}\;\tau} as a product of powers of \ensuremath{\Varid{g}},
\ensuremath{\Varid{m}} and \ensuremath{\tau}, consistently with the Pi theorem
\DONE{I suggest to use much shorter names for the metavariables - perhaps h1, h2, h3 (for ``hole 1'', etc.), or m1, m2, m3, or ...}
\begin{hscode}\SaveRestoreHook
\column{B}{@{}>{\hspre}l<{\hspost}@{}}%
\column{3}{@{}>{\hspre}l<{\hspost}@{}}%
\column{E}{@{}>{\hspre}l<{\hspost}@{}}%
\>[3]{}\Varid{penLen}\;\alpha\;\Varid{g}\;\Varid{m}\;\tau\mathrel{=}\Varid{pow}\;\Varid{g}\;\mathbf{?h_1}\ensuremath{\cdot}\Varid{pow}\;\Varid{m}\;\mathbf{?h_2}\ensuremath{\cdot}\Varid{pow}\;\tau\;\mathbf{?h_3}\ensuremath{\cdot}\Conid{Psi}\;\alpha{}\<[E]%
\ColumnHook
\end{hscode}\resethooks
and fill in the holes with exponents that match the type of \ensuremath{\Varid{penLen}}: 1,
0 and 2.
The function \ensuremath{\Conid{Psi}\ \mathop{:}\ \Conid{Q}\;\Conid{DimLess}\to \Conid{Q}\;\Conid{DimLess}} remains undefined,
but is the only part left to deduce from experiments. The type checker
will not accept other implementations of \ensuremath{\Varid{penLen}} but notice that we
are solving the system of equations \ensuremath{\mathbf{?h_1}\mathrel{=}\mathrm{1}}; \ensuremath{\mathbin{-}\mathrm{2}\ensuremath{\cdot}\mathbf{?h_1}\mathbin{+}\mathbf{?h_3}\mathrel{=}\mathrm{0}}; and \ensuremath{\mathbf{?h_2}\mathrel{=}\mathrm{0}} by hand, with little help from the
type checker!

A better approach would be to ask the type checker to solve the system
for us, e.g., by searching for suitable values of \ensuremath{\mathbf{?h_1}}, \ensuremath{\mathbf{?h_2}} and \ensuremath{\mathbf{?h_3}}
in certain ranges.
Perhaps more importantly, one would like the type checker to detect
situations in which the system has no solutions and recommend possible
decompositions of the arguments of physical relationships into lists of
dimensionally independent and dimensionally dependent components: the
\ensuremath{\Varid{as}} and the \ensuremath{\Varid{bs}} parameters of the Pi theorem.
As discussed in \cref{subsection:quantities,subsection:piexplained1.3},
this requires formulating a fragment of linear algebra in type theory
and modifications to the Idris type checker. But we think that it is an
effort that would be worth pursuing: it would provide domain experts
with alternative, dimensionally consistent views of data sets and help
practitioners reduce the complexity of data-based studies.

\subsection{Related and future work}
\label{subsection:relatedwork}

The best account we have found of the Pi theorem from a linear algebra
perspective is by \citet{CURTIS1982117}. In a compact 10-page paper
they both explain informally, and prove more formally, the Pi theorem
in a classical mathematical style. Unfortunately, there is no
connection to programming languages or types, only vector spaces,
linear transformations, and how to model dimensional analysis in this
setting.

In the key reference on the programming languages side,
\citet{10.1145/263699.263761} describes one way of combining
relational parametricity and units of measure.
Our paper shares some of the main ideas: the use of a typed
functional programming language,
integers as exponents of the base dimensions, and
parametric polymorphism. But there are also differences: where their
types are indexed by units, ours are indexed by dimensions; and where
they prove results using parametricity, we formulate the Pi theorem
using dependent types.
An interesting avenue for future work could be to combine these
approaches and perhaps use parametricity for dependent types
\citep{DBLP:journals/jfp/BernardyJP12} to gain further understanding
of the interplay between dimension analysis and strongly typed
functional programming.

\Citet{dimension-models} provide a categorical framework for semantic
models of a type theory that has special types for physical
quantities. Their starting point is \citep{10.1145/263699.263761} and
they provide a general notion programming language with physical
dimension types. They provide dimension polymorphism, but not regular
System-F-style polymorphism. The paper provides an impressive
collection of definitions, theorems, and examples which explain how
Abelian groups, fibrations, and groupoid actions can be used to build
models for languages with dimensions. Even though the setting is
different (and much more general), their key type \ensuremath{\Conid{Quantity}\;(\Conid{X})} is
basically the same as our \ensuremath{\Conid{Q}\;\Varid{d}}. They briefly mention an instance of
the Pi theorem, but do not formalize it.

The most recent related work is that by
\citet{doi:10.1142/9789811242380_0020} which uses dependent types to
extend type systems for units of measure from scalars to matrices.
Their approach is more algebraic, with graded semirings over a group
of physical dimensions, and it would be interesting to further extend
this to tensor calculus, a domain where we are also developing DSLs
\citep{bernardy_jansson_2025_tensors}.

\subsection{Physical laws revisited}
\label{subsection:covariance}

We conclude this section by going back to \cref{section:dimensions}
where we have argued that equations like Newton's second principle,
\cref{eq:newton1}, or the ideal gas law, \cref{eq:gas0}, summarize
empirical facts about measurements (or put forward axioms about such
measurements) of physical quantities.
Specifically, we have argued that \cref{eq:newton1} posits that
measurements of $F$ (force) are equal to the product of measurements of
$m$ (mass) and measurements of $a$ (acceleration).
In \cref{subsection:quantities} we have formalized the notions of
physical quantity and measurement and in \cref{section:piexplained1} we
have applied these notions to formulate the covariance principle for a
generic function between physical quantities, see \cref{figure:covariance}.

With this understanding, we can now give a clearer meaning to equations
\cref{eq:newton1,eq:gas0} from \cref{section:equations} and, more
generally, to equations that represent physical laws.
The idea is that these equations represent both functions between
physical quantities, for example
\begin{hscode}\SaveRestoreHook
\column{B}{@{}>{\hspre}l<{\hspost}@{}}%
\column{E}{@{}>{\hspre}l<{\hspost}@{}}%
\>[B]{}\Conid{F}\ \mathop{:}\ \Conid{Q}\;\Conid{Mass}\to \Conid{Q}\;\Conid{Acceleration}\to \Conid{Q}\;\Conid{Force}{}\<[E]%
\\
\>[B]{}\Conid{F}\;\Varid{m}\;\Varid{a}\mathrel{=}\Varid{m}\ensuremath{\cdot}\Varid{a}{}\<[E]%
\ColumnHook
\end{hscode}\resethooks
and also instances of the covariance principle as encoded generically in
\cref{figure:covariance}
\begin{equation*}
\ensuremath{\mu_u} \ (m \ensuremath{\cdot} a) = \ensuremath{\mu_u} \ m \ensuremath{\ensuremath{\cdot}} \ensuremath{\mu_u} \ a
\end{equation*}
or, with \ensuremath{\rho_{F}\ \mathop{:}\ \Real\to \Real\to \Real}, \ $\rho_F = (\ensuremath{\cdot})$
\begin{equation*}
\ensuremath{\mu_u} \ (F \ m \ a) = \rho_F \ (\ensuremath{\mu_u} \ m) \ (\ensuremath{\mu_u} \ a)
\end{equation*}
In this special form the covariance principle can actually be proved, as
discussed in \cref{subsection:quantities}.

\section{Conclusions}
\label{section:conclusions}

Specialization and the pervasive usage of computer-based modelling and
simulation in the physical sciences have widened the gap between the
languages of mathematical physics and modelling and those of mathematics
and functional programming.
This gap is a major obstacle to fruitful communication and to
interdisciplinary collaborations: computer-based modelling critically needs
precise specifications and dependently typed programming languages have
enough expressive power to support formulating such specifications.
But dependently typed programming languages are not (yet) well equipped
for encoding the ``grammar of dimensions'' which rules the languages of
mathematical physics and modelling.
Our contribution is a first step towards making FP more suitable for
developing applications in these domains.

We have studied the role of equations, laws and dimensions on the basis
of established examples from the physical sciences and from seminal works
in modelling.
We have analyzed the notions of dimension function, physical quantity
and units of measurement and we have provided an account of the theory
of physical similarity and of Buckingham's Pi theorem from the point of
view of computer science and FP.
Finally, we have proposed a small DSL that encodes these notions in
Idris, supports dimensional judgments, and leverages the type system
of the host language to provide tests of dimensional consistency,
dependence, and independence to ensure the consistency of expressions
involving physical quantities.

\DONE{Perhaps add:} We have formalized the covariance principle (the
requirement that physical laws be independent of the chosen system of
units) as a homomorphism between the algebra of physical quantities
and the algebra of real numbers (in \cref{figure:covariance}). We have
proved that elementary arithmetic operations preserve this property,
providing a rigorous type-theoretic foundation for dimensionally
consistent programming.

The DSL also supports classical, non-implementable formulations of
Buckingham's Pi theorem and we have derived one such formulation.
In \cref{subsection:piexplained1.3}, we have introduced a constructive
direction. This allows us to go beyond merely checking existing
equations: it enables the definition of functions that fulfil the
covariance principle by construction. Consequently, we have shown that
dependently typed languages can support DA-driven program derivation,
where the type checker assists in identifying valid physical laws from
empirical data constraints.

From this perspective, our work is also a contribution towards
understanding relativity principles through formalization.
In the physical sciences these principles are well understood and appreciated.
They have led to important applications in engineering and data science.
%
But it is not clear how relativity principles could be formulated in the
economic or biological sciences, and thus also in climate
science.
We believe that type theory and FP can contribute towards answering this
question.


\section*{Acknowledgments}
We are grateful to Prof.\ Jeremy Gibbons, Dr.\ Julian Newman, the
JFP editors and the anonymous reviewers whose comments and questions
have led to significant improvements of the original manuscript.
Guilherme da Silva contributed to the initial discussions and
developed a prototype Agda library for linear algebra.
The work presented in this paper heavily relies on free software, among others on Idris, Agda, Rocq, GHC, git, vi, Emacs, \LaTeX\ and on the FreeBSD and Debian GNU/Linux operating systems.
It is our pleasure to thank all developers of these excellent products.
This is TiPES contribution No 231. This project has received funding from
the European Union’s Horizon 2020 research and innovation programme
under grant agreement No 820970.

\bibliographystyle{ACM-Reference-Format}
\bibliography{references}
\label{lastpage01}
\end{document}